\documentclass[12pt]{article}

\usepackage{amssymb}
\usepackage{amsmath} 
\usepackage{mathrsfs} 
\usepackage{latexsym}

\usepackage{scalerel}
 
\usepackage{comment}
\usepackage{mathtools}
\usepackage{amsthm}
\usepackage{sectsty}
\sectionfont{\fontsize{13.5pt}{16.8pt}\selectfont}
\subsectionfont{\fontsize{12.5pt}{14.4pt}\selectfont}
\subsubsectionfont{\normalfont\itshape\fontsize{11pt}{13.2pt}\selectfont}

\usepackage{float}
\usepackage{enumitem}
\usepackage{graphicx}
\usepackage{subcaption}
\usepackage{caption}
\usepackage{arydshln}
\usepackage{changepage}
\usepackage[bottom]{footmisc}
\usepackage{booktabs, siunitx}
\pdfoutput=1
\usepackage{xcolor}
\definecolor{realnavy}{RGB}{0,0,102}
\definecolor{refred}{RGB}{153,0,0} 
\usepackage{hyperref}
\hypersetup{
  colorlinks=true,
  linkcolor=refred,
  citecolor=realnavy,
  filecolor=black,
  urlcolor=black
}
\usepackage{xr-hyper}
\usepackage{amsthm}
\usepackage{bbm}



\theoremstyle{plain}
\newtheorem{theorem}{Theorem}
\newtheorem{lemma}{Lemma}
\newtheorem{proposition}{Proposition}
\newtheorem{corollary}{Corollary}

\theoremstyle{definition}
\newtheorem{assumption}{Assumption}

\usepackage[T1]{fontenc}

\usepackage{setspace}
\usepackage[longnamesfirst]{natbib}
\usepackage{ulem}
\title{\textbf{A Joint Analysis of Sensitivity to Anticipation and Parallel Trends Violations\thanks{I thank Matt Masten, Arnaud Maurel, Michael Pollmann, Adam Rosen, and Chris Walker for their guidance on this project, as well as Michael Dinerstein and participants at the 2026 North American Summer Meeting of the Econometric Society (Atlanta, GA) and the Duke microeconometrics breakfast group for valuable comments and suggestions.}}}
\author{
Gianna Fenaroli\thanks{Department of Economics, Duke University, \texttt{gianna.fenaroli@duke.edu}}
}
\date{August 5, 2026}

\begin{document}
\maketitle

\begin{abstract}
\noindent Two key identifying assumptions used to justify difference-in-differences are parallel trends and no anticipation, yet both may fail in practice. I propose a class of assumptions that constitute deviations from no anticipation and derive closed-form, sharp bounds for several common treatment effect parameters while simultaneously relaxing parallel trends. Deviations from both assumptions are jointly disciplined using observed pre-trends. When some anticipation is imposed, the identified set under joint deviations can be shorter than under parallel trends violations alone. These bounds inform a sensitivity analysis assessing the robustness of qualitative conclusions to anticipation and parallel trends violations. For settings with a clean announcement window, I develop a benchmarking procedure to calibrate which values of the anticipation sensitivity parameters are most relevant for assessing robustness, making the practical interpretation of the multidimensional sensitivity analysis closer to that of a familiar one-dimensional sensitivity analysis. I illustrate with two empirical applications. 
\end{abstract}



\allowdisplaybreaks

\newpage
\onehalfspacing
\normalsize
\section{Introduction}
Difference-in-differences (DiD) is one of the most common methods for identifying and estimating causal effects with observational data \citep{ckz2014, dcdhtextbook, trendingdid, bccgbs2025}. Two of the main identifying assumptions used to justify DiD are parallel trends and no anticipation effects. Several recent papers have focused on evaluating the plausibility of the parallel trends assumption \citep{sgw2024, mtt2024} as well as assessing the robustness of conclusions to relaxations of parallel trends \citep{manskipepper2018, rambachanroth}. However, the no anticipation assumption, characterized by \citet{trendingdid} as an ``important and often hidden assumption required for identification,'' has received comparatively less attention. This assumption can be difficult to justify in observational studies because agents are generally forward-looking and may adjust their behavior based on beliefs about future treatment. Anticipation arises frequently in empirical settings, particularly when policies are announced prior to implementation\footnote{Examples include tort reform \citep{malanireif2015}, tax reform \citep{scholesetal1992, mertens2012empirical, cfj2022}, environmental regulation \citep{rittenhousezaragozawatkins2018}, active labor market policy \citep{cfjvdb2018}, in-work benefits reform \citep{bfvdk2011}, health care policy reform \citep{alpert2016}, and incentives-based counter-narcotics policy \citep{pvm2023}}. Therefore, it is important to explicitly consider no anticipation and assess the sensitivity of results to potential violations of this assumption.

This paper proposes a general class of assumptions on anticipation and derives closed-form, sharp bounds for several common treatment effect parameters under this class and relaxations of parallel trends. A central feature of the analysis is that it allows for simultaneous violations of the no anticipation and parallel trends assumptions. This is motivated by the observational equivalence (e.g. \citealt{malanireif2015}) of anticipation effects and parallel trends violations in the pre-treatment period: observed pre-trends may reflect anticipation effects, parallel trends violations, or some combination of the two. Rather than treating anticipation effects and parallel trends violations as unrelated, I use the observed pre-trends to discipline their joint violations. I use the resulting bounds to construct a multidimensional sensitivity analysis that assesses the robustness of qualitative conclusions to anticipation effects and parallel trends violations. I additionally develop a benchmarking exercise that leverages policy announcements to calibrate which values of the anticipation sensitivity parameters are most relevant for assessing robustness.

To relax the parallel trends assumption, I adopt the bounded variation approach developed by \cite{manskipepper2018} and later extended by \cite{rambachanroth}, who refer to it as the relative magnitude bounds approach. Both papers maintain no anticipation, so the observed pre-trends identify pre-treatment parallel trends violations. An additional challenge in the setting considered here is that anticipation effects are also allowed, so pre-treatment parallel trends violations are no longer identified by the observed pre-trends. To address this, I use the assumed bounds on anticipation effects to recover the set of pre-treatment parallel trends violations consistent with the observed pre-trends and the maintained assumptions. I then use this set to bound parallel trends violations in the post-treatment period. Allowing for anticipation thus affects the implied magnitude of parallel trends violations needed to rationalize the observed pre-trends. When the set of allowable anticipation effects excludes zero, allowing for simultaneous violations of no anticipation and parallel trends can produce a shorter identified set than relaxing parallel trends alone. This result shows that studying departures from the two assumptions jointly can yield implications that do not arise when each violation is studied in isolation.

I propose a joint sensitivity analysis, based on the breakdown frontier approach of \cite{mastenpoirier2020}, as a practical tool for assessing the robustness of empirical conclusions. Under simultaneous violations of the two assumptions, the qualitative conclusion of interest may hold for some combinations of anticipation effects and parallel trends violations but fail for others. The breakdown frontier partitions the assumption space into the region where a conclusion (e.g.\ $\text{ATT}_{1} > 0$) holds and the region where it fails. Intuitively, a result is more robust when it continues to hold under larger departures from both assumptions.

I illustrate the sensitivity analysis using two empirical applications. The first application studies prescription drug expenditures under the rollout of Medicare Part D using the data and empirical setting in \cite{engelhardtgruber2011}. There is scope for both anticipation and parallel trends violations: eligible individuals may defer prescription drug purchases until coverage begins \citep{alpert2016}, while drug spending may evolve differently for Medicare-eligible individuals than for younger, ineligible individuals even absent the reform. I assess the conclusion $\text{ATT}_{1}>0$ and use the two-year period between announcement and implementation to apply the benchmarking procedure. The procedure yields a negative calibrated range of anticipation effects, consistent with prior empirical evidence in \cite{alpert2016}. Throughout this range, smaller post-treatment parallel trends violations suffice to overturn the conclusion than under no anticipation, although the conclusion remains robust when these violations are no larger than the largest pre-treatment parallel trends violation.

The second application is from \cite{dinersteinsmith021}, which studies the impact of New York City’s Fair Student Funding reform on private school supply. Concerns about differential trends in untreated outcomes across neighborhoods create scope for parallel trends violations, while the extended litigation preceding the reform creates scope for anticipation. Because the setting does not feature a clean announcement window, I use it to illustrate how the economic context can guide interpretation of the sensitivity analysis. I evaluate whether $\text{ATT}_{1}$ can be bounded below by some value $\tau<0$. Relative to a baseline that allows only for parallel trends violations, allowing for increasingly negative anticipation effects implies that progressively smaller parallel trends violations suffice to overturn the conclusion; that is, the conclusion becomes less robust.

\subsection{Related Literature}

This paper builds on several literatures, including the extensive literature on DiD; see \cite{trendingdid}; \cite{dcdhtextbook}; \cite{bccgbs2025}. In particular, this paper contributes to the growing literature on assessing the DiD identifying assumptions, most of which focuses on parallel trends. To the best of my knowledge, this is the first paper to study identification under joint departures from the parallel trends and no anticipation assumptions. One of the earliest recognitions of the fragility of these assumptions is \cite{ashenfelter1978}, who observed that future participants in job training programs experienced a decline in earnings just before enrolling. \cite{heckmansmith1999} further analyzed this phenomenon, commonly known as ``Ashenfelter’s dip.'' More broadly, Ashenfelter's dip may in principle reflect both anticipatory behavior and violations of parallel trends.

A small but growing literature studies anticipation effects. \cite{malanireif2015} note that anticipation effects and parallel trends violations are observationally equivalent in observed pre-trends. A common approach to address anticipation is to impose a limited anticipation assumption, under which the researcher specifies when anticipation begins and allows anticipation only from that point until treatment (e.g.\ \citealt{malanireif2015}; \citealt{callawaysantanna2021}). Identification then requires parallel trends to hold over the periods affected by anticipation, as well as in the post-treatment period. In contrast, I study continuous, data-driven departures from no anticipation while also allowing for violations of parallel trends, nesting limited anticipation as a special case. \cite{gong2021} studies partial identification in the presence of anticipation effects, but maintains parallel trends throughout the pre- and post-treatment periods and interprets pre-trends as a ``result of unobservable anticipation activities.'' They also bound anticipation differently, using a mixture framework that treats the share of treated units that anticipate as a sensitivity parameter. \cite{agc2025} treat anticipation as a form of treatment misclassification and propose a bias-corrected estimator that relies on a parallel trends assumption. They also develop a two-step testing procedure for anticipation and parallel trends violations, which is used as a specification check rather than to relax these assumptions. \cite{fangliebl2025} study simultaneous inference in event studies that can accommodate \textit{a priori} researcher-specified bounds on bias from violations of parallel trends and/or no anticipation, but do not derive new identification results to obtain such bounds. My paper derives identification results for bounds under joint violations, which could in principle serve as inputs into their inferential procedure, making the two papers complementary.

Recent work has studied the plausibility of the parallel trends assumption and identification when the assumption may not hold. \cite{sgw2024} provide guidance for justifying the parallel trends assumption under various selection mechanisms, and \cite{mtt2024} examine the consistency of the parallel trends assumption with dynamic choice behavior. \cite{manskipepper2018} and \cite{rambachanroth} relax parallel trends by bounding post-treatment violations relative to pre-treatment violations, an approach I extend to allow for simultaneous departures from parallel trends and no anticipation.

This paper also relates to the literature on sensitivity analysis, particularly in DiD settings. For instance, \cite{manskipepper2018}, \cite{rambachanroth}, and \cite{bkkms2025} construct sensitivity analyses for DiD, but they all solely focus on violations of the parallel trends assumption. More broadly, my paper contributes to research extending the notion of a one-dimensional breakdown point \citep{horowitzmanski1995} to a two-dimensional breakdown frontier \citep{mastenpoirier2020}. For further background on sensitivity analysis and breakdown frontiers, see Appendix A of \cite{mastenpoirier2020} and \cite{sensitivitysurvey} for a general survey.

\subsection{Outline of Paper}
\noindent The rest of this paper is organized as follows. Section \ref{sec:bias} outlines the setup and notation, and also provides a stylized model to motivate the analysis of joint violations. Section \ref{sec:partialident} presents the assumptions on anticipation and violations of parallel trends, as well as the main identification results. Section \ref{sec:extensions} presents extensions. Section \ref{sec:sensitivity} outlines the joint sensitivity analysis and the benchmarking exercise. Section \ref{sec:empiricalillustration} illustrates the joint sensitivity analysis in the previously mentioned empirical applications to Medicare Part D and the FSF Reform. Section \ref{sec:conclusion} concludes. Auxiliary results and proofs are found in the Appendix.

\section{Setup and Motivation}
\label{sec:bias}

\subsection{Notation}\label{subsec:notation}
I begin by introducing the notation for the simultaneous adoption setting used in the main results. Section \ref{sec:extensions} extends the approach to staggered adoption and introduces the additional notation required for that setting. Time is indexed by $t \in \{-S, \ldots, T\}$ where $S \geq 1$ (so that we have at least one pre-trend) and $T \geq 1$. Periods $\{-S, \ldots , 0\}$ are pre-treatment and $\{1, \ldots , T\}$ are post-treatment. Treatment is binary: let $X_i \in \{0,1\}$ indicate whether unit $i$ belongs to the treated cohort (i.e.\ receives treatment starting at $t=1$). At the population level, I drop the unit subscript and let $X \in \{0,1\}$ denote the corresponding random variable. Treatment is an absorbing state, so treatment status at time $t$ is given by: $X_t \coloneqq X \cdot \mathbbm{1}[t \geq 1]$.

Let $Y_{it}(x_{-S}, \ldots , x_{T})$ denote the potential outcome of unit $i$ at time $t$ under a given treatment path $(x_{-S}, \ldots ,x_{T})$. Throughout, I maintain that a unit's potential outcomes depend only on its own treatment path and not on the treatment assignments of other units. Since there is a single treated cohort and a common adoption date, I simplify the notation as follows: $Y_{it}(1)$ is the potential outcome of unit $i$ at time $t$ under the path where treatment begins at $t=1$, and $Y_{it}(0)$ is the potential outcome of unit $i$ at time $t$ under receiving no treatment. I impose the following sufficient condition for the framework developed in this paper: comparison units realize the fully untreated outcome in each period used in the analysis. That is, for units with $X_i=0$, $Y_{it}=Y_{it}(0)$. This rules out incorrect anticipation of treatment among comparison units, a possibility studied in \cite{gong2021}. At the population level, I drop the subscript $i$ to denote the random variables $(Y_{t}(0),Y_{t}(1))$ whose distributions are the population distribution across units. 

The population-level average anticipation effect is given by
\begin{equation*}
    \varphi_{t} \coloneq \mathbb{E}[Y_{t}(1) - Y_{t}(0) \mid X = 1], \qquad t \in \{-S, \ldots ,0\}.
\end{equation*}
For deviations from parallel trends, define
$$\delta_{t} \coloneqq \mathbb{E}[Y_{t}(0) - Y_{t-1}(0) \mid X = 1] - \mathbb{E}[Y_{t}(0) - Y_{t-1}(0) \mid X = 0].$$
Intuitively, $\delta_{t}$ denotes the difference in consecutive-period outcome trends between the treated and comparison groups in the absence of treatment. In the standard $2 \times 2$ DiD model, the parallel trends assumption imposes that $\delta_{\text{1}} = 0$. 

For $t \in \{-(S-1), \ldots , 0 \} \eqqcolon \mathcal{S}$, consecutive pre-trends are denoted as follows:
\begin{align*}
    \Delta_t &\coloneqq \mathbb{E}[Y_t - Y_{t-1} \mid X=1] - \mathbb{E}[Y_t - Y_{t-1} \mid X=0].
\end{align*}
Lastly, denote the DiD estimand corresponding to post-treatment period $t$ by
$$\theta_t \coloneqq \mathbb{E}[Y_t - Y_{0} \mid X=1] - \mathbb{E}[Y_t - Y_{0} \mid X=0], \quad t \geq 1.$$ 

For exposition, when presenting the main results in Section \ref{sec:partialident}, I focus on the canonical average effect of treatment on the treated units at $t=1$:
$$\text{ATT}_{1} \coloneqq \mathbb{E}[Y_{1}(1) - Y_{1}(0) \mid X = 1].$$ 
This baseline case conveys the central intuition of the framework. Section \ref{sec:extensions} discusses how the results extend to other parameters of interest, including treatment effects further out in the post-treatment period (i.e. $\text{ATT}_{h}$ for $h > 1$), averages across such post-treatment effects, and staggered adoption parameters.

\subsection{Theoretical Motivation: A Stylized Model of Joint Violations}\label{sec:model}

To illustrate how anticipation effects and violations of parallel trends may arise simultaneously, consider as a simple example a job training program with periods $t \in \{-1,0,1\}$. In period $t=-1$, the policymaker assigns workers to training based on their current earnings. At the end of that same period (i.e.\ after the screening earnings are realized), they announce the policy and treatment assignment. Assigned workers receive training at $t=1$. Since assignment is fixed before workers learn about the program, selection and anticipation operate through distinct channels.

Suppose training raises workers' $t=1$ earnings by a common amount $\tau > 0$. Workers have rational expectations about this gain. After learning her future treatment status $x \in \{0,1\}$, the worker chooses $t=0$ search effort $e \geq 0$. Assume that one unit of search effort increases earnings at $t=0$ by one unit, measure utility in earnings units, and let $e^{2}\slash2$ denote the cost of effort. Define
\[ u_{0}(e;x) \coloneqq e - \frac{e^2}{2} + \beta \tau x(1-\kappa e), \qquad e_{0}^{\star}(x) \in \arg \max_{e \geq 0} u_{0}(e;x),  \]
where $\beta \in (0,1)$ is the discount factor and $\kappa \geq 0$. The model imposes that valuable future training reduces current search effort. To capture this mechanism, the term $\beta \tau x$ gives the present value of the future training gain, while the  term $-\beta \tau x \kappa e$ reduces the marginal payoff from current search when training is forthcoming. The parameter $\kappa$ governs the responsiveness of current search incentives to the value of future training: a one unit increase in the discounted value of future training ($\beta \tau x$) lowers the marginal payoff from searching today by $\kappa$ units. This mechanism of reduced job search effort upon learning of upcoming training is documented empirically by \cite{cfjvdb2018}, who find that notification of future training lowers the exit rate from unemployment.

The payoff is strictly concave in effort. Assume that $\beta \kappa \tau < 1$, which rules out the extreme response in which a worker who learns she will receive training stops searching entirely. The non-negativity constraint then does not bind and
\[  e^{\star}_{0}(x) = 1- \beta \kappa \tau x, \qquad e^{\star}_{0}(1) - e^{\star}_{0}(0) = - \beta \kappa \tau.\]

I next introduce the parallel trends violations. Adapting the earnings and selection structure in \citet{ashenfeltercard1985}, suppose fully untreated earnings satisfy
\[ Y_{it}(0) = A_i + \lambda_t + \varepsilon_{it}, \quad \varepsilon_{it} = \rho \varepsilon_{i,t-1} + \eta_{it}, \quad \eta_{it} \overset{\text{i.i.d.}}{\sim} \mathcal{N}(0,\sigma_\eta^2),\quad \rho\in(0,1),  \]
where $A_i$ is a worker fixed effect (e.g.\ permanent productivity), $\lambda_t$ is a common time effect, and $\varepsilon_{it}$ is a stationary AR(1) transitory earnings shock. Assume $A_i$ is independent of the transitory shock process. The policymaker assigns workers with sufficiently low earnings in $t=-1$ to training:\ $X_i = \mathbbm{1}[Y_{i,-1}(0) \leq c],$ where $c$ is a threshold determined by the policymaker\footnote{\cite{sgw2024} study a more general version of this selection rule, and Example 1 in \cite{mtt2024} takes the form $X_{i} = \mathbbm{1}[Y_{i0}(0) = 0]$: workers with a ``bad'' past outcome take up treatment. In their Appendix C, \cite{mtt2024} show how selection on past outcomes can arise from a dynamic optimization problem in which past outcomes enter the information set used to evaluate gains from treatment.}. Assume that threshold $c$ is chosen such that both treated and untreated workers are present in the population, i.e. $0 < \mathbb{P}(X=1) < 1$. Define $d \coloneqq \mathbb{E}[\varepsilon_{-1} \mid X=1] - \mathbb{E}[\varepsilon_{-1} \mid X=0]$. Under the stated conditions, one can show that $d < 0$ and 
\[  \delta_0 = -(1-\rho)d >0, \quad \delta_1 = -\rho(1-\rho)d >0. \]
Moreover, $\delta_1=\rho\delta_0$, so the model satisfies $|\delta_1|\leq M|\delta_0|$ for any $M\geq\rho$, including the common benchmark $M=1$. This provides a simple stylized justification for the relative magnitude restriction that will be used in the main identification analysis in Section \ref{sec:partialident}.

Finally, suppose that search at $t=0$ affects current earnings, but does not alter the $t=1$ earnings gain from training:
\[ Y_{i0}(x) = Y_{i0}(0) + e_{0}^{\star}(x) - e_{0}^{\star}(0), \qquad Y_{i1}(x) = Y_{i1}(0) + x\tau.  \]
Hence, $\text{ATT}_{1} = \tau$. Since assignment is announced only after period $t=-1$ earnings are realized, $\varphi_{-1}=0$, while 
\[ \varphi_{0} = -\beta \kappa \tau = k_0 \text{ATT}_{1}, \qquad k_0 \coloneqq -\beta \kappa.\]
Therefore, $\varphi_0 < 0$ if $\kappa >0$ (the possibility of $\kappa=0$ nests no anticipation): the model generates negative anticipation because the value of forthcoming training reduces current search effort and therefore lowers $t=0$ earnings for eventually treated workers. The parameter $k_0$ expresses the $t=0$ anticipation effect as a proportion of the treatment effect, one of the parameterizations studied in Section \ref{sec:partialident}. In this model, its magnitude depends on how much workers value the future gain and how responsive current search incentives are to that value. 

\section{Identification}
\label{sec:partialident}

In this section, I introduce and discuss the assumptions governing departures from no anticipation and parallel trends and derive sharp bounds on the $\text{ATT}_{1}$ under these assumptions. I also present the key parameter decompositions that inform these results and propose several parameterizations of the bounds on anticipation.

\subsection{Decompositions}
\begin{lemma}\label{lemma:attbiasdecomp23}
    The $\text{ATT}_{1}$ can be written in terms of anticipation effects $(\varphi_0)$ and post-treatment parallel trends violations $(\delta_1)$ as:
    $\text{ATT}_{1} = \theta_1 + \varphi_0 - \delta_1.$
\end{lemma}

\begin{corollary}\label{corr:ptvaenotident}
    Absent any restrictions on $\varphi_{0}$ and $\delta_{1}$, neither parameter is identified, and consequently the $\text{ATT}_{1}$ is completely not identified; i.e. the identified set for the $\text{ATT}_{1}$ is $\mathbb{R}$.
\end{corollary}

We can provide an analogous decomposition to Lemma \ref{lemma:attbiasdecomp23}, shifted back to an earlier period in the pre-treatment window. This result, Corollary \ref{prop:pretrenddecopmpsim}, implies that the data alone cannot distinguish whether differences in pre-treatment outcome trends between treated and untreated groups arise from violations of parallel trends, the presence of anticipation effects, or both. In other words, anticipation effects and violations of parallel trends in the pre-treatment period are observationally equivalent \citep{malanireif2015}. For this reason, considering simultaneous deviations from both assumptions is natural.

\begin{corollary}\label{prop:pretrenddecopmpsim}
   A pre-trend $\Delta_{t}$ can be decomposed as $\Delta_{t}  = \delta_t + \varphi_t - \varphi_{t-1}$, where $t \in \mathcal{S}$.
\end{corollary}

The observation in Corollary \ref{corr:ptvaenotident} motivates the need to impose restrictions on $\varphi_0$ and $\delta_1$ in order to obtain nontrivial identification of the $\text{ATT}_{1}$. Therefore, I next introduce a formal structure for deviations from parallel trends and no anticipation.

\subsection{Assumptions on Anticipation and Parallel Trends Violations}
A key element of this paper’s approach to simultaneously allowing for anticipation effects and parallel trends violations is the observation from Corollary \ref{prop:pretrenddecopmpsim} that a pre-trend can be decomposed into two distinct components:\ one coming from parallel trends violations and the other coming from anticipation. I begin by restricting anticipation through bounds on each increment $(\varphi_t - \varphi_{t-1})$. Together with an initial condition, these restrictions yield bounds on $\varphi_t$, and in particular $\varphi_0$, in terms of known objects (see Lemma \ref{aeexpression} in Appendix \ref{appendix:auxresults}).

\begin{assumption}[Initial Condition]\label{initialize}
In the first period in the data, anticipation effects are zero: $\varphi_{-S} = 0.$
\end{assumption}

 \begin{assumption}[Bounded Anticipation Increments]\label{abounds}
    $$(\varphi_{t} - \varphi_{t-1}) \in [\underline{A}_{t}, \overline{A}_{t}] \; \text{ for } \; t \in \mathcal{S},$$
where, for each $t \in \mathcal{S}$, $\underline{A}_{t}, \overline{A}_{t} \in \mathbb{R}$ are known and satisfy $\underline{A}_t \leq \overline{A}_t$.
\end{assumption}

\noindent It is convenient to directly bound the difference $(\varphi_t - \varphi_{t-1})$ rather than levels since I use the pre-trends to discipline how much I deviate from one assumption relative to the other. With Assumption \ref{initialize}, $\underline{A}_t = \overline{A}_t = 0$ for all $t \in \mathcal{S}$ yields the no anticipation baseline. 

Assumption \ref{abounds} defines a general class of assumptions on anticipation that nests several benchmark cases when combined with a parallel trends relaxation including the baseline case in which both assumptions hold, cases where one assumption holds and other may fail, simultaneous \textit{relaxations} of parallel trends and no anticipation (i.e. $0 \in [\underline{A}_t, \overline{A}_t]$ for all $t \in \mathcal{S}$), and deviations from both parallel trends and no anticipation, where some anticipation is assumed to be present (i.e. $0 \notin [\underline{A}_t, \overline{A}_t]$ for some $t \in \mathcal{S}$). The latter two cases are new cases studied in this paper. The discussion following Theorem \ref{thm:attmultpretrendsa} describes how each case affects the identified set for the $\text{ATT}_{1}$ and Section \ref{subsec:discussion} examines the last case in more detail.

Moreover, since the bounds in Assumption \ref{abounds} are allowed to vary across pre-treatment periods, the framework is flexible enough to accommodate simple modeling choices. For example, if one is willing to take a stance on the direction of anticipation, monotone anticipation in levels can be imposed by restricting the period-specific bounds on the increments $\varphi_t-\varphi_{t-1}$ to have the appropriate sign. 

For departures from parallel trends, the approach I use is inspired by the bounded variation/relative magnitude bounds approach from \cite{manskipepper2018} and \citet[hereafter RR]{rambachanroth} which rests on the premise that the pre-treatment period is informative about the post-treatment period. 

\begin{assumption}[Relative Magnitude Bounds]\label{assump:rmmultpretrends}
\begin{equation}\label{rmdelta1}
    | \delta_1 | \leq   M | \delta_{s^{\star}} |,
\end{equation}
    where $M \in \mathbb{R}_{\geq 0}$ is known and $s^{\star} \in \arg \max_{s \in \mathcal{S}} |\delta_s |$.
\end{assumption}
\noindent This assumption implies if the largest pre-treatment parallel trends violation is zero  ($\delta_{s^{\star}} = 0$), then the post-treatment parallel trends violation is zero ($\delta_1 = 0$). 

As detailed on pg.\ 2563 of RR, this approach bounds ``the maximum post-treatment violation of parallel trends between consecutive periods by $M$ times the maximum pre-treatment violation of parallel trends''. Therefore, the parameter $M$ maintains the same interpretation in my paper as it does in RR\footnote{Under RR’s no anticipation assumption, observed pre-trends coincide with pre-treatment parallel trends violations, so $M$ can be interpreted as bounding post-treatment parallel trends violations relative to either the largest observed pre-trend or the largest pre-treatment parallel trends violation. With anticipation, these objects need not coincide; throughout, I use the latter interpretation.}. However, implementing Assumption \ref{assump:rmmultpretrends} differs in the setting considered in my paper compared to RR. Since RR assume no anticipation, the pre-trends identify the pre-treatment parallel trends violations, so the right side of the inequality in \eqref{rmdelta1} is identified. However, since my paper allows for anticipation, the pre-trends can come from parallel trends violations, anticipation effects, or a mixture of both, so the pre-treatment parallel trends violations are not identified by the data. 

To illustrate how I address the fact that pre-treatment parallel trends violations are not identified by the pre-trends, consider a simple setting where $t \in \{-1,0,1\}$, so we have two pre-treatment periods and one post-treatment period. By Assumption \ref{initialize}, $\varphi_{-1} = 0$. Therefore, by Corollary \ref{prop:pretrenddecopmpsim}, we have $\Delta_{0} = \delta_0 + \varphi_0.$ Then, by Assumption \ref{assump:rmmultpretrends}:
$$|\delta_1| \leq M |\delta_0| = M |\Delta_0 - \varphi_0|.$$
By Assumption \ref{abounds}, we have $\varphi_0 \in [\underline{A}_{0}, \overline{A}_{0}],$ which in turn implies that $\delta_0$ lies in the following set, whose bounds are known objects
\begin{equation*}
    \delta_0 \in \{ \Delta_0 - a : a \in [\underline{A}_0, \overline{A}_0]   \} = [\Delta_0 - \overline{A}_0, \Delta_0 - \underline{A}_0].
\end{equation*}

\noindent Therefore, instead of using the pre-trends to directly bound the deviation from one assumption, here they are used to discipline simultaneous deviations from two assumptions. 

\subsection{Results}

Using the approach outlined above, I obtain sharp bounds on the $\text{ATT}_{1}$.

\begin{theorem}\label{thm:attmultpretrendsa}
    Suppose $(\Delta_{-(S-1)}, \ldots, \Delta_0, \theta_1 )$ are known and that Assumptions \ref{initialize}, \ref{abounds}, and \ref{assump:rmmultpretrends} hold. Then the identified set for the $\text{ATT}_{1}$ is given by
    \begin{align*}
        \mathcal{I}_{\text{ATT}_{1}}^{A} \coloneqq \biggl[ &\min_{r \in \mathcal{S}} \operatorname{LB}(r), \;  \max_{r \in \mathcal{S}} \operatorname{UB}(r)  \biggr],
    \end{align*}
    where
    \begin{align*}
        \text{LB}(r) \coloneqq \theta_1 + \sum_{j \in \mathcal{S} \setminus \{r \}} \underline{A}_{j} + \min_{a \in \{\underline{A}_r, \overline{A}_r \}} \bigl\{ a - M |\Delta_r - a| \bigr\}, \\
        \text{UB}(r) \coloneqq \theta_1 + \sum_{j \in \mathcal{S} \setminus \{r \}} \overline{A}_{j} + \max_{a \in \{\underline{A}_r, \overline{A}_r \}} \bigl\{ a + M |\Delta_r - a| \bigr\}.
    \end{align*}
\end{theorem}

Theorem \ref{thm:attmultpretrendsa} provides a closed-form characterization of the identified set that only requires optimization over small finite sets. To provide intuition for the bounds in Theorem \ref{thm:attmultpretrendsa}, I illustrate how they recover some of the cases outlined after Assumption \ref{abounds}. If we set $\underline{A}_t = \overline{A}_t = 0$ for all $t \in \mathcal{S}$, but allow $M>0$, then we are in the setting where no anticipation holds and parallel trends are possibly violated. Notice that this yields the RR setup where post-treatment parallel trends violations are bounded with respect to the largest pre-treatment parallel trends violations, which corresponds to the largest pre-trend (denoted $\Delta_{s^{\star}}$):
\begin{equation}\label{ptonly}
    \mathcal{I}_{\text{ATT}_{1}}^{A} = \bigl[ \theta_1 - M |\Delta_{s^{\star}}|, \theta_1 + M |\Delta_{s^{\star}}|    \bigr].
\end{equation}

The bounds in Theorem \ref{thm:attmultpretrendsa} nest limited anticipation by fixing $\ell \in \{0,\ldots,S\}$ and setting $M=0$ (which enforces $\delta_1 = 0$). If $\ell >0$, set $\underline{A}_t=\overline{A}_t=\Delta_t$ for $t\in\{-\ell+1,\ldots,0\}$; if $\ell<S$, set $\underline{A}_t=\overline{A}_t=0$ for $t\in\{-(S-1),\ldots,-\ell\}$. This point identifies $\text{ATT}_1$:
\begin{equation}\label{eqn:limitedanticip}
    \mathcal{I}_{\text{ATT}_{1}}^{A}
    =
    \begin{cases}
        \{\theta_1\}, & \ell=0,\\[0.4em]
        \left\{\theta_1 + \displaystyle\sum_{j=-\ell+1}^{0} \Delta_j\right\},
        & \ell\in\{1,\ldots,S\}.
    \end{cases}
\end{equation}
Note that $\ell = 0$ along with $M=0$ imposes the baseline case where both assumptions hold.

I next consider parameterizations in terms of data-driven sensitivity parameters. In each of the parameterizations the sensitivity parameters are unit-free proportions. In general, Theorem \ref{thm:attmultpretrendsa} applies directly to any parameterization of Assumption \ref{abounds} where, for fixed values of the sensitivity parameters, $\{\underline{A}_t, \overline{A}_t\}_{t \in \mathcal{S}}$ are known functions of observables and the admissible set for the anticipation increments $(\varphi_t - \varphi_{t-1})_{t \in \mathcal{S}}$ takes a product form: $\prod_{t \in \mathcal{S}} [\underline{A}_t , \overline{A}_t ]$. 

Consider the following parameterization of Assumption \ref{abounds}:
\begin{equation}\label{eq:pbounds}
    (\varphi_{t} - \varphi_{t-1}) = p_{t}^{\star} \Delta_{t} 
\end{equation}
where $p_{t}^{\star} \in [\underline{p}, \overline{p}]$ for all $t \in \mathcal{S}$ and $\underline{p}, \overline{p} \in \mathbb{R}$ are known with $\underline{p} \leq \overline{p}$. This implies Assumption \ref{abounds} holds with
$\underline{A}_{t} = \min \{ \underline{p} \Delta_t, \overline{p} \Delta_t \}$ and $\overline{A}_{t} = \max \{ \underline{p} \Delta_t, \overline{p} \Delta_t \}$.
In this setup, $p_t^\star$ represents the share of the pre-trend from period $t-1$ to $t$ that is due to the change in anticipation over that same increment. Bounding the consecutive \textit{difference} in anticipation as a proportion of the pre-trend is natural because $\Delta_t$ is itself a period-to-period difference. Parallel trends violations are similarly defined in terms of changes in selection over time. 

The bounds $(\underline{p}, \overline{p})$ do not vary across $t$. Variation in the bounds on $(\varphi_t - \varphi_{t-1})$ in \eqref{eq:pbounds} across pre-treatment periods instead comes from the observed pre-trends $\Delta_t$. An example of a choice for $[\underline{p}, \overline{p}]$ is $[0,1]$, which implies that the change in anticipation effects from $t-1$ to $t$ and the parallel trends violation are the same sign as the pre-trend, and no larger in magnitude. However, note that $[\underline{p}, \overline{p}]$ need not be $[0,1]$: the anticipation increment and parallel trends violation in Corollary \ref{prop:pretrenddecopmpsim} can be of opposite signs, and therefore potentially larger in magnitude than the pre-trend. 

However, this parameterization has three practical limitations. First, because the sign of the implied bounds on the anticipation increments depends on the sign of the corresponding pre-trend, this parameterization is best suited to settings in which the pre-trends have a stable sign, or where the researcher does not wish to impose a common sign restriction on anticipation increments across all pre-treatment periods. Second, since $p^{\star}_s$ is a ratio relative to $\Delta_s$, moderate values of the anticipation increment can correspond to large values of $p^{\star}_s$ when some pre-trends are close to zero, which is common in practice. Third, if $\Delta_t = 0$, then both $(\varphi_t - \varphi_{t-1}) = 0$ and $\delta_t = 0$, so this assumption class rules out cases where anticipation effects and parallel trends violations both exist but exactly offset one another.

A complementary parameterization instead benchmarks anticipation increments to the magnitude of the largest observed pre-trend:
\begin{equation}\label{eq:qbounds}
    (\varphi_{t} - \varphi_{t-1}) = q_{t}^{\star} \Delta^{\text{all}}
\end{equation}
where $\Delta^{\text{all}} \coloneqq \max_{s\in\mathcal S}|\Delta_s|$, $q_{t}^{\star} \in [\underline{q}, \overline{q}]$ for all $t \in \mathcal{S}$ and $\underline{q}, \overline{q} \in \mathbb{R}$ are known with $\underline{q} \leq \overline{q}$. This implies Assumption \ref{abounds} holds with $\underline{A}_{t} = \underline{q} \Delta^{\text{all}}$ and  $\overline{A}_{t} = \overline{q} \Delta^{\text{all}}.$ This parameterization has several practical advantages. First, the sign of anticipation is transparent:\ the sign of $q_{t}^{\star}$ directly determines the sign of the anticipation increment. Second, the denominator of the ratio $q_{t}^{\star} = (\varphi_{t} - \varphi_{t-1})\slash\Delta^{\text{all}}$ is chosen to reduce the risk of the near-zero denominator problem that can arise with ratios:\ $\Delta^{\text{all}}$ is the largest observed pre-trend in magnitude. However, if all observed pre-trends are close to zero, then $\Delta^{\text{all}}$ will also be close to zero. I use this parameterization in an empirical illustration in Section \ref{subsec:medicare}. 

Continuing to use sensitivity parameters that are proportions relative to a meaningful benchmark, another option is calibrating the anticipation effects relative to the treatment effect itself. This parameterization is particularly useful when researchers have institutional knowledge about how anticipation effects relate to the actual treatment effect. However, unlike the parameterizations in \eqref{eq:pbounds} and \eqref{eq:qbounds}, this parameterization bounds anticipation relative to an \textit{unknown} object, the $\text{ATT}_{1}$, and therefore Theorem \ref{thm:attmultpretrendsa} no longer immediately applies. Nonetheless, I show below we can still obtain an identified set for the $\text{ATT}_{1}$ under this parameterization, and therefore that the framework presented in this paper is not restricted to cases where the bounds on anticipation are known.

\begin{assumption}[Relative Anticipation Bounds: Treatment Effect Calibration]\label{assump:krelaxmultpre}
 $$\varphi_t = k_{t}^{\star} \cdot \text{ATT}_{1},$$
where $k_{t}^{\star} \in [\underline{k}_{t}, \overline{k}_{t}]$ for all $t \in \{-S, \ldots , 0\}$, and $\underline{k}_{t}, \overline{k}_{t} \in \mathbb{R}$ are known and satisfy $\underline{k}_t \leq \overline{k}_t$ for all $t \in \{-S, \ldots , 0\}$. 
\end{assumption}
Under this assumption class, we bound the anticipation effect at a particular point in time to align with the fact that the $\text{ATT}_{1}$ is itself defined in levels. An implication of Assumption \ref{assump:krelaxmultpre} is that if the treatment effect is zero ($\text{ATT}_{1} = 0$), then there are no anticipation effects ($\varphi_{t} = 0$ for all $t \in \{-S, \ldots, 0\}$). Notice that Assumption \ref{assump:krelaxmultpre} renders the initial condition in Assumption \ref{initialize} unnecessary since it specifies restrictions on $\varphi_t$ for every pre-treatment period, including $t=-S$. Moreover, if one is only interested in studying violations of the no anticipation assumption while maintaining parallel trends, this parameterization does not require observed pre-trends and can therefore be applied in a two-period DiD setting with a single pre-treatment period. 

In this setup, $k_{t}^{\star}$ represents the ratio of the anticipation effect in that period to the $\text{ATT}_{1}$. Assumption \ref{assump:krelaxmultpre} is in a similar spirit to Assumption 3.2 in \cite{gong2021}, which restricts the magnitude of the anticipation effect relative to the treatment effect. By introducing the proportion parameters $k^{\star}_{t}$, Assumption \ref{assump:krelaxmultpre} extends this idea to allow for flexible, period-specific calibration of anticipation effects. Assumption \ref{assump:krelaxmultpre} also exhibits a symmetry with Assumption \ref{assump:rmmultpretrends}: Assumption \ref{assump:rmmultpretrends} uses pre-treatment parallel trends violations to bound the post-treatment violation, while Assumption \ref{assump:krelaxmultpre} uses the post-treatment effect to bound (pre-treatment) anticipation effects.

Define the set of indices $J(r) \coloneqq \{0, r, r-1\}$, when $r=0$, then $J(0) = \{0,-1\}$. Under Assumption \ref{assump:krelaxmultpre}, the expression for the $\text{ATT}_{1}$ is a fraction. To obtain finite bounds on the $\text{ATT}_{1}$, I impose the following additional assumption:
\begin{assumption}[Nonzero Denominator Condition]\label{assump:nonzerodenomk}
    $$1 - k_0 - m(k_r - k_{r-1}) \neq 0$$
for every $r \in \mathcal{S}$ and every $\bigl( (k_j)_{j \in J(r)}, m   \bigr) \in \prod_{j \in J(r)} [\underline{k}_j, \overline{k}_j] \times [-M,M]$.
\end{assumption}

If Assumption \ref{assump:nonzerodenomk} did not hold, then one or both of the bounds of the identified set in Theorem \ref{thm:attmultpretrendsk} below would equal infinity. It is also important to note that Assumption \ref{assump:nonzerodenomk} is not innocuous; it imposes nontrivial restrictions on the feasible set one can consider for the sensitivity parameters. For additional discussion of this assumption, see Appendix \ref{appendix:assumpnonzero}.

Sign-based conclusions (e.g.\ $\text{ATT}_{1}>0$ or $\text{ATT}_{1}<0$) yield breakdown values of $M$ that do not vary with the bounds imposed on the $k_t$ parameters, except insofar as those bounds change the admissible range of $M$. Under Assumption \ref{assump:nonzerodenomk}, the denominator in the expressions for the bounds in Theorem \ref{thm:attmultpretrendsk} below has a constant sign over the feasible set, as it cannot cross zero. Therefore, for fixed admissible values of $M$, the sign of the bounds is determined entirely by the numerator, which does not depend on the $k_t$ parameters.

This invariance, however, is a special case. For conclusions of the form $\text{ATT}_{1} > \tau$, for example, with $\tau \neq 0$, the bounds in Theorem \ref{thm:attmultpretrendsk} below do vary with the choice of bounds on the $k_s$ terms and can therefore be used to construct a meaningful sensitivity analysis. For instance, they allow one to assess the robustness of conclusions that rule out large negative effects, which I illustrate in Section \ref{subsec:privateschools}.

\begin{theorem}\label{thm:attmultpretrendsk}
Suppose $(\Delta_{-(S-1)},\ldots,\Delta_0,\theta_1)$ are known and that 
Assumptions \ref{assump:rmmultpretrends}, \ref{assump:krelaxmultpre}, and \ref{assump:nonzerodenomk} hold. Then the identified set for $\text{ATT}_{1}$ is
\[
\mathcal{I}^{k}_{\text{ATT}_{1}}
\;\coloneqq\;
\left[
\min_{r\in\mathcal{S}} \text{LB}(r),\;
\max_{r\in\mathcal{S}} \text{UB}(r)
\right],
\]
where for each $r\in\mathcal{S}$,
\begin{align*}
\text{LB}(r)
&\coloneqq
\min_{\substack{
k_j \in \{\underline{k}_j, \overline{k}_j \}, \; j \in J(r)\\
m\in\{-M,M\}}}
\frac{\theta_1 - m\,\Delta_r}{
1 - k_0 - m\,(k_r - k_{r-1})
},
\\[1em]
\text{UB}(r)
&\coloneqq
\max_{\substack{
k_j \in \{\underline{k}_j, \overline{k}_j \}, \; j \in J(r)\\
m\in\{-M,M\}}}
\frac{\theta_1 - m\,\Delta_r}{
1 - k_0 - m\,(k_r - k_{r-1})
}.
\end{align*}
\end{theorem}

As in Theorem \ref{thm:attmultpretrendsa}, it suffices to evaluate the objective at the corner points of the feasible set. Theorem \ref{thm:attmultpretrendsk} delivers a closed-form characterization of the identified set by reducing the problem to optimization over small finite sets. I use this parameterization in an empirical illustration in Section \ref{subsec:privateschools}.

\subsection{Special Case}\label{subsec:discussion}
If we rule out no anticipation by assumption (i.e.\ the set of allowable anticipation effects excludes zero) simultaneous deviations from parallel trends and no anticipation can produce shorter identified sets than relaxing parallel trends alone. To illustrate, consider the identified set presented in Theorem \ref{thm:attmultpretrendsa}, and suppose that we are in the three-period case (one pre-trend) where $\mathcal{S} = \{0 \}$. Then the identified set in Theorem \ref{thm:attmultpretrendsa} is given by
\begin{equation}\label{eqn:23aset}
    \mathcal{I}_{\text{ATT}_{1}}^{A} = \bigl[ \min \{ \text{LB}(\underline{A}_{0}), \text{LB}(\overline{A}_{0}) \}, \max \{ \text{UB}(\underline{A}_{0}), \text{UB}(\overline{A}_{0}) \}   \bigr],
\end{equation}
where $\text{LB}(a) = \theta_1 + a - M |\Delta_0 - a|$ and $\text{UB}(a) = \theta_1 + a + M |\Delta_0 - a|$. If we consider the case where we only allow for parallel trends violations $(\underline{A}_{0} = \overline{A}_{0} = 0)$, then the width of the identified set in (\ref{eqn:23aset}) is
\begin{equation}\label{ptonlywidth}
    2M|\Delta_0|.
\end{equation}
Suppose the same endpoint yields the lower and upper bound in (\ref{eqn:23aset}), say $\overline{A}_{0}$ (see Appendix \ref{appendix:discussion23} for additional detail on when this can happen). If we allow for both anticipation effects and parallel trends violations, the width of the set in (\ref{eqn:23aset}) is then
\begin{equation}\label{ptaewidth}
    2M|\Delta_0 - \overline{A}_{0}|.
\end{equation}
Suppose $M>0$. If $\Delta_0 > 0$, then (\ref{ptaewidth}) is smaller than (\ref{ptonlywidth}) if and only if $\overline{A}_{0} \in (0, 2\Delta_0 )$. Furthermore, in the three-period case with $\Delta_{0}>0$, if $\overline{A}_{0}\in(0,2\Delta_{0})$ and the DGP satisfies the conditions in Lemma \ref{lem:feasiblesets23} under which $\overline{A}_0$ attains both the upper and lower bounds of the identified set in Theorem \ref{thm:attmultpretrendsa}, then it follows that $\underline{A}_{0}>0$; hence, the admissible anticipation effects exclude zero (i.e., some anticipation is enforced).

\section{Extensions}\label{sec:extensions}

\subsection{Multiple Post-Treatment Periods}\label{subsec:multpostperiods}

The analysis in Section \ref{sec:partialident} focused on the first post-treatment period. I now extend the framework to treatment effects at post-treatment horizons $h \in \{1, \ldots , T\}$ in the simultaneous adoption setting. As in the baseline setting, anticipation enters the treatment effect decomposition through the last pre-treatment period ($\varphi_0$). The difference is that treatment effects at later horizons depend on the cumulative contribution of the consecutive-period parallel trends violations $(\delta_1,\ldots,\delta_h)$.

\begingroup
\renewcommand{\thelemma}{\ref*{lemma:attbiasdecomp23}$'$}
\begin{lemma}\label{lemma:attKdecomp}
Let $h \in \{1, \ldots , T\}$.  The $\text{ATT}_{h}$ can be written in terms of anticipation effects $(\varphi_0)$ and post-treatment parallel trends violations $(\delta_1,\ldots,\delta_h)$ as: $\text{ATT}_h = \theta_h + \varphi_0 - \sum_{j=1}^{h} \delta_j.$
\end{lemma}
\addtocounter{lemma}{-1}
\endgroup

While the $\text{ATT}_h$ captures the treatment effect at a particular post-treatment horizon, researchers may also be interested in averages of treatment effects over the first $h$ post-treatment periods. The next result provides the decomposition for $\overline{\text{ATT}}_h \coloneqq \frac{1}{h}\sum_{j=1}^h \text{ATT}_j.$

\begin{corollary}\label{corr:avgdecomp}
Let $h \in \{1, \ldots , T\}$. The $\overline{\text{ATT}}_h$ can be written in terms of anticipation effects $(\varphi_0)$ and post-treatment parallel trends violations $(\delta_1,\ldots,\delta_h)$ as follows: \\
$
  \overline{\text{ATT}}_h
  \;=\; \frac{1}{h}\sum_{j=1}^h\theta_j
         + \varphi_0
         - \frac{1}{h}\sum_{j=1}^{h}(h-j+1)\,\delta_j.
$
\end{corollary}

Assumption \ref{assump:rmmult} extends Assumption \ref{assump:rmmultpretrends} from a single post-treatment period to multiple post-treatment periods by imposing a common relative magnitude bound on each post-treatment parallel trends violation. The parameter $M$ maintains the same interpretation here as in Assumption \ref{assump:rmmultpretrends}, and, as in RR, I maintain a common $M$ across horizons. 

\begingroup
\renewcommand{\theassumption}{\ref*{assump:rmmultpretrends}$'$}
\begin{assumption}[Relative Magnitude Bounds: Multiple Post-Treatment Periods]
\label{assump:rmmult}
\[
  |\delta_t| \leq M\,|\delta_{s^\star}|
  \quad \text{for all } t \in \{1,\ldots,T\},
\]
where $M \in \mathbb{R}_{\geq 0}$ is known and
$s^\star \in \operatorname{argmax}_{s\in\mathcal{S}}|\delta_s|$.
\end{assumption}
\addtocounter{assumption}{-1}
\endgroup

The next result characterizes sharp bounds on $\text{ATT}_{h}$ and $\overline{\text{ATT}}_{h}$. As before, computing the bounds only requires optimization over finite sets.

\begingroup
\renewcommand{\thetheorem}{\ref*{thm:attmultpretrendsa}$'$}
\begin{theorem}\label{thm:multpost}
Suppose $(\Delta_{-(S-1)},\ldots,\Delta_0,\theta_1,\ldots,\theta_h)$ are known and
Assumptions~\ref{initialize},~\ref{abounds}, and~\ref{assump:rmmult} hold.
Define
\[
  \text{LB}_h(r) \coloneqq \theta_h + \sum_{j\in\mathcal{S}\setminus\{r\}} \underline{A}_j
    + \min_{a\in\{\underline{A}_r,\,\overline{A}_r\}}\bigl\{a - hM|\Delta_r - a|\bigr\},
\]
\[
  \text{UB}_h(r) \coloneqq \theta_h + \sum_{j\in\mathcal{S}\setminus\{r\}} \overline{A}_j
    + \max_{a\in\{\underline{A}_r,\,\overline{A}_r\}}\bigl\{a + hM|\Delta_r - a|\bigr\},
\]
and analogously
\[
  \text{LB}_{\text{avg}}(r) \coloneqq \frac{1}{h}\sum_{j=1}^h\theta_j
    + \sum_{j\in\mathcal{S}\setminus\{r\}} \underline{A}_j
    + \min_{a\in\{\underline{A}_r,\,\overline{A}_r\}}\Bigl\{a - \tfrac{h+1}{2}M|\Delta_r - a|\Bigr\},
\]
\[
  \text{UB}_{\text{avg}}(r) \coloneqq \frac{1}{h}\sum_{j=1}^h\theta_j
    + \sum_{j\in\mathcal{S}\setminus\{r\}} \overline{A}_j
    + \max_{a\in\{\underline{A}_r,\,\overline{A}_r\}}\Bigl\{a + \tfrac{h+1}{2}M|\Delta_r - a|\Bigr\}.
\]
\begin{enumerate}[label=(\roman*)]

\item The identified set for $\text{ATT}_h$ is
\[
  \mathcal{I}_{\text{ATT}_h}
  \;=\;
  \Bigl[
    \min_{r\in\mathcal{S}}\,\text{LB}_h(r),\;
    \max_{r\in\mathcal{S}}\,\text{UB}_h(r)
  \Bigr].
\]

\item The identified set for $\overline{\text{ATT}}_h = \frac{1}{h}\sum_{j=1}^h\text{ATT}_j$
is
\[
  \mathcal{I}_{\overline{\text{ATT}}_h}
  \;=\;
  \Bigl[
    \min_{r\in\mathcal{S}}\,\text{LB}_{\text{avg}}(r),\;
    \max_{r\in\mathcal{S}}\,\text{UB}_{\text{avg}}(r)
  \Bigr].
\]

\end{enumerate}
\end{theorem}
\addtocounter{theorem}{-1}
\endgroup

\subsection{Staggered Adoption}\label{subsec:staggeredadoption}

The preceding results all assume a common treatment adoption date. This section extends the analysis to the staggered adoption setting, where different units may be treated at different calendar times. I begin by extending the notation and assumptions to accommodate this setting, and then I present the results. 

\subsubsection*{Setup and Notation}\label{subsubsec:stagsetup}

Let $\mathcal{G}$ be a finite set of treatment cohorts. Cohort $g \in \mathcal{G}$ consists of all units whose last pre-treatment period is calendar period $g$, so that treatment begins at calendar period $g + 1$ for cohort $g$. Each unit $i$ has a cohort indicator $G_i \in \mathcal{G} \cup \{\infty\}$, where $G_i = \infty$ denotes a never-treated unit. Because there is no single treatment date relative to which calendar time can be normalized in the staggered adoption setting, I denote the common calendar time range by $t \in \{t_{\text{min}}, \ldots , t_{\text{max}}\}$. To have at least one pre-trend and one observed treatment period per treated cohort, I require $\mathcal{G} \subseteq \{t_{\text{min}} +1, \ldots, t_{\text{max}} -1 \}$. Treatment is an absorbing state: a unit in cohort $g$ is untreated in all periods $t \leq g$ and treated in all periods $t \geq g + 1$. Throughout this section, the comparison group consists of never-treated units, and I maintain throughout that this group is non-empty. 

As before, potential outcomes are indexed by the complete treatment path $(x_{t_{\text{min}}}, \ldots, x_{t_{\text{max}}})$. Because treatment is an absorbing state, treatment paths are uniquely determined by the last pre-treatment period $g$. Write $Y_{it}(g)$ for the potential outcome of unit $i$ at calendar time $t$ under the treatment path $(0, \ldots , 0, 1, \ldots ,1)$, where the switch occurs at period $g+1$, and $Y_{it}(\infty)$ for the potential outcome under the never treated path. Observed outcomes are $Y_{it} = Y_{it}(G_i)$. 

For cohort $g$, define event time $r \coloneqq t - g.$ Analogous to the simultaneous adoption setting, $r = 0$ is the last pre-treatment period for cohort $g$ and $r \geq 1$ are post-treatment event times. Let $S_g \coloneqq g - t_{\text{min}}$ so that $-S_g$ is the first event time at which cohort $g$ is observed. Define $\mathcal{S}_g \coloneqq \{-(S_g - 1), \ldots, 0\}$ as the set of event times for which cohort-$g$ pre-trends are observed, and $g \geq t_{\text{min}}+1$ implies $S_g = |\mathcal{S}_g| \geq 1$. 

Anticipation effects for cohort $g$ at event time $r \in \{-S_g, \ldots , 0\}$ are given by
\[\varphi_{g,r} \;\coloneqq\; \mathbb{E}[Y_{g+r}(g) - Y_{g+r}(\infty) \mid G = g],\]
and consecutive parallel trends violations for cohort $g$ at event time $r \in \mathcal{S}_g \cup \{1, \ldots , t_{\text{max}} - g\}$ are given by
\[\delta_{g,r} \;\coloneqq\; \mathbb{E}[Y_{g+r}(\infty) - Y_{g+r-1}(\infty) \mid G = g] - \mathbb{E}[Y_{g+r}(\infty) - Y_{g+r-1}(\infty) \mid G = \infty].\]
For the pre-trend in $r \in \mathcal{S}_g$ for cohort $g$, define
\[\Delta_{g,r} \;\coloneqq\; \mathbb{E}[Y_{g+r} - Y_{g+r-1} \mid G = g] - \mathbb{E}[Y_{g+r} - Y_{g+r-1} \mid G = \infty],\]
and similarly for cohort $g$ and $r \in \{1, \ldots , t_{\text{max}}-g\}$, the DiD estimand is given by
\[\theta_{g,r} \;\coloneqq\; \mathbb{E}[Y_{g+r} - Y_g \mid G = g] - \mathbb{E}[Y_{g+r} - Y_g \mid G = \infty],\]
which uses calendar period $g$ (event time $r = 0$, the last pre-treatment period) as the baseline. 

To begin, the parameter of interest is the average effect of treatment on the treated for cohort $g$ at horizon $r=h \in \{1, \ldots , t_{\text{max}} - g\}$:
\[\text{ATT}(g,h) \coloneqq \mathbb E[Y_{g+h}(g)-Y_{g+h}(\infty)\mid G=g].
\]
In addition to studying identification of the $\text{ATT}(g,h)$, later in this section I study identification of aggregations of these parameters in the spirit of \cite{callawaysantanna2021}. For each $g \in \mathcal{G}$, the decomposition of $\text{ATT}(g,h)$ is the direct analog of Lemma \ref{lemma:attKdecomp}:
\begin{align}
    \text{ATT}(g,h) &= \theta_{g,h} + \varphi_{g,0} - \sum_{j=1}^{h} \delta_{g,j} \qquad h \in \{1, \ldots , t_{\text{max}}-g\}. \label{eq:attghdecomp} 
\end{align}
Next, I introduce the analogs of Assumptions \ref{initialize}--\ref{assump:rmmultpretrends} for the staggered adoption setting.

\begingroup
\renewcommand{\theassumption}{\ref*{initialize}$''$}
\begin{assumption}[Cohort-Level Initial Condition]
\label{assump:staginit}
For each $g \in \mathcal{G}$: $\varphi_{g,-S_g} = 0$.
\end{assumption}
\addtocounter{assumption}{-1}
\endgroup

\begingroup
\renewcommand{\theassumption}{\ref*{abounds}$''$}
\begin{assumption}[Bounded Cohort-Level Anticipation Increments]
\label{assump:stagbounds}
For each $g \in \mathcal{G}$,
\[(\varphi_{g,r} - \varphi_{g,r-1}) \in [\underline{A}_r,\,\overline{A}_r] \quad \text{ for } r \in \mathcal{S}_g,\]
where the event time bounds $\underline{A}_r$ and $\overline{A}_r$ are common across cohorts and, for each relevant event time $r$, $\underline{A}_r,\overline{A}_r \in \mathbb{R}$ are known and satisfy $\underline{A}_r \leq \overline{A}_r$.
\end{assumption}
\addtocounter{assumption}{-1}
\endgroup

\begingroup
\renewcommand{\theassumption}{\ref*{assump:rmmultpretrends}$''$}
\begin{assumption}[Cohort-Level Relative Magnitude Bounds]
\label{assump:stagrm}
For each $g \in \mathcal{G}$ and each $r \in \{1, 2, \ldots, t_{\text{max}}-g\}$:
$$|\delta_{g,r}| \;\leq\; M |\delta_{g, s_{g}^{\star}}|,$$
where $M \in \mathbb{R}_{\geq 0}$ is known and $s_{g}^{\star} \in \arg \max_{s \in \mathcal{S}_g} |\delta_{g,s}|$.
\end{assumption}
\addtocounter{assumption}{-1}
\endgroup

\subsubsection*{Results}

The first result is a direct analog of Theorem \ref{thm:multpost}. Fixing cohort $g$ and using never-treated units as the comparison group reduces the staggered adoption problem to a single-cohort problem in event time. As a result, the identification argument from Theorem~\ref{thm:multpost} applies cohort-by-cohort.

\begin{corollary}\label{corr:stagcohort}
Fix $g \in \mathcal{G}$ and $h \in \{1, \ldots , t_{\text{max}}-g\}$. Suppose the never-treated group is non-empty,
$(\Delta_{g,-(S_g-1)}, \ldots, \Delta_{g,0}, \theta_{g,1}, \ldots, \theta_{g,h})$ are known, and Assumptions~\ref{assump:staginit},~\ref{assump:stagbounds}, and~\ref{assump:stagrm} hold. Define
\[l_{g,h}(r) \;\coloneqq\; \theta_{g,h} + \sum_{j \in \mathcal{S}_g \setminus \{r\}} \underline{A}_j + \min_{a \in \{\underline{A}_r,\, \overline{A}_r\}}\bigl\{a - hM|\Delta_{g,r} - a|\bigr\},\]
\[u_{g,h}(r) \;\coloneqq\; \theta_{g,h} + \sum_{j \in \mathcal{S}_g \setminus \{r\}} \overline{A}_j + \max_{a \in \{\underline{A}_r,\, \overline{A}_r\}}\bigl\{a + hM|\Delta_{g,r} - a|\bigr\}.\]
The identified set for $\text{ATT}(g,h)$ is
\[\mathcal{I}_{g,h} \;\coloneqq\; \Bigl[\min_{r \in \mathcal{S}_g} l_{g,h}(r),\; \max_{r \in \mathcal{S}_g} u_{g,h}(r)\Bigr].\]
\end{corollary}

While the cohort-specific effects $\text{ATT}(g,h)$ may be of interest in their own right, researchers are often interested in summary measures that aggregate treatment effects across cohorts. Following \cite{callawaysantanna2021}, I define aggregate parameters through weighted averages of the cohort-specific effects. 

For a fixed event time $h \geq 1$, let $\mathcal{G}_h \coloneqq \{g \in \mathcal{G} : g + h \leq t_{\text{max}} \}$ be the set of cohorts who receive treatment for at least $h$ periods in the observed data, and define
\begin{equation}\label{eq:attkw}
  \text{ATT}_{h}^{w} \coloneqq \sum_{g \in \mathcal{G}_h} w_g\,\text{ATT}(g,h),
\end{equation}
where $w_g \geq 0$ for all $g \in \mathcal{G}_h$ and $\sum_{g \in \mathcal{G}_h} w_g = 1$. Particular choices of $w_g$ recover several commonly used aggregate parameters; for example, the event study and balanced event study aggregations from \cite{callawaysantanna2021}, which I outline below.

The event study aggregation at event time $h$ weights each cohort by its
share among cohorts observed for at least $h$ treated periods: $w_g^{ES}(h)\coloneqq\mathbb{P}(G=g \mid G \in \mathcal G_h)$ for $g\in\mathcal G_h$. The resulting parameter 
\[ \text{ATT}_{h}^{ES} \coloneqq \sum_{g\in\mathcal G_h} w_g^{ES}(h)\text{ATT}(g,h) \]
is the average treatment effect $h$ periods after treatment among cohorts for which event time $h$ is observed.

When the goal is to analyze an event study path over $h = 1,\ldots,h_{\max}$, the event study aggregation above allows the composition of cohorts to vary with $h$. To hold cohort composition fixed across event times $h = 1,\ldots,h_{\max}$, one can instead use balanced event study weights. Define $\mathcal G_{h_{\max}}$ analogously to $\mathcal{G}_h$. This is the set of cohorts who have received treatment for at least $h_{\max}$ periods in the observed data. Define
$w_g^{ES,BAL}(h_{\max})\coloneqq\mathbb P(G=g \mid G\in\mathcal G_{h_{\max}})$, for $g\in\mathcal G_{h_{\max}}.$ The resulting parameter
\[\text{ATT}_{h, h_{\text{max}}}^{ES,BAL}\coloneqq\sum_{g\in\mathcal G_{h_{\max}}}w_g^{ES,BAL}(h_{\max})\text{ATT}(g,h)
\]
averages the event-time-$h$ treatment effects over this common set of cohorts. Therefore, comparisons of $\text{ATT}_{h, h_{\text{max}}}^{ES,BAL}$ across values of $h \in \{ 1,\ldots,h_{\max}\}$ are not driven by changes in cohort composition \citep{callawaysantanna2021}.

\begin{theorem}\label{thm:stagagg}
Fix $h \geq 1$ such that $\mathcal{G}_h \neq \varnothing$ and suppose that, for each $g\in\mathcal{G}_h$, \\ $(\Delta_{g,-(S_g-1)}, \ldots, \Delta_{g,0}, \theta_{g,1}, \ldots, \theta_{g,h})$ are known. Suppose also that the never-treated group is non-empty. Let $\{w_g\}_{g \in \mathcal{G}_h}$ be weights satisfying
\[
w_g \geq 0
\quad \text{ for all } g \in \mathcal{G}_h,
\qquad
\sum_{g \in \mathcal{G}_h} w_g = 1.
\]
For each $g \in \mathcal{G}_h$, write $l_{g,h}
\coloneqq
\min_{r \in \mathcal{S}_g} l_{g,h}(r)$ and $u_{g,h}
\coloneqq
\max_{r \in \mathcal{S}_g} u_{g,h}(r),$ so that
$\mathcal{I}_{g,h} = [l_{g,h},u_{g,h}].$ Under Assumptions~\ref{assump:staginit}--\ref{assump:stagrm}, the identified set for $\text{ATT}_h^w$ is
\[
\mathcal{I}_{\text{ATT}_h^w}
\coloneqq
\sum_{g \in \mathcal{G}_h} w_g\,\mathcal{I}_{g,h}
=
\left[
\sum_{g \in \mathcal{G}_h} w_g\,l_{g,h},
\;
\sum_{g \in \mathcal{G}_h} w_g\,u_{g,h}
\right].
\]
\end{theorem}

The results in this section show that the analysis in Section \ref{sec:partialident} extends naturally to staggered adoption when never-treated units serve as the comparison group. In particular, the analysis yields closed-form sharp bounds for both cohort-specific treatment effects and aggregate treatment effect parameters, including the event study and balanced event study aggregations of \cite{callawaysantanna2021} commonly used in empirical applications. One could alternatively use not-yet-treated cohorts as the comparison group. In that case, the analysis would need to account for anticipation effects in the comparison group.

\section{Practical Implications for Sensitivity Analysis}\label{sec:sensitivity}
The identified sets enable global sensitivity analysis of qualitative conclusions to joint deviations from parallel trends and no anticipation. For instance, for what values of the sensitivity parameters does the conclusion $\text{ATT}_{1} > 0$ hold?  The set of values for which the conclusion just fails constitutes the breakdown frontier. In this section, I outline how I use the identification results to construct this joint sensitivity analysis.

\subsection{Overview}

I briefly review breakdown frontiers and adapt the approach of \cite{mastenpoirier2020} to the setting considered in this paper. A breakdown frontier starts with a conclusion of interest, such as $\text{ATT}_{1} > 0$. Under violations of parallel trends and no anticipation, the $\text{ATT}_{1}$ is generally partially identified, but qualitative conclusions may still be robust over some region of the sensitivity parameter space. 

Let $\boldsymbol{a} \in \mathbb{R}^{d_a}$ denote the vector of parameters that indexes the maintained restriction on anticipation, where $d_a$ is the dimension of the chosen parameterization of Assumption \ref{abounds}. For a given $\boldsymbol{a}$, the corresponding parameterization determines the bounds on anticipation increments: $\{\underline{A}_t (\boldsymbol{a}), \overline{A}_t (\boldsymbol{a}) \}_{t \in \mathcal{S}}$. For example, the parameterization in \eqref{eq:pbounds} corresponds to $\boldsymbol{a} = (\underline{p}, \overline{p})$, and the parameterization in \eqref{eq:qbounds} corresponds to $\boldsymbol{a} = (\underline{q}, \overline{q})$. 

Consider the conclusion $\text{ATT}_{1} > 0$ and let $\text{LB}(\boldsymbol{a}, M)$ denote the lower bound on the identified set for the $\text{ATT}_{1}$ under sensitivity parameters $(\boldsymbol{a}, M)$. The conclusion is robust whenever $\text{LB}(\boldsymbol{a}, M) > 0$. Define the breakdown value of $M$ as\footnote{Here, I allow $M \geq 0$ in defining the breakdown frontier and robust region. However, the admissible range of $M$ may depend on the chosen parameterization and accompanying assumptions. For example, under the parameterization in Assumption \ref{assump:krelaxmultpre} that calibrates anticipation relative to the treatment effect, Assumption \ref{assump:nonzerodenomk} may restrict the admissible range of $M$. See Appendices \ref{appendix:assumpnonzero} and \ref{appendix:fsf} for more details.}
\begin{equation}\label{eq:mbp}
    M^{bp}(\boldsymbol{a}) \coloneqq \inf \{ M \geq 0 : \text{LB}(\boldsymbol{a}, M) \leq 0\}.
\end{equation}
Thus, $M^{bp}(\boldsymbol{a})$ is the smallest value of $M$ at which the conclusion breaks down. Equation \eqref{eq:mbp} therefore defines the breakdown frontier
\begin{equation}\label{eq:breakdownfrontier}
    \text{BF} \coloneqq \{ (\boldsymbol{a}, M) : M = M^{bp}(\boldsymbol{a}) \}.
\end{equation}
The breakdown frontier partitions the space of sensitivity parameters into the region where the conclusion of interest holds and the region where it fails. The robust region is defined as
\begin{equation}\label{eq:robustregion}
    \text{RR} \coloneqq \{(\boldsymbol{a}, M) : M < M^{bp}(\boldsymbol{a})  \}.
\end{equation}
Analytical derivations of the breakdown frontiers used in the empirical applications in Section \ref{sec:empiricalillustration} are provided in Appendix \ref{appendix:ksensitivity}.

\subsection{Estimation and Inference}

The identified sets in Section \ref{sec:partialident} and the breakdown frontier are functionals of the reduced-form parameter $\gamma \coloneqq (\Delta_{-(S-1)}, \ldots , \Delta_0, \theta_1)^{\prime}$ but the mappings from $\gamma$ to the identified sets and the breakdown frontier need not be Hadamard differentiable, which precludes regular frequentist asymptotic inference \citep{hiranoporter2012}. Recent work provides frequentist methods for Hadamard directionally differentiable functionals (e.g.\ \citealt{fangsantos2019}). In the present setting, however, the breakdown frontier can fail even to be Hadamard directionally differentiable: under some parameterizations of anticipation, the denominator in the breakdown frontier expression can be zero at meaningful values of the sensitivity parameters. Thus, existing frequentist methods do not directly apply in these cases.

To estimate and conduct inference on the breakdown frontier, I therefore use the Bayesian bootstrap \citep{ferguson1973, rubin1981}. Since the identified sets in Section \ref{sec:partialident}, and therefore the breakdown frontier, are determined by the identifiable reduced-form parameter $\gamma$, the Bayesian inference framework of \cite{klinetamer2016} offers a principled approach to inference in this setting. I use the Bayesian bootstrap as a tractable nonparametric Bayesian approach to inference on the data-generating distribution, which simultaneously yields inference on functionals of the data-generating distribution such as the identified sets and breakdown frontier \citep{chamberlainimbens2003, klinetamer2016}.

There are several benefits to using the Bayesian bootstrap for inference in this setting. First, the Bayesian bootstrap is nonparametric. Under the topology of convergence in distribution, the Bayesian bootstrap posterior is supported on the set of probability distributions whose support is contained in the support of the empirical distribution \citep{ghosalvandervaart}. Second, the Bayesian bootstrap is easy to implement, as it amounts to a simple reweighting of the data. Third, the Bayesian bootstrap still delivers a valid Bayesian interpretation of uncertainty quantification even when regular frequentist approximations are not available. Like the frequentist nonparametric bootstrap, the Bayesian bootstrap need not achieve frequentist coverage for non-Hadamard differentiable functionals \citep{kmpv2020}. I therefore interpret the bands below as Bayesian credible bands, rather than frequentist confidence bands. Recent work also emphasizes the usefulness of Bayesian bootstrap procedures for communicating uncertainty and minimizing posterior regret when normal approximations perform poorly \citep{andrewsshapiro2025}.

Using the outputs from the Bayesian bootstrap, I construct simultaneous lower credible bands\footnote{Several papers have considered various versions of simultaneous Bayesian credible bands, including \cite{crcjg2007} and \cite{kkc2010}.}. Let $\Pi (\cdot \mid Z)$ denote the posterior probability measure over the breakdown frontier, given data $Z$. A simultaneous lower credible band is any function $\hat{L}(\boldsymbol{a})$ implicitly defined as
\begin{equation}\label{eqn:simullowercredbands}
   \Pi (\hat{L}(\boldsymbol{a}) \leq M^{bp}(\boldsymbol{a}) \text{ for all } \boldsymbol{a} \in \mathcal{H} \mid Z) \geq 1- \alpha. 
\end{equation}
where $\mathcal{H} \subseteq \mathbb{R}^{d_a}$ denotes a finite grid of admissible values of the anticipation sensitivity parameters $\boldsymbol{a}$. The interpretation is as follows: conditional on the data, with at least $100(1-\alpha) \%$ probability the breakdown frontier lies above $\hat{L}(\boldsymbol{a})$. For details on the Bayesian bootstrap and construction of the simultaneous lower credible bands, see Appendix \ref{appendix:inference}.

\subsection{Benchmarking Sensitivity Parameters for Assessing Robustness}\label{subsec:benchmark}

Motivated by empirical settings in which a policy is announced before implementation, I outline an approach for benchmarking plausible values of the sensitivity parameters. Such implementation lags arise in many settings, including health care policy, minimum wage changes, taxes, and welfare benefits \citep{alpert2016}. Suppose the policy is announced at the end of period $t_{\text{ann}}$, where $-(S-1)\leq t_{\text{ann}}\leq -1$. Before the announcement, units have not yet learned about the policy, so there is no scope for anticipation. By Corollary \ref{prop:pretrenddecopmpsim}, the corresponding observed pre-trends identify the parallel trends violations: $\Delta_t = \delta_t$ for $t \in \{-(S-1), \ldots, t_{\text{ann}}\}$. The benchmarking strategy uses these identified pre-announcement violations to restrict the magnitude of post-announcement, pre-treatment parallel trends violations. Combined with the observed post-announcement pre-trends, these restrictions yield empirically motivated bounds on the unobserved anticipation increments.

The benchmarking strategy rests on the idea that the identified pre-announcement parallel trends violations are informative about post-announcement, pre-treatment violations. The stylized model in Section \ref{sec:model} provides one rationale: when selection is based on earnings in period $t_{\text{ann}}$, eventually treated units tend to have unusually low transitory earnings in that period \citep{ashenfeltercard1985}, generating parallel trends violations that are largest near the screening date and decline with distance from it. This motivates benchmarking post-announcement, pre-treatment parallel trends violations by the largest pre-announcement violation, which in turn determines the bounds on anticipation increments.

\begin{proposition}\label{prop:aboundsbenchmark}
    Suppose Assumption \ref{initialize} holds. Suppose there is no anticipation prior to period $t_{\text{ann}} +1$: $\varphi_t = 0$ for $t \in \{-S, \ldots, t_{\text{ann}}\}$. Let $\Delta^{\text{ann}} \coloneqq \max_{t \in \{-(S-1), \ldots , t_{\text{ann}}\}} |\Delta_t|$. If $|\delta_t| \leq \Delta^{\text{ann}}$ for $t \in \{t_{\text{ann}}+1, \ldots , 0\}$, then Assumption \ref{abounds} holds with
\begin{equation}\label{eq:cal2bounds}
[\underline{A}_t,\overline{A}_t]
=
\begin{cases}
[0,0], & t \in \{-(S-1), \ldots , t_{\text{ann}}\},\\
[\Delta_t - \Delta^{\text{ann}}, \Delta_t + \Delta^{\text{ann}}], & t \in \{t_{\text{ann}}+1, \ldots , 0\}.
\end{cases}
\end{equation}
\end{proposition}
\noindent One could also consider other benchmarks, such as using the range of pre-announcement parallel trends violations to bound post-announcement, pre-treatment violations.  

For the sensitivity analysis, I use a breakdown frontier to summarize how the identified set changes with the sensitivity parameters. The full collection of post-announcement, pre-treatment anticipation parameters,
$\{(\underline{A}_t,\overline{A}_t):t=t_{\text{ann}}+1,\ldots,0\}$,
may be high-dimensional, so I adopt a lower-dimensional parameterization that yields an interpretable two-dimensional frontier. Specifically, for illustration I adapt the parameterization in \eqref{eq:qbounds} to incorporate the announcement timing, imposing no anticipation before the announcement and a common scaled increment $q\Delta^{\text{all}}$ afterward:
\begin{equation}\label{eq:atq}
    \underline{A}_t(q)=\overline{A}_t(q)= \begin{cases} 0, & t \in \{-(S-1),\ldots,t_{\text{ann}}\},\\ q\Delta^{\text{all}}, & t\in\{t_{\text{ann}}+1,\ldots,0\}. \end{cases} 
\end{equation}

For each $q$, \eqref{eq:atq} imposes constant post-announcement anticipation increments and hence a linear anticipation path in levels. To determine which values of $q$ are empirically plausible, I find those consistent with the calibrated anticipation bounds in \eqref{eq:cal2bounds}. Let $[\underline{A}_t^{\text{cal}},\overline{A}_t^{\text{cal}}]$ denote the implied bounds on the anticipation increment in period $t\in\{t_{\text{ann}}+1,\ldots,0\}$ from \eqref{eq:cal2bounds}. These yield the period-specific calibrated intervals
\[ \mathcal Q_t^{\text{cal}} \coloneqq \left[ \frac{\underline{A}_t^{\text{cal}}}{\Delta^{\text{all}}}, \frac{\overline{A}_t^{\text{cal}}}{\Delta^{\text{all}}} \right], \qquad \Delta^{\text{all}} > 0, \; t\in\{t_{\text{ann}}+1,\ldots,0\}. \]
Because \eqref{eq:atq} imposes a common $q$ across these periods, its calibrated region is
\begin{equation}\label{eq:qcal}
    \mathcal{Q}^{\text{cal}} \coloneqq \bigcap_{t=t_{\text{ann}}+1}^{0}\mathcal Q_t^{\text{cal}}.
\end{equation}
For the common-$q$ specification in \eqref{eq:atq}, the benchmarking exercise proceeds in four steps. First, compute $\Delta^{\text{ann}}$ from the pre-announcement pre-trends. Second, use \eqref{eq:cal2bounds} to construct calibrated bounds on each post-announcement anticipation increment. Third, divide these bounds by $\Delta^{\text{all}}$ to obtain the period-specific intervals $\mathcal Q_t^{\mathrm{cal}}$. Finally, intersect these intervals as in \eqref{eq:qcal}. The resulting set contains exactly the values of the common $q$ whose implied anticipation path satisfies the benchmark in every post-announcement period. If the intersection is empty, no path under the common-$q$ specification is compatible with all period-specific bounds, although other parameterizations may be.

Once the calibrated region for $q$ is obtained, the remaining task is similar to the one-dimensional sensitivity analysis in RR: within this region, the researcher assesses the breakdown values of $M$\footnote{Alternatively, one could use Proposition \ref{prop:aboundsbenchmark} to choose the bounds on anticipation increments in Assumption \ref{abounds} and then apply Theorem \ref{thm:attmultpretrendsa} to obtain sharp bounds on $\text{ATT}_1$ that depend only on $M$. This yields a sensitivity analysis for parallel trends violations in which anticipation is calibrated at the outset. The approach detailed in this section instead begins with a parameterization of anticipation, constructs the corresponding breakdown frontier, and then uses Proposition \ref{prop:aboundsbenchmark} to determine which values of the anticipation sensitivity parameters are most relevant for assessing robustness.}. This benchmarking exercise thus reduces the interpretation of a multi-dimensional sensitivity analysis to a familiar one-dimensional sensitivity analysis. The calibration is not required for the identification results; it guides which values of the sensitivity parameters are most relevant for assessing robustness in a given application. Section \ref{subsec:medicare} applies this procedure to the rollout of Medicare Part D.

\section{Empirical Illustrations}
\label{sec:empiricalillustration}

I illustrate the proposed sensitivity analysis in two empirical settings. The Medicare Part D application has a clean announcement window, allowing me to illustrate the benchmarking procedure in Section \ref{subsec:benchmark}. The FSF reform does not have a clean announcement window, allowing me to show how the sensitivity analysis can still be interpreted without one.

\subsection{Prescription Drug Expenditures under Medicare Part D}\label{subsec:medicare}
Medicare is a federal health insurance program covering adults aged 65 and older, as well as certain younger individuals who qualify through disability. Medicare Part D added prescription drug coverage to the program, representing the largest expansion of Medicare since its establishment in 1965. It was part of the Medicare Prescription Drug, Improvement, and Modernization Act (MMA), which was signed into law in December 2003, but Part D did not take effect until January 2006. 

This illustration draws on the data and empirical setting in \cite{engelhardtgruber2011} and on \cite{alpert2016}, who motivates the clean announcement window and scope for anticipation. I compare Medicare-eligible individuals aged 65--70 to a near-elderly comparison group aged 60--64, with total prescription drug expenditures as the outcome. Using data from 2000--2007, I estimate dynamic DiD coefficients for this outcome. This specification differs from those in \cite{engelhardtgruber2011}; the goal is to illustrate the proposed sensitivity analysis and benchmarking procedure rather than replicate or extend their analysis.

The clean announcement structure makes Part D well suited to the benchmarking procedure in Section \ref{subsec:benchmark}. The December 2003 passage of the MMA received broad media coverage and specified the January 2006 implementation date, giving beneficiaries information about both the future benefit and its timing \citep{alpert2016}. Because the announcement was largely unexpected, the pre-announcement periods can plausibly be treated as free of anticipation.

\begin{figure}[t]
    \centering

    \caption{Event Study Estimates for Medicare Part D}
    \label{fig:medicareeventstudy}

    \begin{subfigure}[t]{0.48\textwidth}
        \centering
        \makebox[\linewidth][c]{\hspace*{0.06\linewidth}{\small Conventional Event Study}}\par\medskip
        \includegraphics[width=\linewidth]{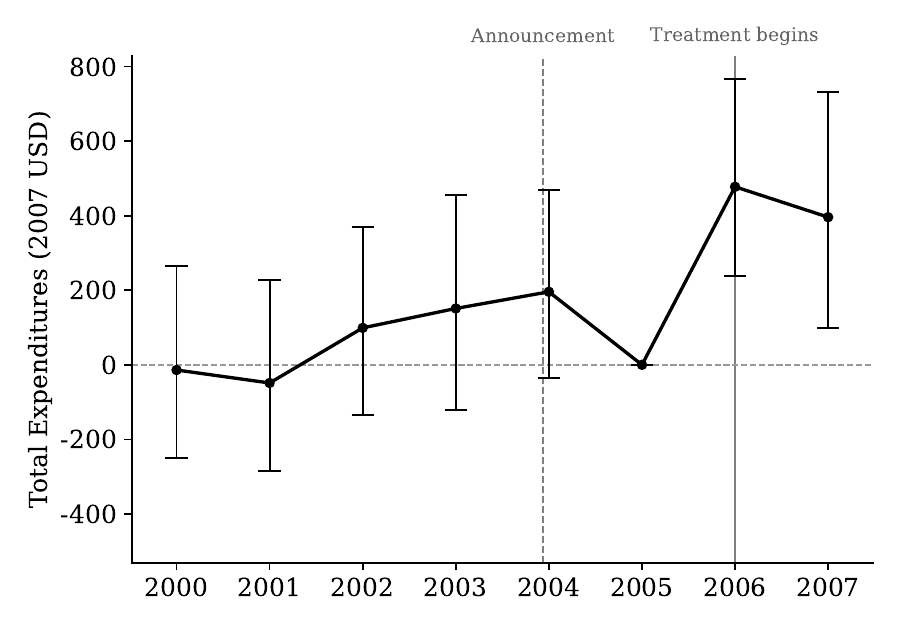}
        \caption{}
    \end{subfigure}\hfill
    \begin{subfigure}[t]{0.48\textwidth}
        \centering
        \makebox[\linewidth][c]{\hspace*{0.06\linewidth}{\small Asymmetric Event Study}}\par\medskip
        \includegraphics[width=\linewidth]{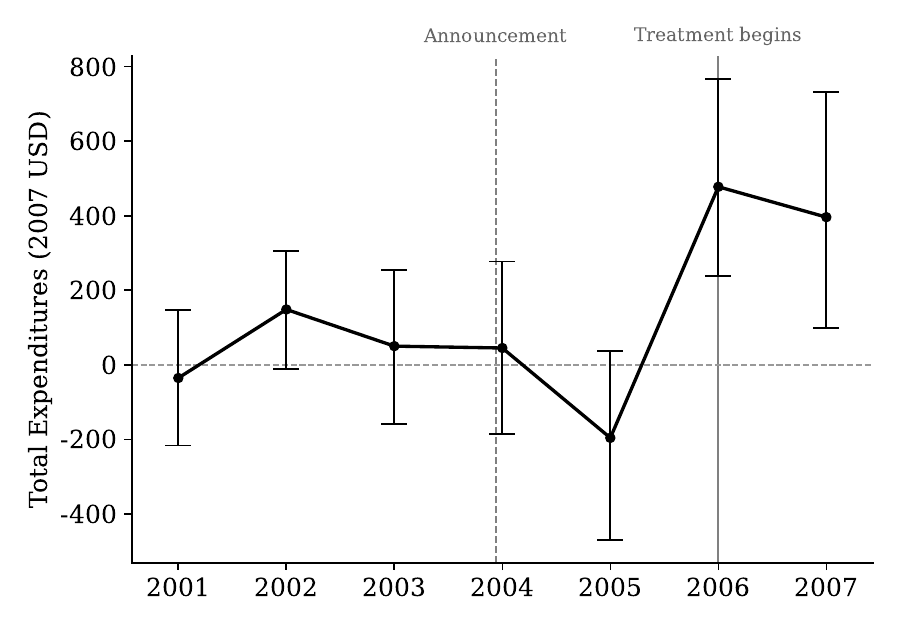}
        \caption{}
    \end{subfigure}

    \caption*{\footnotesize \textit{Notes:} Panel (a) reports conventional nonparametric event study coefficients, each calculated relative to the 2005 baseline. Panel (b) reports an asymmetric event study specification: consecutive differences, $\Delta_t$, in the pre-treatment period and differences relative to the 2005 baseline, $\theta_t$, in the post-treatment period. The treated group consists of individuals aged 65--70, and the comparison group consists of individuals aged 60--64. Estimates are posterior medians with 95\% equitailed probability intervals based on 20,000 Bayesian bootstrap draws. Within each draw, a common weight is assigned to observations belonging to the same household and age group, in line with the clustered inference in \cite{engelhardtgruber2011}. Data Source: \cite{engelhardtgruber2011data}. }
\end{figure}

Prior studies typically compare outcomes before and after the 2006 implementation of Part D, implicitly ruling out anticipation \citep{alpert2016}. Anticipation is plausible because the MMA was signed two years before Part D's implementation, informing beneficiaries of a future reduction in prescription drug prices. Ex ante, its sign is ambiguous: individuals may defer drug utilization until coverage begins, generating negative anticipation, or increase utilization through income effects, generating positive anticipation \citep{alpert2016}. \cite{alpert2016} finds that utilization declined after announcement and rose in the implementation year, suggesting that intertemporal substitution dominated. Although I study total expenditures, one would expect utilization and expenditures to move together. Moreover, one would expect this form of anticipation to be concentrated near implementation, since elderly beneficiaries may briefly delay prescription fills but are unlikely to defer needed medications for an extended period. Consistent with this mechanism, Panel (b) of Figure \ref{fig:medicareeventstudy} shows that $\widehat{\Delta}_{2005}$ is lower than the earlier pre-trends\footnote{In both Figures \ref{fig:medicareeventstudy} and \ref{fig:dseventstudy}, Panels (a) and (b) use different event study constructions. Panel (a) uses ``long differences'' throughout the pre- and post-treatment periods, whereas Panel (b) uses ``short differences'' in the pre-treatment period and long differences in the post-treatment period; see the notes below each figure for details. Panel (b) contains the objects relevant for the sensitivity analysis, but should not be interpreted in the same way as Panel (a); see \cite{rotheventstudies}.}. Furthermore, there is also scope for parallel trends violations. Even absent Part D, developments such as new drugs or changing drug prices may have caused prescription drug expenditures to evolve differently for the population aged 65--70 than for the population aged 60--64.

\begin{figure}[t]
    \centering

    \caption{Estimated Breakdown Frontier for $\text{ATT}_{2006} > 0$ and Identified Sets}
    \label{fig:medicarebfidentifiedsets}

    \begin{subfigure}{0.48\textwidth}
        \centering
        \includegraphics[width=\textwidth]{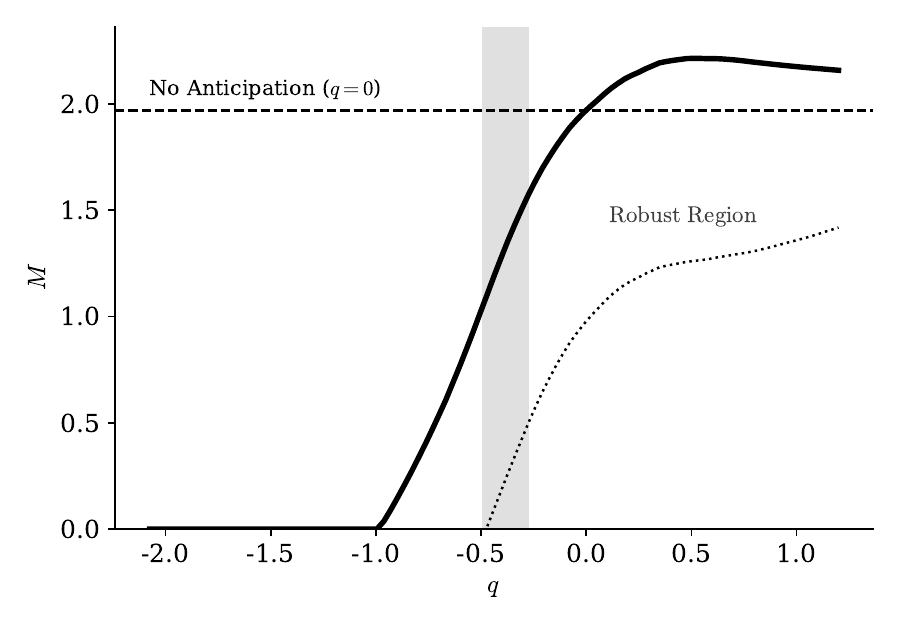}
        \caption{}
    \end{subfigure}
    \hspace{0.6em}
    \begin{subfigure}{0.48\textwidth}
        \centering
        \includegraphics[width=\textwidth]{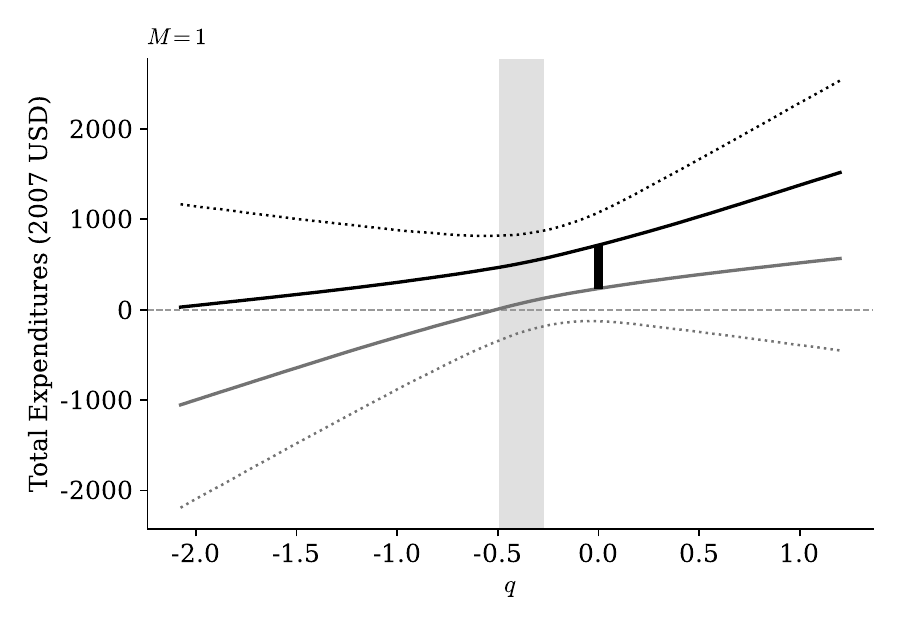}
        \caption{}
    \end{subfigure}

    \caption*{\footnotesize \textit{Notes:} In Panel (a), the estimated breakdown frontier for $\text{ATT}_{2006} > 0$ is denoted by the solid line and the 95\% simultaneous lower credible band is denoted by the dotted line. See Appendix \ref{subsec:simullowercredbands} for additional detail on the simultaneous lower credible band and its construction. In Panel (b), the estimated bounds of the identified sets are denoted by the solid lines and 95\% pointwise credible sets are denoted by the dashed lines. The darker lines correspond to the upper bound and the lighter lines correspond to the lower bound. See Appendix \ref{subsec:credints} for additional detail on the pointwise credible sets and their construction. The identified set under deviations from parallel trends only $(q=0)$ is shown in bold. In both figures, the gray shaded region denotes the calibrated region for $q$ using the benchmarking procedure in Section \ref{subsec:benchmark}. Data Source: \cite{engelhardtgruber2011data}.}
\end{figure}
Panel (a) of Figure \ref{fig:medicarebfidentifiedsets} shows the breakdown frontier for the conclusion $\text{ATT}_{2006} >0$ using the specification in \eqref{eq:atq} with $t_{\text{ann}} = 2003$. If we impose no anticipation and only allow for violations of parallel trends, the estimated breakdown value of $M$ is about 1.97. Therefore, in order for the conclusion $\text{ATT}_{2006} > 0$ to be robust (i.e.\ the identified set lies strictly above 0), post-treatment parallel trends violations must be less than 1.97 times the largest pre-treatment parallel trends violation. 
Once we add anticipation as an additional axis in the sensitivity analysis, we see that for certain values of $q$ the conclusion is more robust than the $q=0$ case, and for other values it is less robust. 

The benchmarking procedure in Section \ref{subsec:benchmark} provides guidance on the values of $q$ most relevant for assessing robustness. The calibrated region, shown by the gray shaded area, ranges from approximately $-0.50$ to $-0.27$, consistent with the finding in \cite{alpert2016} of negative anticipation effects. Focusing on this region reduces the task to one similar to the one-dimensional sensitivity analysis in RR: assessing the breakdown values of $M$ within this region. These values range from about 1 to 1.6. Thus, the conclusion is less robust than under no anticipation, but remains robust throughout the calibrated region relative to the common benchmark $M=1$.

Panel (b) of Figure \ref{fig:medicarebfidentifiedsets} shows identified sets for $\text{ATT}_{2006}$ with $M$ fixed at 1. The identified set under deviations from parallel trends only, corresponding to $q=0$, is highlighted in bold. For approximately $q \in (-0.59, 0)$, the identified set at $M = 1$ is shorter than under violations of parallel trends alone. The results align with the breakdown frontier in Panel (a): near $q=-0.51$, the lower bound in Panel (b) just crosses zero, corresponding to a breakdown value of approximately $M=1$. For values of $q$ greater than -0.51, the identified set lies above zero and the breakdown value exceeds one, as shown in Panel (a), whereas for values of $q$ less than -0.51, the identified set contains zero and the breakdown value is below one.

\subsection{Private School Supply Response to a Funding Reform}\label{subsec:privateschools}

\cite{dinersteinsmith021} study the effect of New York City's Fair Student Funding (FSF) reform on private school supply. The reform followed a November 2006 New York Court of Appeals decision reaffirming the \textit{Campaign for Fiscal Equity, Inc.\ vs.\ New York} ruling and determined the allocation of up to \$3.2 billion in additional public school funding. ``Winning'' schools began receiving funds in the 2007--2008 school year \citep{dinersteinsmith021}. For this illustration, I focus on the event study in Figure 6A of \cite{dinersteinsmith021}, which estimates the effect of projected FSF funding on the number of private schools within one mile of a public school. If a nearby public school receives additional funding, some students may switch from private to public school. The resulting enrollment losses reduce tuition revenue and, if sufficiently large, may lead to private school closure.

The setting in \cite{dinersteinsmith021} is well suited to assessing robustness to joint violations of no anticipation and parallel trends. \cite{dinersteinsmith021} explicitly highlight a potential violation of parallel trends: neighborhoods near public schools receiving additional funding may have followed different untreated trends from those farther away because of unobservables correlated with the reform's funding changes. Anticipation is also plausible given the prolonged litigation preceding implementation: the Campaign for Fiscal Equity filed suit in 1999, the trial court ruled for the plaintiffs in 2001, and the Court of Appeals reaffirmed the ruling in 2006 \citep{fruchtermr}. This extended timeline may have allowed private schools to anticipate the reform.

\begin{figure}[t]
    \centering

    \caption{Event Study Estimates for the FSF Reform}
    \label{fig:dseventstudy}

    \begin{subfigure}[t]{0.48\textwidth}
        \centering
        \makebox[\linewidth][c]{\hspace*{0.06\linewidth}{\small Conventional Event Study}}\par\medskip
        \includegraphics[width=\linewidth]{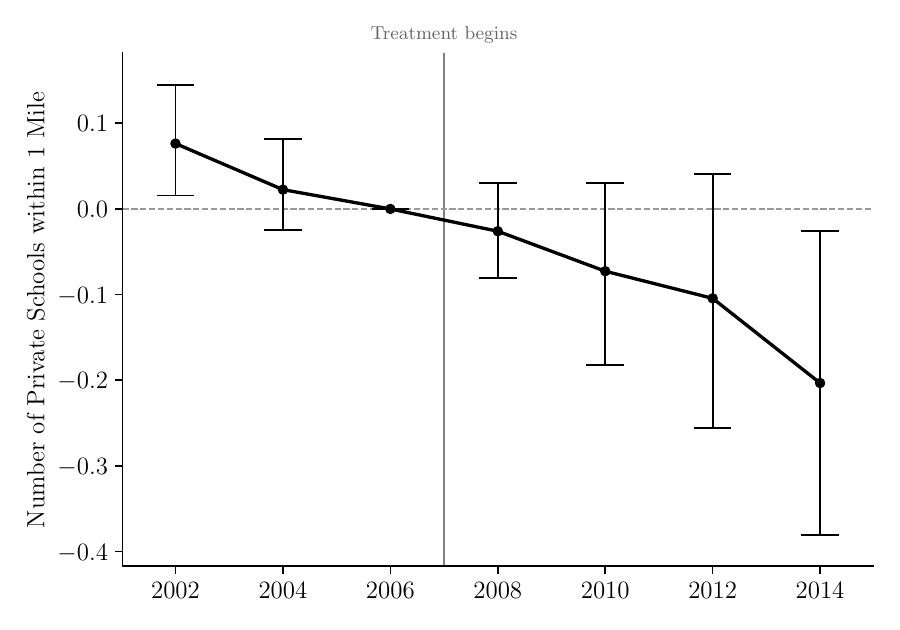}
        \caption{}
    \end{subfigure}\hfill
    \begin{subfigure}[t]{0.48\textwidth}
        \centering
        \makebox[\linewidth][c]{\hspace*{0.06\linewidth}{\small Asymmetric Event Study}}\par\medskip
        \includegraphics[width=\linewidth]{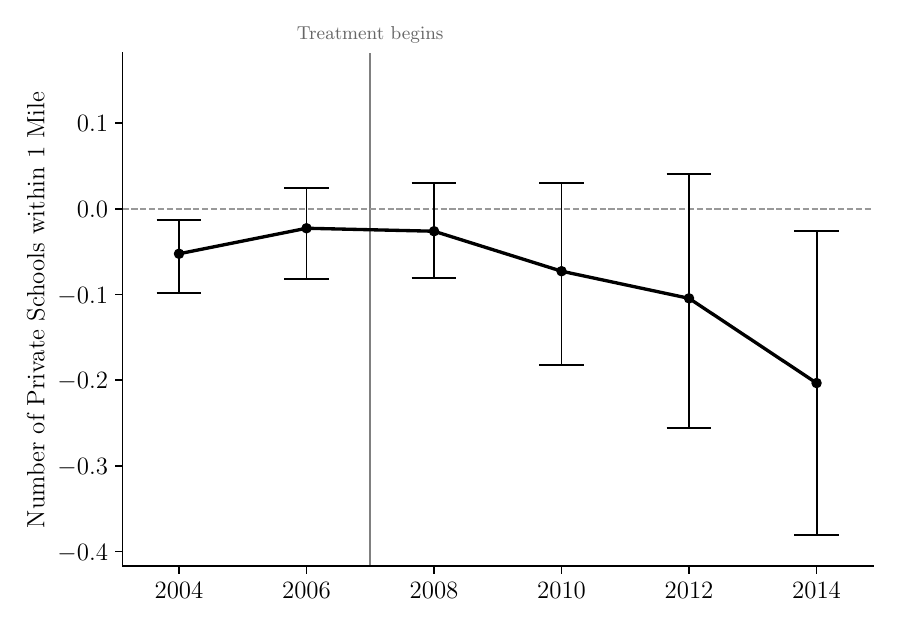}
        \caption{}
    \end{subfigure}

    \caption*{\footnotesize \textit{Notes:} Panel (a) reports conventional nonparametric event study coefficients, each calculated relative to the 2006 baseline. Panel (b) reports an asymmetric event study specification: consecutive differences, $\Delta_t$, in the pre-treatment period and differences relative to the 2006 baseline, $\theta_t$, in the post-treatment period. Estimates are posterior medians and 95\% equitailed probability intervals computed from a Bayesian bootstrap with 20,000 draws that re-weights at the ZIP level (one weight per ZIP applied to all observations in that ZIP). This ZIP–level weighting is intended to be analogous to the ZIP–clustered inference in
\cite{dinersteinsmith021}. Data Source: \cite{DinersteinSmith2021Data}.}
\end{figure}

I define treatment as an indicator equal to one for schools with a strictly positive projected funding change under the FSF reform and zero for schools whose projected change is exactly zero. The FSF had a ``hold harmless'' provision that precluded negative funding changes. The treated and untreated groups each comprise approximately 50\% of the sample. The private school data are biennial, so I follow \cite{dinersteinsmith021} in pooling adjacent years. Therefore, I index these objects by the leading even calendar year; for example, $\widehat{\Delta}_{2004}$ denotes the estimated pre-trend between the 2003-2004/2004-2005 bin and the preceding 2001-2002/2002-2003 bin. The horizontal axis in Figure \ref{fig:dseventstudy} follows this convention. In Panel (b) of Figure \ref{fig:dseventstudy}, we see that the posterior medians are negative for both pre-trends. The 95\% equitailed probability interval for $\widehat{\Delta}_{2004}$ lies entirely below zero.

For illustration, suppose we are interested in ruling out large negative effects. Before the reform, public schools had about 3.4 private schools within one mile, so the hypothesis $\text{ATT}_{2008}>-0.1$ corresponds to ruling out reductions exceeding roughly 3 percent of the pre-reform average. For comparison, \cite{dinersteinsmith021} estimate that a \$1000 per-student funding increase reduced the number of nearby private schools by 0.21 over six years, about 6 percent of the pre-reform average. The conclusion breaks down when the lower bound of the identified set reaches $-0.1$.

For this application, I construct the sensitivity analysis using the identified sets in Theorem \ref{thm:attmultpretrendsk}, which calibrate anticipation effects relative to the actual treatment effect. This parameterization is natural in this setting because anticipation and the post-treatment effect likely operate through the same behavioral channel: expected future enrollment losses may induce exposed private schools to close before implementation. If the treatment effect is negative, values of $k_t$ between zero and one capture anticipation also being negative but smaller in magnitude, reflecting uncertainty among exposed private schools about the extent of future enrollment losses. I do not apply the benchmarking procedure because expectations about the reform may have evolved over an extended period, leaving no clean pre-announcement window.

\begin{figure}[t]
    \centering

    \caption{Estimated Breakdown Frontier for $\text{ATT}_{2008} > -0.1$ and Identified Sets}
    \label{fig:bfkparam}

    \begin{subfigure}{0.46\textwidth}
        \centering
        \includegraphics[width=\textwidth]{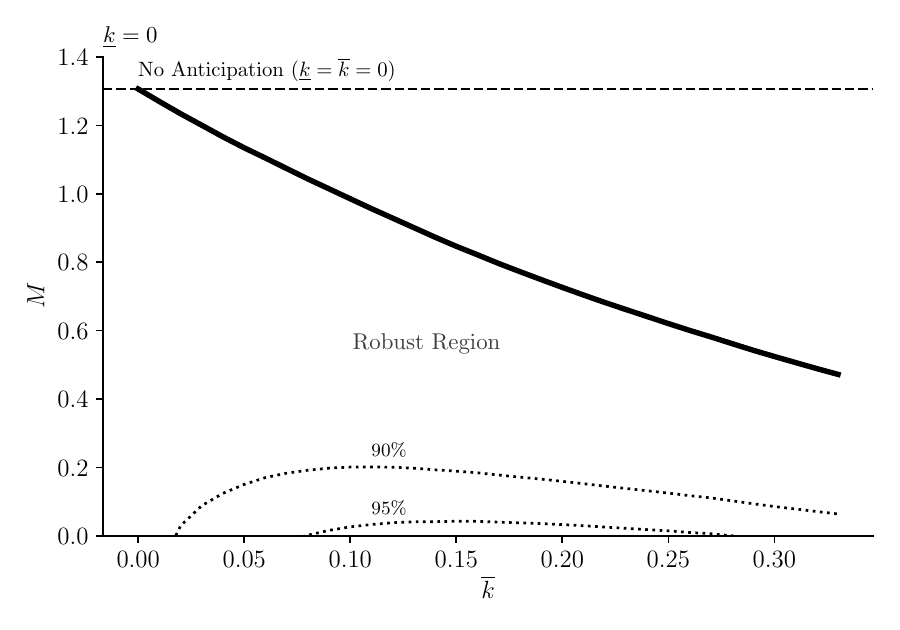}
        \caption{}
    \end{subfigure}
    \hspace{0.6em}
    \begin{subfigure}{0.46\textwidth}
        \centering
        \includegraphics[width=\textwidth]{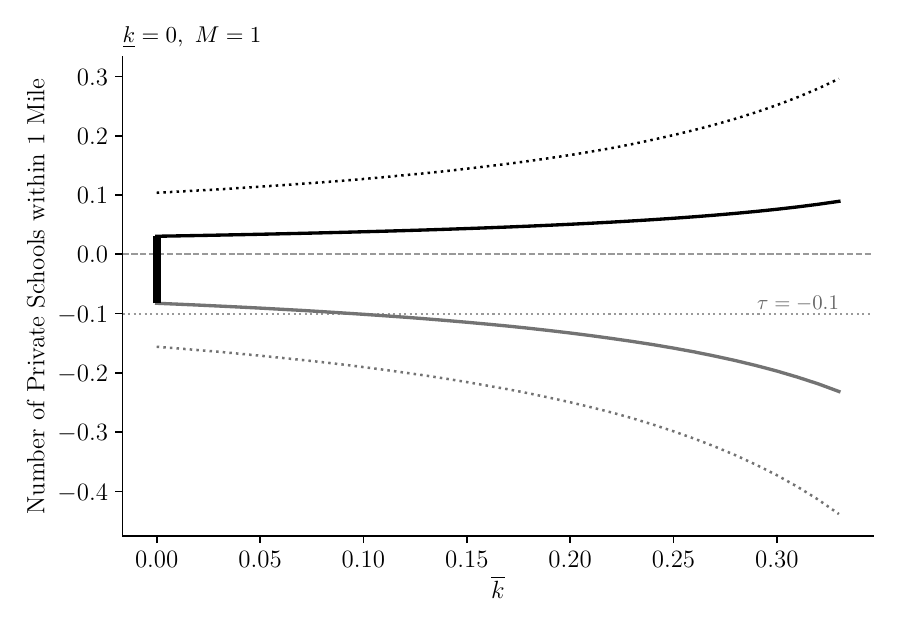}
        \caption{}
    \end{subfigure}

    \caption*{\footnotesize \textit{Notes:} In Panel (a), the estimated breakdown frontier for $\text{ATT}_{2008} > -0.1$ is denoted by the solid line and the 90\% and 95\% simultaneous lower credible bands are denoted by dotted lines. See Appendix \ref{subsec:simullowercredbands} for additional detail on the simultaneous lower credible band and its construction. In Panel (b), the estimated bounds of the identified sets are denoted by the solid lines and 95\% pointwise credible sets are denoted by the dashed lines. The darker lines correspond to the upper bound and the lighter lines correspond to the lower bound. See Appendix \ref{subsec:credints} for additional detail on the pointwise credible sets and their construction. The identified set under deviations from parallel trends only $(\underline{k} = \overline{k} = 0)$ is shown in bold.  Data Source: \cite{DinersteinSmith2021Data}.}

\end{figure}
Figure \ref{fig:bfkparam} presents the breakdown frontier for this conclusion, along with the identified sets at $M=1$. To simplify the setup for this illustration and allow for two-dimensional visualization, I set $\underline{k}_t = \underline{k}$ and $\overline{k}_{t} = \overline{k}$ for all $t \in \{-S, \ldots ,0\}$. In the figure, I hold $\underline{k}$ fixed at 0 and consider values of $\overline{k} \in [0,0.33]$. Appendix \ref{appendix:fsf} provides additional details on how this sensitivity analysis was constructed. 

Under no anticipation, the conclusion $\text{ATT}_{2008}>-0.1$ breaks down at approximately $M=1.3$. As the admissible range of negative anticipation expands, the breakdown value of $M$ falls, reaching about $0.52$ when $\overline{k}=0.3$. However, values of $M$ above one may be empirically relevant because the post-treatment period overlaps with the financial crisis, and \cite{dinersteinsmith021} note that recession effects may have differed across neighborhoods, potentially making post-treatment parallel trends violations larger than their pre-treatment counterparts.

\section{Conclusion}
\label{sec:conclusion}

Motivated by the observational equivalence of anticipation effects and parallel trends violations in the pre-treatment period, this paper develops a novel approach for conducting sensitivity analysis under simultaneous deviations from the no anticipation and parallel trends assumptions in the DiD framework. I propose a general class of assumptions on anticipation and derive closed-form sharp bounds for several common treatment effect parameters under simultaneous violations of both assumptions. Using these bounds, I construct a joint sensitivity analysis based on the breakdown frontier approach of \cite{mastenpoirier2020}. I also develop a benchmarking procedure for settings with a clean announcement window to calibrate the range of anticipation sensitivity parameters most relevant for assessing robustness. Once this range is obtained, the practical interpretation of the multi-dimensional sensitivity analysis becomes similar to that of a familiar one-dimensional sensitivity analysis.

More broadly, this paper shows that studying deviations from no anticipation and parallel trends jointly can reveal insights that are not apparent when contemplating each assumption separately. Rather than treating departures from the two assumptions as unrelated, the analysis links anticipation effects and parallel trends violations through the pre-trends, which discipline how much the researcher can deviate from one assumption relative to the other. This approach highlights meaningful robustness trade-offs that do not arise when each assumption is studied in isolation. Extending this perspective to other settings by using identified objects to discipline joint departures from multiple identifying assumptions offers a promising direction for future research. Another direction for future work is to extend the joint sensitivity framework to more complex designs in which comparison units may also anticipate, such as staggered adoption settings where not-yet-treated units comprise the comparison group or settings in which units in the untreated group incorrectly anticipate treatment. Overall, this paper's proposed sensitivity framework extends existing approaches that study deviations from a single assumption and provides a more nuanced and comprehensive picture of sensitivity when multiple identifying assumptions may fail.

\singlespacing
\bibliographystyle{econometrica}
\bibliography{references}

\makeatletter\@input{suppaux.tex}\makeatother

\end{document}


\maketitle

\appendix
{\fontsize{10pt}{12pt}\selectfont

\makeatletter
\@addtoreset{lemma}{section}
\@addtoreset{proposition}{section}
\@addtoreset{corollary}{section}
\@addtoreset{table}{section}
\@addtoreset{equation}{section}

\renewcommand{\thelemma}{\Alph{section}.\arabic{lemma}}
\renewcommand{\theproposition}{\Alph{section}.\arabic{proposition}}
\renewcommand{\thecorollary}{\Alph{section}.\arabic{corollary}}
\renewcommand{\thetable}{\Alph{section}.\arabic{table}}
\renewcommand{\theequation}{\Alph{section}.\arabic{equation}}
\makeatother

\onehalfspacing

\setcounter{tocdepth}{2} 
{\hypersetup{linkcolor=black}
\tableofcontents
}
\vspace{1em} 

\section{Additional Results: Shorter Identified Sets in a Three-Period Case}\label{appendix:discussion23}

This section presents additional results related to the discussion in Section \ref{subsec:discussion}. Throughout this section, I focus on the three-period case with a single pre-trend, i.e. $\mathcal{S}=\{0\}$, and consider the corresponding identified set in Theorem \ref{thm:attmultpretrendsa}.

Specifically, I establish three supporting results. First, I characterize the feasible identified sets that arise under Theorem \ref{thm:attmultpretrendsa} in the $\mathcal{S}=\{0\}$ case and the DGPs under which they obtain. Second, I characterize the DGPs under which the identified set is shorter under joint violations of parallel trends and no anticipation compared to when we only consider departures from parallel trends. Third, I formally state that a shorter identified set under joint deviations from parallel trends and no anticipation relative to relaxing parallel trends alone obtains only when anticipation is required to be present, i.e., when $0 \notin [\underline{A}_0, \overline{A}_0]$. \smallskip 

\noindent The first result characterizes feasible configurations of $\mathcal{I}_{\text{ATT}_{1}}^{A}$ when $\mathcal{S} = \{0\}$. 

\begin{lemma}\label{lem:feasiblesets23}
Suppose $\mathcal{S}=\{0\}$ and let $\underline{A}_{0}\le \overline{A}_{0}$. For $a\in\{\underline{A}_{0},\overline{A}_{0}\}$, define
\[
\text{LB}(a) \coloneqq \theta_1 + a - M\lvert \Delta_0 - a\rvert,
\qquad
\text{UB}(a) \coloneqq \theta_1 + a + M\lvert \Delta_0 - a\rvert.
\]
The identified set in Theorem \ref{thm:attmultpretrendsa} is
\[
\mathcal{I}^{A}_{\text{ATT}_1}
=
\Bigl[
\min\{\text{LB}(\underline{A}_{0}),\text{LB}(\overline{A}_{0})\},\ 
\max\{\text{UB}(\underline{A}_{0}),\text{UB}(\overline{A}_{0})\}
\Bigr].
\]
If $M>0$, define the thresholds
\[
c_L \coloneqq \frac{\underline{A}_{0}(1+M)-\overline{A}_{0}(1-M)}{2M},
\qquad
c_U \coloneqq \frac{\overline{A}_{0}(1+M)-\underline{A}_{0}(1-M)}{2M}.
\]
Then $\mathcal{I}^{A}_{\text{ATT}_1}$ can take the following forms:
\begin{enumerate}[label=(\roman*)]
    \item $\bigl[ \theta_1 + \underline{A}_{0} - M \lvert \Delta_0 - \underline{A}_{0}\rvert,\ 
                  \theta_1 + \underline{A}_{0} + M \lvert \Delta_0 - \underline{A}_{0}\rvert \bigr]$
    if
    \begin{itemize}
        \item $\overline{A}_{0} \leq \Delta_0$ and $M \geq 1$, or
        \item $\underline{A}_{0} \leq \Delta_0 \leq \overline{A}_{0}$, $\Delta_0 > c_U$, and $M > 1$.
    \end{itemize}

    \item $\bigl[ \theta_1 + \overline{A}_{0} - M \lvert \Delta_0 - \overline{A}_{0}\rvert,\ 
                  \theta_1 + \overline{A}_{0} + M \lvert \Delta_0 - \overline{A}_{0}\rvert \bigr]$
    if
    \begin{itemize}
        \item $\underline{A}_{0} \geq \Delta_0$ and $M \geq 1$, or
        \item $\underline{A}_{0}\le \Delta_0 \le \overline{A}_{0}$, $\Delta_0 < c_L$, and $M > 1$.
    \end{itemize}

    \item $\bigl[ \theta_1 + \underline{A}_{0} - M \lvert \Delta_0 - \underline{A}_{0}\rvert,\ 
                  \theta_1 + \overline{A}_{0} + M \lvert \Delta_0 - \overline{A}_{0}\rvert \bigr]$
    if
    \begin{itemize}
        \item $M=0$, or
        \item $\overline{A}_{0} \leq \Delta_0$ and $M < 1$, or
        \item $\underline{A}_{0} \geq \Delta_0$ and $M < 1$, or
        \item $\underline{A}_{0} \leq \Delta_0 \leq \overline{A}_{0}$ and $c_L \leq \Delta_0 \leq c_U$.
    \end{itemize}

    \item If $\underline{A}_{0}<\overline{A}_{0}$, there does not exist a DGP such that
    \[
    \mathcal{I}^{A}_{\text{ATT}_1}
    =
    \bigl[
    \theta_1 + \overline{A}_{0} - M \lvert \Delta_0 - \overline{A}_{0}\rvert,\ 
    \theta_1 + \underline{A}_{0} + M \lvert \Delta_0 - \underline{A}_{0}\rvert
    \bigr].
    \]
\end{enumerate}
\end{lemma} \smallskip

\noindent Now that the feasible sets have been characterized in Lemma \ref{lem:feasiblesets23}, I state the width of the identified set in (\ref{eqn:23aset}) under parallel trends violations only, and joint violations of parallel trends and no anticipation. \smallskip

\noindent The width of (\ref{eqn:23aset}) under parallel trends violations only $(\underline{A}_{0} = \overline{A}_0 = 0)$ is given by:
    \begin{equation}\label{width:ptv}
        W_{PT} \coloneqq 2 M |\Delta_0|,
    \end{equation}
    and the width of (\ref{eqn:23aset}) under joint violations of parallel trends and no anticipation is given by:
\begin{equation}\label{width:ptae}
W_{\text{PTAE}} \;:=\;
\begin{cases}
2M\,\lvert \Delta_0 - \widetilde{A} \rvert
& \text{(same endpoint)}\\[4pt]
(\overline{A}_{0}-\underline{A}_{0})
+ M\!\bigl(\lvert\Delta_0-\overline{A}_{0}\rvert+\lvert\Delta_0-\underline{A}_{0}\rvert\bigr)
& \text{(opposite endpoints)}
\end{cases}
\end{equation}
where $\widetilde{A}  \in \{ \underline{A}_{0}, \overline{A}_{0} \}$ denotes the endpoint that yields the lower and upper bounds in the same endpoint case. The next result characterizes the DGPs under which $W_{PTAE} < W_{PT}$.  \smallskip 

\begin{proposition}\label{corr:Awidth}
Suppose $M>0$. Then $W_{PTAE}<W_{PT}$ holds in the following cases.

\smallskip
\noindent \textup{\textbf{(i) Same-endpoint bounds.}} Suppose the lower and upper bounds are attained at the same endpoint
$\widetilde{A}\in\{\underline{A}_{0},\overline{A}_{0}\}$
(this includes the singleton case $\underline{A}_{0}=\overline{A}_{0}$). Then
\[
W_{PTAE}< W_{PT}
\ \Longleftrightarrow\
\widetilde{A} \in
\begin{cases}
(0,\,2\Delta_0), & \text{if } \Delta_0 > 0,\\
(2\Delta_0,\,0), & \text{if } \Delta_0 < 0,
\end{cases}
\]
(and the condition is never satisfied when $\Delta_0=0$).

\smallskip
\noindent \textup{\textbf{(ii) Cross-endpoint bounds.}} Suppose the lower and upper bounds are attained at different endpoints. By Lemma \ref{lem:feasiblesets23}, the only feasible cross-endpoint configuration is
\[
\bigl[\text{LB}(\underline{A}_{0}),\ \text{UB}(\overline{A}_{0})\bigr].
\]
In this case, $W_{PTAE}<W_{PT}$ holds in the following feasible sub-cases:
\begin{enumerate}[label=\textit{(ii.\alph*)}, leftmargin=2.1em]
\item $\overline{A}_{0}\le \Delta_0$ and $M<1$: $W_{PTAE}< W_{PT}
\ \Longleftrightarrow\
\Delta_0-|\Delta_0|\ <\ \frac{\underline{A}_{0} (1+M) - \overline{A}_{0} (1-M)}{2M}.$

\item $\underline{A}_{0}\ge \Delta_0$ and $M<1$: $W_{PTAE}< W_{PT}
\ \Longleftrightarrow\
\Delta_0+|\Delta_0|\ >\ \frac{\overline{A}_{0} (1+M)-\underline{A}_{0} (1-M)}{2M}.$

\item $\underline{A}_{0}\le \Delta_0 \le \overline{A}_{0}$ and
$\frac{\underline{A}_{0} (1+M)-\overline{A}_{0} (1-M)}{2M} \le \Delta_0 \le \frac{\overline{A}_{0} (1+M)-\underline{A}_{0} (1-M)}{2M}$: \\
$W_{PTAE}< W_{PT}
\ \Longleftrightarrow\
|\Delta_0|\ >\ \frac{(\overline{A}_{0}-\underline{A}_{0})(1+M)}{2M}\,.$
\end{enumerate}
\end{proposition}

Proposition \ref{corr:Awidth} characterizes the DGPs under which the identified set is strictly shorter under parallel trends violations and anticipation (where $0 \notin [\underline{A}_0, \overline{A}_0]$), compared to when we only consider departures from parallel trends. The following corollary highlights a direct implication of this result that is particularly useful in practice.

\begin{corollary}\label{corr:0inAset23}
    Suppose $M>0$. If $0 \in [\underline{A}_{0}, \overline{A}_{0}]$, then it cannot be the case that $W_{PTAE} < W_{PT}$.
\end{corollary}

\section{Additional Discussion of Assumption \ref{assump:nonzerodenomk}}
\label{appendix:assumpnonzero}

By Assumption \ref{assump:nonzerodenomk}, in order to have finite bounds in the identified set in Theorem \ref{thm:attmultpretrendsk} we need to ensure the denominator of the below $\text{ATT}_{1}$ objective (see proof of Theorem \ref{thm:attmultpretrendsk} in Appendix \ref{appendix:proofs} for derivation of this expression) is nonzero across the entire feasible set:
$$f\big((k_j)_{j\in J(r)},m, r \big) = \frac{\theta_1 - m \Delta_{r}}{1 - k_{0} - m(k_{r} - k_{r-1})}$$
To provide some intuition for this condition, consider the simple case where $\underline{k}_s = -K$ and $\overline{k}_s = K$ for some $K \geq 0$ for all $s \in \{-S, \ldots , 0 \}$. The next result provides a necessary and sufficient condition for ensuring that the nonzero denominator condition holds under this set-up.

\begin{lemma}\label{lemma:nonzerokmultpretrends}
    Let $K \geq 0$ and $M \geq 0$. Then, for every $r \in \mathcal{S}$, $1 - k_0 - m(k_r - k_{r-1}) \neq 0$
for every feasible 
$\bigl( (k_j)_{j \in J(r)}, m   \bigr) \in \prod_{j \in J(r)} [-K,K] \times [-M,M]$ if and only if $K < \frac{1}{1+2M}.$
\end{lemma}

The necessary and sufficient condition for a nonzero denominator in Lemma \ref{lemma:nonzerokmultpretrends} is not innocuous, as it places meaningful restrictions on the admissible choice of $K$ relative to $M$. For instance, the condition requires that $1 \notin [-K,K]$. This condition arises because $0 \in [-M,M]$ by construction, so if $k_{0}=1$ was feasible, then at $m=0$ the denominator of the $\text{ATT}_{1}$ objective would equal $1-k_0=0$ for $k_0 =1$. In the symmetric $[-K,K]$ specification, this therefore requires $K<1$. Intuitively, this restriction corresponds to ruling out anticipation effects that are as large as, or larger than, the treatment effect itself. Nonetheless, such a restriction could be plausible in many applied settings.

More broadly, the condition highlights a tradeoff between allowable deviations from parallel trends in the post-treatment period and allowable anticipation effects: as $M$ increases, the upper bound on $K$ necessarily decreases. For example, when $M= 1/2$ the condition requires $K<1/2$, when $M=1$ it requires $K<1/3$, and when $M=2$ it requires $K<1/5$. However, the symmetric restriction $\underline{k}_s = -K$ and $\overline{k}_s = K$ for all $s \in \{-S, \ldots , 0 \}$ represents one particular choice of the admissible set for the $k_{s}$ parameters. Alternative specifications of this set can in principle make the nonzero denominator condition less restrictive\footnote{\label{fn:kexample}For example, if one considers the set-up $\underline{k}_s = 0$ and $\overline{k}_s = K$ for all $s \in \{-S, \ldots , 0\}$, the necessary and sufficient condition for a nonzero denominator across the entire feasible set becomes $K < 1 \slash (1+M)$.}.

Consider the three-period case where $\mathcal{S} = \{0 \}$ and to enforce $\varphi_{-1} = 0$, set $\underline{k}_{-1} = \overline{k}_{-1} = 0$. Then, the denominator in the $\text{ATT}_{1}$ objective becomes $1 - k_{0}(1+m)$. If Assumption \ref{assump:nonzerodenomk} did not hold, then the $\text{ATT}_{1}$ objective would be undefined for values  $(k_0,m)\in [\underline{k}_0,\overline{k}_0] \times [-M,M]$ that yield a zero denominator. Whether both the upper and lower bounds equal infinity depends on whether the values of $(k_0, m)$ that yield a zero denominator occur on the interior or only the boundary of the feasible region $[\underline{k}_0, \overline{k}_0] \times [-M,M]$. If the singularity occurs only on the boundary, then the resulting identified set for the $\text{ATT}_{1}$ may only have one infinite bound, which depends on the sign of the numerator of the $\text{ATT}_{1}$ objective on the boundary and the denominator on the interior. If the sign of the numerator varies along the singularity or the singularity occurs on the interior of the feasible region, then both bounds of the identified set for the $\text{ATT}_{1}$ equal infinity.

\section{Additional Details on Estimation and Inference}
\label{appendix:inference}

\subsection{Bayesian Bootstrap}\label{subsec:bayesianbootstrap}
Let $P$ denote the joint distribution of $(X,\;Y_{-S}, \ldots , Y_{1})$. We observe an i.i.d. sample $Z = \{Z_i\}_{i=1}^{n}$ drawn from $P$, where $Z_i \coloneqq (X_i, Y_{i,-S}, \ldots , Y_{i,1})$. \smallskip 

\noindent Define the following functionals of $P$: $\mathbb{E}_P\!\big[(Y_t - Y_{t-1})X\big], \; \mathbb{E}_P\!\big[(Y_t - Y_{t-1})(1-X)\big], \; \mathbb{E}_P[X].$
Knowing $P$ therefore pins down all objects needed to construct $\gamma $.

\begin{enumerate}[label=\textbf{Step \arabic*:}, leftmargin=*, align=left]
    \item Beginning with the sampling logic, for each bootstrap draw $b = 1, \ldots B$, generate the weights as follows:
    \begin{itemize}
        \item Generate $(V_1^{(b)}, \ldots , V_n^{(b)}) \stackrel{iid}{\sim}\text{Exp}(1)$.
        \item Using the $V_i$, construct the weights\footnote{Note that Dirichlet distributions can be represented through Gamma random variables, and that $Exp(1)$ is equivalent to $Gamma(1,1)$.}: 
        $w_i^{(b)}=\frac{V_i^{(b)}}{\sum_{j=1}^n V_j^{(b)}}.$
        
        \item Compute $\mathbb{E}_{P^{(b)}}\!\big[(Y_t - Y_{t-1})X\big], \; \mathbb{E}_{P^{(b)}}\!\big[(Y_t - Y_{t-1})(1-X)\big], \; \mathbb{E}_{P^{(b)}}[X]$ as follows:
        \begin{align*}
            \mathbb{E}_{P^{(b)}}\!\big[(Y_t - Y_{t-1})X\big] &= \sum_{i=1}^{n} w_{i}^{(b)} (Y_{i,t} - Y_{i,t-1}) X_i \\
            \mathbb{E}_{P^{(b)}}\!\big[(Y_t - Y_{t-1})(1-X)\big] &= \sum_{i=1}^{n} w_{i}^{(b)} (Y_{i,t} - Y_{i,t-1}) (1-X_i) \\
            \mathbb{E}_{P^{(b)}}[X] &= \sum_{i=1}^{n} w_{i}^{(b)} X_i
        \end{align*}
        \item Compute $\gamma^{(b)} = (\Delta_{-(S-1)}^{(b)}, \ldots , \Delta_{0}^{(b)}, \theta_{1}^{(b)} )^{\prime} $. 
        \item Repeat $B$ times to get a sample of posterior draws $\{\gamma^{(b)} \}_{b=1}^{B}$. 
    \end{itemize}
    \item Fix values of the sensitivity parameters $(\boldsymbol{a}, M)$. For each posterior draw \\
    $\gamma^{(b)} = (\Delta_{-(S-1)}^{(b)}, \ldots , \Delta_{0}^{(b)}, \theta_{1}^{(b)})^{\prime},$
    compute the corresponding identified set, which we will denote by
    $\mathcal{I}^{\boldsymbol{a},(b)}_{\text{ATT}_{1}}$. Section \ref{subsec:credints} provides additional detail on how we use $\{ \mathcal{I}^{\boldsymbol{a},(b)}_{\text{ATT}_{1}} \}_{b=1}^{B}$ to conduct inference on the identified sets.
    
    \item For each posterior draw $\gamma^{(b)}$, compute the corresponding draw of the breakdown frontier (see Appendix \ref{appendix:ksensitivity}): $M^{bp, (b)}(\boldsymbol{a}) \equiv M^{bp}(\boldsymbol{a}; \gamma^{(b)}).$

    To construct a two-dimensional breakdown frontier, I consider a finite gride of anticipation sensitivity values $\{ \boldsymbol{a}_j \}_{j=1}^{J}$, holding fixed any components of $\boldsymbol{a}$ that are not varied in the figure. For example, in Figure \ref{fig:bfkparam} with $\underline{k} = 0$, $\boldsymbol{a}_j = (0, \overline{k}_j)$. For each grid point $\boldsymbol{a}_j$, I compute the posterior draws $\{M^{bp, (b)}(\boldsymbol{a}_j)\}_{b=1}^{B}$ and estimate the breakdown frontier by the posterior median:
   $$\hat m_j \;=\; \operatorname{Med}\left( \!\left\{\, M^{bp,(b)}\!(\boldsymbol{a}_j) \right\}_{b=1}^{B} \right).$$
Plotting the scalar component of $\boldsymbol{a}_j$ varied in the figure against $\hat{m}_j$ and connecting adjacent grid points yields the estimated two-dimensional breakdown frontier.

\end{enumerate}

\subsection{Simultaneous Lower Credible Band for the Breakdown Frontier}
\label{subsec:simullowercredbands}

Using the outputs from the Bayesian bootstrap, I now outline how to construct simultaneous lower credible bands for the breakdown frontier. We want to construct \textit{simultaneous} credible bands rather than \textit{pointwise} credible bands, because we are conducting inference over an entire function---i.e.\ we want to make statements about the degree of certainty that the entire curve lies above the one-sided lower band. If we instead relied on a sequence of pointwise statements, this could lead to overly optimistic inferential statements. We are constructing lower bands rather than upper bands because we are interested in claims regarding the \textit{robustness} of a conclusion. \smallskip

\noindent Below, I outline the construction of $\hat{L}(\boldsymbol{a})$.

\begin{enumerate}[label=\textbf{Step \arabic*:}, leftmargin=*, align=left]
    \item Consider a finite grid $\mathcal{H} = \{\boldsymbol{a}_j\}_{j=1}^{J}$ of anticipation sensitivity values. The grid may vary all components of $\boldsymbol{a}$ or it may vary one component while holding the others fixed. From Step 3 of the Bayesian bootstrap procedure, we have:
    $$M^{bp, (b)}(\boldsymbol{a}_j) \quad \text{for } b = 1, \ldots , B; \; j = 1, \ldots , J.$$
    For each $\boldsymbol{a}_j$, let $\overline{m}_j$ denote the approximated posterior mean of $M^{bp, (b)}(\boldsymbol{a}_j)$:
    $$\overline{m}_j = \frac{1}{B} \sum_{b=1}^{B} M^{bp, (b)}(\boldsymbol{a}_j).$$
    Let $s_j$ denote the approximated posterior standard deviation:
    $$s_j = \sqrt{\frac{1}{B} \sum_{b=1}^{B} (M^{bp, (b)}(\boldsymbol{a}_j) - \overline{m}_j)^{2}}.$$

    \item Calculate the largest downward deviation for each simulated draw $b$:
    $$D^{(b)} = \max_{j = 1, \ldots , J} \left\{ \max \left\{ \frac{\hat{m}_j  - M^{bp, (b)}(\boldsymbol{a}_j) }{s_j}, 0 \right\} \right\}.$$

    \item The critical value $c^{D}_{1-\alpha}$ is the $(1-\alpha)$ posterior quantile of $D$.:
    $$c^{D}_{1-\alpha} \coloneqq \inf \{ c \in \mathbb{R} : \Pi(D \leq c \mid Z) \geq 1 - \alpha\}.$$
    In practice, this is approximated by the $(1-\alpha)$ empirical quantile of $\{ D^{(b)} \}_{b=1}^{B}$.

    \item Construct the simultaneous lower credible band on the grid $\mathcal{H}$:
    $$\hat{L}(\boldsymbol{a}) \coloneqq \hat m_j  - c^{D}_{1-\alpha} \cdot s_j \quad \text{ for each } j = 1, \ldots , J.$$
\end{enumerate}

Then, as stated in (\ref{eqn:simullowercredbands}), we have:
\begin{equation*}
   \Pi (\hat{L}(\boldsymbol{a}) \leq M^{bp}(\boldsymbol{a}) \text{ for all } \boldsymbol{a} \in \mathcal{H} \mid Z) \geq 1- \alpha.
\end{equation*}
This holds with equality if the posterior distribution of $D \mid Z$ is continuous at $c^{D}_{1-\alpha}$. The proof of this posterior coverage statement can be found in Appendix \ref{appendix:proofs}. \smallskip

\noindent I use the convention $M^{bp}(\boldsymbol{a}_j)=+\infty$ to denote cases in which no finite breakdown point exists over the admissible range of $M$. The simultaneous lower credible band is constructed using posterior moments computed from the finite draws of $M^{bp}(\boldsymbol{a}_j)$. Because the band is one-sided and lower, it is constructed from downward deviations; an infinite draw is an upward deviation and therefore contributes zero. Studentized deviations are evaluated only when the finite-draw posterior standard deviation is positive. When at least half of the posterior draws are infinite, I treat the posterior median breakdown frontier as infinite and report no finite band.

\subsection{Pointwise Credible Sets for the Identified Sets}\label{subsec:credints}

Outputs from the Bayesian bootstrap can also be used to construct (pointwise) credible sets for the identified sets. Here, inference is pointwise in $(\boldsymbol{a}, M)$: for a fixed choice of $(\boldsymbol{a}, M)$, we are interested in the posterior probability that the credible set contains the corresponding identified set, rather than in simultaneous coverage over the full lower- and upper-bound functions. \smallskip

\noindent Section \ref{subsec:bayesianbootstrap} outlines how the Bayesian bootstrap posterior on $P$ induces a posterior for $\gamma$. Pushing each posterior draw $\gamma^{(b)}$ through the mapping that defines a given identified set induces a posterior distribution over the identified set endpoints. For a fixed choice of $(\boldsymbol{a}, M)$ write
$$\mathcal{I}^{\boldsymbol{a}}_{\text{ATT}_{1}}(\boldsymbol{a}, M;P) = \bigl[ L(\boldsymbol{a}, M;P), U(\boldsymbol{a}, M;P) \bigr], \quad P \mid Z \sim \Pi(\cdot \mid Z),$$
where each bootstrap draw $P^{(b)}$ yields endpoints $L^{(b)} = L(\boldsymbol{a}, M;P^{(b)})$ and $U^{(b)} = U(\boldsymbol{a}, M;P^{(b)})$. \smallskip

\noindent The idea behind the construction of the credible sets is to symmetrically expand outward from an estimate of the identified set until the Bayesian credibility level reaches $1-\alpha$ \citep{klinetamer2016}. Below, I describe how this idea is implemented using Bayesian bootstrap draws. \smallskip

\begin{enumerate}[label=\textbf{Step \arabic*:}, leftmargin=*, align=left]
    \item For a given choice of sensitivity parameters $(\boldsymbol{a}, M)$ and each Bayesian bootstrap draw \(b=1,\ldots,B\), compute the identified set: $\bigl[L^{(b)},U^{(b)}\bigr]$.

\item Take the median of the lower and upper bounds across the approximated posterior:
$$\hat{L}
=
\operatorname{Med}\bigl(\{L^{(b)}\}_{b=1}^B\bigr),
\qquad
\hat{U}
=
\operatorname{Med}\bigl(\{U^{(b)}\}_{b=1}^B\bigr).$$

\item To construct the credible set, symmetrically expand out from $[\hat{L}, \hat{U}]$:
$$[\hat{L} - c, \hat{U} + c], \qquad c \geq 0.$$
Note that:
\begin{align*}
    [L,U] \subseteq [\hat{L} - c, \hat{U} + c] &\Longleftrightarrow L \geq \hat{L} - c \text{ and } U \leq \hat{U} + c \\
    &\Longleftrightarrow c \geq \hat{L} - L \text{ and } c \geq U - \hat{U}.
\end{align*}
Hence, we have $[L,U] \subseteq [\hat{L} - c, \hat{U} + c]$ if and only if:
$$c \geq \max \{\hat{L} - L, U - \hat{U}, 0 \}.$$
Define $C \coloneqq \max \{\hat{L} - L, U - \hat{U}, 0 \}$, where $C$ is the minimum symmetric enlargement required for the interval $[\hat{L} - c, \hat{U} + c]$ to contain the identified set $[L,U]$ (i.e., $[L,U] \subseteq [\hat{L} - c, \hat{U} + c] \iff C \leq c$). For each draw, compute:
$$C^{(b)} \coloneqq \max \{\hat{L} - L^{(b)}, U^{(b)} - \hat{U}, 0 \}.$$

\item Let $c^{C}_{1-\alpha}$ denote the $(1-\alpha)$ posterior quantile of $C \mid Z$:
$$c^{C}_{1-\alpha} \coloneqq \inf \{ c \in \mathbb{R}: \Pi(C \leq c \mid Z) \geq 1- \alpha  \}.$$
In practice, $c^{C}_{1-\alpha}$ is approximated by the $(1-\alpha)$ empirical quantile of $\{C^{(b)}\}_{b=1}^{B}$.

\item The $(1-\alpha)$-credible set for $\mathcal{I}^{\boldsymbol{a}}_{\text{ATT}_{1}}(\boldsymbol{a}, M;P)$ is given by:
$$\hat{I}_{1-\alpha}(\boldsymbol{a}, M) \coloneqq \bigl[ \hat{L} - c^{C}_{1-\alpha}, \hat{U} + c^{C}_{1-\alpha}  \bigr],$$
where:
\begin{equation}\label{eqn:credsets}
    \Pi(\mathcal{I}^{\boldsymbol{a}}_{\text{ATT}_{1}}(\boldsymbol{a}, M;P) \subseteq \hat{I}_{1-\alpha}(\boldsymbol{a}, M) \mid Z) \geq 1 -\alpha,
\end{equation}
with equality if the posterior distribution of $C \mid Z$ is continuous at $c^{C}_{1-\alpha}$. The proof of this posterior coverage statement can be found in Appendix \ref{appendix:proofs}.
\end{enumerate}

\section{Deriving the Breakdown Frontier}\label{appendix:ksensitivity}

\subsection{Medicare Part D \citep{engelhardtgruber2011, alpert2016}}
The conclusion of interest is $\text{ATT}_{2006}>0$, so the relevant endpoint of the identified set is the lower bound. The announcement occurred in December 2003, so I treat 2001--2003 as the pre-announcement window and 2004--2005 as the post-announcement, pre-treatment window. Let $A_t \coloneqq \varphi_t - \varphi_{t-1}$. Then, \eqref{eq:atq} used in the application imposes
\[
A_t(q) =
\begin{cases}
0, & t\in\{2001,2002,2003\},\\
q\Delta^{\text{all}}, & t\in\{2004,2005\},
\end{cases}
\]
where $\Delta^{\text{all}}=\max_{t\in\{2001,\ldots,2005\}}|\Delta_t|.$ Thus, by Lemma \ref{aeexpression} the implied anticipation effect in the baseline period is \[\varphi_0(q)=\sum_{t=2001}^{2005} A_t(q)=2q\Delta^{\text{all}}.\]
By Corollary \ref{prop:pretrenddecopmpsim}, the corresponding pre-treatment parallel trends violations are $\Delta_t-A_t(q)$, so define 
\[B(q)\coloneqq \max_{t\in\{2001,\ldots,2005\}}|\Delta_t-A_t(q)|.\]
Equivalently, $
B(q)
=
\max\left\{
|\Delta_{2001}|,
|\Delta_{2002}|,
|\Delta_{2003}|,
|\Delta_{2004}-q\Delta^{\text{all}}|,
|\Delta_{2005}-q\Delta^{\text{all}}|
\right\}.
$ \smallskip 

\noindent Applying the identified set from Theorem \ref{thm:attmultpretrendsa}, the identified set for the $\text{ATT}_{2006}$ under sensitivity parameters $(q,M)$ is
\[
\left[
\theta_{2006}+2q\Delta^{\text{all}}-M \cdot B(q),
\;
\theta_{2006}+2q\Delta^{\text{all}}+M \cdot B(q)
\right].
\]
The conclusion $\text{ATT}_{2006}>0$ holds whenever the lower bound is strictly positive. Therefore, the breakdown value of $M$ is
\[
M^{bp}(q)
\coloneqq
\inf\left\{
M\geq 0:
\theta_{2006}+2q\Delta^{\text{all}}-M \cdot B(q)\leq 0
\right\}.
\]
Solving this expression yields
\[
M^{bp}(q)
=
\begin{cases}
\displaystyle
\max\left\{0,\frac{\theta_{2006}+2q\Delta^{\text{all}}}{B(q)}\right\},
& \text{if } B(q)>0,\\[1.1em]
0,
& \text{if } B(q)=0 \text{ and } \theta_{2006}+2q\Delta^{\text{all}}\leq 0,\\[0.4em]
+\infty,
& \text{if } B(q)=0 \text{ and } \theta_{2006}+2q\Delta^{\text{all}}>0.
\end{cases}
\]

\subsection{FSF Reform \citep{dinersteinsmith021}}\label{appendix:fsf}

\subsubsection*{Setup}
For simplicity, let $\underline{k}_s = \underline{k}$ and $\overline{k}_s = \overline{k}$ for all $s \in \{-S, \ldots, 0\}$. For the breakdown frontier, I will hold $\underline{k}$ fixed at zero, and consider a range of values of $\overline{k}$ from 0 to 0.33.  \smallskip

\noindent We are interested in conclusions of the form $\text{ATT}_{2008} > \tau$, where $\tau \neq 0$ (see discussion following Assumption \ref{assump:nonzerodenomk} in the main text). Throughout this appendix section, I retain the relative-time indexing from Theorem \ref{thm:attmultpretrendsk} for the pre-treatment sensitivity parameters, where $t=-2,-1,0$ correspond to the 2002, 2004, and 2006 periods, respectively. Accordingly, $r-1$ denotes the preceding observed biennial period rather than the preceding calendar year. Consistent with Section \ref{subsec:privateschools}, the target parameter and post-treatment DiD estimand are indexed directly by calendar year, 2008.

I define the breakdown value of $M$ as the following:
$$M^{bp}(\underline{k}, \overline{k}) = \inf \{ M \in [0, M_{\text{max}}(\underline{k}, \overline{k})): \text{LB}^{k}(\underline{k}, \overline{k}, M) \leq \tau \},$$
where $\text{LB}^{k}(\underline{k}, \overline{k}, M)$ is the lower bound of the identified set in Theorem \ref{thm:attmultpretrendsk}:
$$\text{LB}^{k}(\underline{k}, \overline{k}, M) = \min_{r \in \mathcal{S}} \min_{\substack{
k_j \in \{\underline{k}_j, \overline{k}_j \}, \; j \in J(r)\\
m\in\{-M,M\}}}
\frac{\theta_{2008} - m\,\Delta_r}{
1 - k_0 - m\,(k_r - k_{r-1})
}$$
and the threshold $M_{\text{max}}(\underline{k}, \overline{k})$ enforces that Assumption \ref{assump:nonzerodenomk} holds across the entire feasible set. I derive $M_{\text{max}}(\underline{k}, \overline{k})$ in the next section. I use the convention that $\inf \emptyset = + \infty$. The value $+ \infty$ should not necessarily be interpreted as robust to arbitrarily large violations, but rather as having no breakdown of the conclusion within the admissible set.

\subsubsection*{Enforcing Assumption \ref{assump:nonzerodenomk}}

First, we need to ensure that the feasible set respects Assumption \ref{assump:nonzerodenomk}, otherwise the bounds on the $\text{ATT}_{2008}$ will not be finite. 

A nonzero denominator in the lower bound requires $|1 - k_0 - m(k_r - k_{r-1})| > 0$ for all $( (k_j)_{j \in J(r)}, m   ) \in [\underline{k}, \overline{k}]^{|J(r)|} \times [-M,M].$ Using the fact that we are minimizing a convex function over a compact interval, we obtain: $\min_{m \in [-M,M]} |1 - k_0 - m(k_r - k_{r-1}) | = \max \{0, |1-k_0 | - M |k_r - k_{r-1}| \}$. Therefore, we need:
\begin{equation}\label{nonzerodenomcondition}
|1 - k_0| - M |k_r - k_{r-1}| > 0
\qquad
\text{for all } (k_j)_{j \in J(r)} \in [\underline{k}, \overline{k}]^{|J(r)|}.
\end{equation}
Note that the left-hand side is smallest when $|k_r - k_{r-1}|$ is large, and $|1-k_0|$ is small. Further, note that:
\begin{enumerate}
    \item $|k_r - k_{r-1}| \leq \overline{k} - \underline{k}$
    \item Since we are considering values of $\overline{k}$ less than 1, we have: $|1 - k_0| = 1 - k_0$\footnote{Since $m=0$ is always feasible and we want to be able to consider proper relaxations---i.e. $0 \in [\underline{k}, \overline{k}]$---we must impose $\overline{k} < 1$ to exclude $k_0 = 1$ (which yields a zero denominator in the bounds in Theorem \ref{thm:attmultpretrendsk} when $m=0$).}. Hence, $|1 - k_0| \geq 1 - \overline{k}$. 
\end{enumerate}

\noindent Therefore,  $|1 - k_0| - M |k_r - k_{r-1}| \geq (1-\overline{k}) - M(\overline{k} - \underline{k}).$ Hence, a sufficient condition for (\ref{nonzerodenomcondition}) is: $(1-\overline{k}) - M(\overline{k} - \underline{k}) > 0 \Leftrightarrow M < \frac{1-\overline{k}}{\overline{k} - \underline{k}}.$ \smallskip

\noindent Next, I show necessity:
Conversely, if $\underline{k}<\overline{k}$ and
$M \geq (1-\overline{k})/(\overline{k}-\underline{k})$, then
$k_0=\overline{k}$, $k_r=\overline{k}$, $k_{r-1}=\underline{k}$, and
$m=(1-\overline{k})/(\overline{k}-\underline{k})$ are feasible and yield
\[
|1 - k_0| - m |k_r - k_{r-1}|
=
1-\overline{k}
-
\left( \frac{1-\overline{k}}{\overline{k}-\underline{k}} \right)
(\overline{k}-\underline{k})
=0.
\]
Thus, the nonzero-denominator condition fails whenever
$M \geq (1-\overline{k})/(\overline{k}-\underline{k})$. Therefore, we have:
$$M_{\text{max}}(\underline{k}, \overline{k}) = \begin{cases} 
      \frac{1-\overline{k}}{\overline{k} - \underline{k}} & \text{ if } \underline{k} < \overline{k} \\
      + \infty & \text{ if } \underline{k} = \overline{k}
   \end{cases}$$
where we restrict attention to $M \in [0,M_{\text{max}}(\underline{k}, \overline{k})). $ Notice that the upper endpoint is excluded.

Note that the restriction $M \in [0, M_{\text{max}}(\underline{k},\overline{k}))$ ensures that the denominator of $\text{LB}^{k}(\underline{k}, \overline{k}, M)$ is nonzero throughout the feasible set. Since I want to consider cases where $0 \in [\underline{k}, \overline{k}]$, I restrict $\overline{k} < 1$. This precludes a zero denominator when $m = 0$, which is always feasible (alternatively, one could restrict $\underline{k} > 1$ to prevent a zero denominator at $m=0$ but this makes $0 \in [\underline{k}, \overline{k}]$ infeasible). Under $M < M_{\text{max}}(\underline{k}, \overline{k})$ and $\overline{k} < 1$, the denominator is strictly positive throughout the feasible set.

\subsubsection*{Derivation}

Fix $r \in \mathcal{S}$ and $(k_j)_{j \in J(r)} \in [\underline{k}, \overline{k}]^{|J(r)|}$. Let $M \in [0, M_{\text{max}}(\underline{k}, \overline{k}))$.  \smallskip

\noindent \textbf{Case 1:} $m = M$.
\begin{align}
    &\frac{\theta_{2008} - M \Delta_r}{1 - k_0 - M(k_r - k_{r-1})} \leq \tau \notag \\
    \Longleftrightarrow \quad &\theta_{2008} - \tau(1-k_0) \leq M(\Delta_r - \tau(k_r - k_{r-1})), \label{mbpMpluscondition}
\end{align}
which follows since $1 - k_0 - m(k_r - k_{r-1}) > 0$ throughout the feasible set.

\begin{itemize}
    \item If $\Delta_r - \tau(k_r - k_{r-1} )> 0$, then (\ref{mbpMpluscondition}) is equivalent to:
    $M \geq \frac{\theta_{2008} - \tau(1-k_0)}{\Delta_r - \tau (k_r - k_{r-1})}.$

    \item If $\Delta_r - \tau(k_r - k_{r-1}) = 0$, then (\ref{mbpMpluscondition}) is equivalent to:
    $\theta_{2008} - \tau(1-k_0) \leq 0,$
    which either holds for all $M$ (if $\theta_{2008} \leq \tau(1-k_0)$) or does not hold for all $M$ (otherwise).

    \item If $\Delta_r - \tau(k_r - k_{r-1}) < 0$, then (\ref{mbpMpluscondition}) is equivalent to:
    $M \leq \frac{\theta_{2008} - \tau(1-k_0)}{\Delta_r - \tau (k_r - k_{r-1})}.$
    Then:
    \begin{itemize}
        \item If $\theta_{2008} - \tau(1-k_0) > 0$, then no value of $M \geq 0$ can satisfy (\ref{mbpMpluscondition}).
        \item If $\theta_{2008} - \tau(1-k_0) \leq 0$, then (\ref{mbpMpluscondition}) already holds at $M=0$, so breakdown occurs at $M=0$.
    \end{itemize}
\end{itemize}

Therefore, we have:
\begin{equation*}\label{eq:mu_plus_k}
\mu^{+}_{rk}(\underline{k}, \overline{k})
=
\begin{cases}
\displaystyle
\max\left\{ 0,\ \frac{\theta_{2008} - \tau(1-k_0)}{\Delta_r - \tau(k_r - k_{r-1})} \right\},
& \text{if } \Delta_r - \tau(k_r - k_{r-1}) > 0, \\[1.1em]
0,
& \text{if } \Delta_r - \tau(k_r - k_{r-1}) \le 0\ \text{and}\ \theta_{2008} - \tau(1-k_0) \le 0, \\[0.4em]
+\infty,
& \text{otherwise}.
\end{cases}
\end{equation*}

\noindent \textbf{Case 2:} $m = -M$.
\begin{align}
    &\frac{\theta_{2008} + M \Delta_r}{1 - k_0 + M(k_r - k_{r-1})} \leq \tau \notag \\
    \Longleftrightarrow \quad &\theta_{2008} - \tau(1-k_0) \leq -M(\Delta_r - \tau(k_r - k_{r-1})), \label{mbpMnegcondition}
\end{align}
which follows since $1 - k_0 - m(k_r - k_{r-1}) > 0$ throughout the feasible set.

\begin{itemize}
    \item If $\Delta_r - \tau(k_r - k_{r-1}) < 0$, then (\ref{mbpMnegcondition}) is equivalent to:
    $M \geq \frac{\theta_{2008} - \tau(1-k_0)}{-(\Delta_r - \tau(k_r - k_{r-1}))}.$

    \item If $\Delta_r - \tau(k_r - k_{r-1}) = 0$, then (\ref{mbpMnegcondition}) is equivalent to:
    $\theta_{2008} - \tau(1-k_0) \leq 0,$
    which either holds for all $M$ (if $\theta_{2008} \leq \tau(1-k_0)$) or does not hold for all $M$ (otherwise).

    \item If $\Delta_r - \tau(k_r - k_{r-1}) > 0$, then (\ref{mbpMnegcondition}) is equivalent to:
    $M \leq \frac{\theta_{2008} - \tau(1-k_0)}{-(\Delta_r - \tau(k_r - k_{r-1}))}.$
    Then:
    \begin{itemize}
        \item If $\theta_{2008} - \tau(1-k_0) > 0$, then no value of $M \geq 0$ can satisfy (\ref{mbpMnegcondition}).
        \item If $\theta_{2008} - \tau(1-k_0) \leq 0$, then (\ref{mbpMnegcondition}) already holds at $M=0$, so breakdown occurs at $M=0$.
    \end{itemize}
\end{itemize}

Therefore, we have:

\begin{equation*}\label{eq:mu_minus_k}
\mu^{-}_{rk}(\underline{k}, \overline{k})
=
\begin{cases}
\displaystyle
\max\left\{ 0,\ \frac{\theta_{2008} - \tau(1-k_0)}{-(\Delta_r - \tau(k_r - k_{r-1}))} \right\},
& \text{if } \Delta_r - \tau(k_r - k_{r-1}) < 0, \\[1.1em]
0,
& \text{if } \Delta_r - \tau(k_r - k_{r-1}) \ge 0\ \text{and}\ \theta_{2008} - \tau(1-k_0) \le 0, \\[0.4em]
+\infty,
& \text{otherwise}.
\end{cases}
\end{equation*}

If the candidate is greater than or equal to $M_{\text{max}}(\underline{k}, \overline{k})$, then breakdown does not occur within the admissible set and we set $\mu^{\pm}_{rk} = +\infty$. Putting it all together:

\begin{equation*}\label{eq:mu_min_k}
\mu_{rk}(\underline{k}, \overline{k})
\coloneqq
\min\left\{ \mu^{+}_{rk}(\underline{k}, \overline{k}),\ \mu^{-}_{rk}(\underline{k}, \overline{k}) \right\}.
\end{equation*}

\begin{equation}\label{eq:mbp_k_min}
M^{bp}(\underline{k}, \overline{k})
=
\min_{r \in \mathcal{S}}
\min_{\substack{
k_j \in \{\underline{k}_j, \overline{k}_j\},\ j \in J(r)
}}
\mu_{rk}(\underline{k}, \overline{k}).
\end{equation}

\section{Auxiliary Results}\label{appendix:auxresults}

\subsection{Deriving Anticipation Effects at time $t$ for Multiple Pre-Trends Case}

Corollary \ref{prop:pretrenddecopmpsim} gives us an expression for the pre-trend $\Delta_s$ in terms of the parallel trends violations and \textit{differences} in consecutive anticipation effects. Define $A_s \coloneqq \varphi_s - \varphi_{s-1}.$ \smallskip

\noindent Recall from Lemma \ref{lemma:attbiasdecomp23} that for purposes of identifying the $\text{ATT}_{t}$ for some period $t$ in the post-treatment period, we are not interested in the change in anticipation effects between consecutive periods, but instead the anticipation effects in a given pre-treatment period, typically period $0$. It can be shown that anticipation effects in pre-treatment period are given by the following telescoping sum: $\varphi_{s} = \sum_{j = -(S-1)}^{s} A_j 
    + \varphi_{-S}.$ \smallskip

\noindent By Assumption \ref{abounds}, we have a vector of anticipation terms across the pre-treatment period\footnote{Since $t= -S$ is the first pre-treatment period in the data, the first pre-trend is observed in period $-(S-1)$, which is why the $A_s$ terms are indexed starting at $-(S-1)$.}:
$\boldsymbol{A} = (A_{-(S-1)}, \ldots , A_{0})^{\prime} \in \prod_{s=-(S-1)}^{0} [\underline{A}_{s}, \overline{A}_{s}] \subset \mathbb{R}^{S}.$

\begin{lemma}\label{aeexpression}
    Suppose Assumptions \ref{initialize} and \ref{abounds} hold. For a given $\boldsymbol{A} \in \prod_{j=-(S-1)}^{0} [\underline{A}_{j}, \overline{A}_{j}]$, anticipation effects in pre-treatment period $s$ are given by $\varphi_s = \sum_{j = -(S-1)}^{s} A_j,$ where $A_j \in [\underline{A}_j, \overline{A}_j]$ for $j \in \{-(S-1), \ldots , s\}$.
\end{lemma}

\subsection{Separability Lemma for Results in Section \ref{sec:partialident}}

\begin{lemma}\label{separableresult}
    Let $\mathcal{X}_{1}, \ldots , \mathcal{X}_{J} \subseteq \mathbb{R}$, and define the Cartesian product $\mathcal{X} = \mathcal{X}_1 \times \ldots \times \mathcal{X}_{J} \subseteq \mathbb{R}^{J}$. Let $f: \mathcal{X} \rightarrow \mathbb{R}$ be a separable function of the form: $f(\boldsymbol{x}) = \sum_{j=1}^{J} f_j (x_j),$
    where each $f_j : \mathcal{X}_j \rightarrow \mathbb{R}$ is a real-valued function. Furthermore, suppose that for each $j$, $f_j$ attains its minimum and maximum on $\mathcal{X}_j$. Then:
    $$\min_{\boldsymbol{x} \in \mathcal{X}} \sum_{j=1}^{J} f_j (x_j) = \sum_{j=1}^{J} \min_{x_j \in \mathcal{X}_j} f_j (x_j) \quad \text{ and } \quad \max_{\boldsymbol{x} \in \mathcal{X}} \sum_{j=1}^{J} f_j (x_j) = \sum_{j=1}^{J} \max_{x_j \in \mathcal{X}_j} f_j (x_j).$$
\end{lemma}

\section{Proofs}
\label{appendix:proofs}

\subsection*{Main Text}

\noindent I first prove the more general Lemma \ref{lemma:attKdecomp}; Lemma \ref{lemma:attbiasdecomp23} then follows by specializing the same argument to $h=1$. \smallskip

\ProofOf{Lemma \ref{lemma:attKdecomp}}
\begin{proofnolab}
\noindent We have $\text{ATT}_{h} = \mathbb{E}[Y_{h}(1) - Y_{h}(0) \mid X=1] = \mathbb{E}[Y_h \mid X=1] - \mathbb{E}[Y_h (0) \mid X=1].$ By the definition of $\delta_h$ and $\varphi_0$, some algebra yields 
    \begin{align*}
      \mathbb{E}[Y_h (0) \mid X=1] = \mathbb{E}[Y_0  \mid X=1] + \mathbb{E}[Y_h  - Y_0  \mid X=0] - \varphi_0 + \sum_{j=1}^{h} \delta_j.
    \end{align*}
    Therefore, the $\text{ATT}_{h}$ is given by $\text{ATT}_{h} = \theta_h + \varphi_0 - \sum_{j=1}^{h} \delta_j.$
\end{proofnolab} \smallskip 

\ProofOf{Lemma \ref{lemma:attbiasdecomp23}}
\begin{proofnolab}
\noindent This is the case $h=1$ of Lemma~\ref{lemma:attKdecomp}:
\[
\text{ATT}_1=\theta_1+\varphi_0-\delta_1.
\]
\end{proofnolab}

\ProofOf{Corollary \ref{corr:ptvaenotident}}
\begin{proofnolab}
\noindent Pick any $k\in\mathbb{R}$. Preserve the observed distribution and, for treated units, define
$\widetilde{Y}_0(0)\coloneqq Y_0(0)+2k$ and
$\widetilde{Y}_1(0)\coloneqq Y_1(0)+k$, leaving the remaining potential outcomes unchanged. Then
\[
\widetilde{\mathrm{ATT}}_1=\mathrm{ATT}_1-k,\qquad
\widetilde{\varphi}_0=\varphi_0-2k,\qquad
\widetilde{\delta}_1=\delta_1-k.
\]
These changes leave all observed outcomes unchanged, and, by Lemma
\ref{lemma:attbiasdecomp23},
\[
\theta_1
=(\widetilde{\mathrm{ATT}}_1+k)
-(\widetilde{\varphi}_0+2k)
+(\widetilde{\delta}_1+k)
=\widetilde{\mathrm{ATT}}_1-\widetilde{\varphi}_0+\widetilde{\delta}_1.
\]
Hence the original and transformed DGPs are observationally equivalent. Since $k$ is arbitrary, the identified set for $\mathrm{ATT}_1$ is $\mathbb{R}$, so $\mathrm{ATT}_1$ is completely unidentified.
\end{proofnolab} \smallskip

\ProofOf{Corollary \ref{prop:pretrenddecopmpsim}}
\begin{proofnolab}
\noindent Let $s \in \{-(S-1), \ldots , 0\}$ index a pre-treatment period, where $s=0$ denotes the last period before treatment. The observed pre-trend at event time $s$, denoted $\Delta_{s}$, is given by:
\begin{equation}\label{pretrend}
    \Delta_{s} = \mathbb{E}[Y_{s} - Y_{s-1} \mid X=1] - \mathbb{E}[Y_{s} - Y_{s-1} \mid X=0].
\end{equation}

\noindent Expressing observed outcomes in terms of corresponding potential outcomes and simple algebra yields the following decomposition:
\begin{align*}
    \Delta_{s} &=\mathbb{E}[Y_s(0) - Y_{s-1}(0) \mid X=1]
 - \mathbb{E}[Y_s(0) - Y_{s-1}(0) \mid X=0] \\
 &\qquad + \mathbb{E}[Y_s(1) - Y_s(0) \mid X=1]
 - \mathbb{E}[Y_{s-1}(1) - Y_{s-1}(0) \mid X=1] \\
 &= \delta_s + \varphi_s - \varphi_{s-1}.  
\end{align*}

\end{proofnolab}

\smallskip

\noindent I first prove the more general Theorem \ref{thm:multpost}; Theorem \ref{thm:attmultpretrendsa} then follows by specializing the same argument to $h=1$. \smallskip

\ProofOf{Theorem \ref{thm:multpost}}
\begin{proofnolab}
\noindent  I first show $\text{ATT}_h\in\mathcal{I}_{\text{ATT}_h}$ and $\overline{\text{ATT}}_h\in\mathcal{I}_{\overline{\text{ATT}}_h}$ (Part 1), and then that each set is sharp: for every $\tau$ in the respective set there is a DGP consistent with the maintained assumptions and observed data under which the corresponding parameter of interest equals $\tau$ (Part 2).
\smallskip

\noindent \textbf{Part 1(i):} $\text{ATT}_h \in \mathcal{I}_{\text{ATT}_h}$. \smallskip

\noindent Let $A_s \coloneqq \varphi_s - \varphi_{s-1}$ and $ \boldsymbol{A} \coloneqq (A_{-(S-1)},\ldots,A_0)' \in  \mathcal{A}$, where $\mathcal{A} \coloneqq \prod_{j\in\mathcal{S}}[\underline{A}_j,\overline{A}_j]$. By Lemma~\ref{lemma:attKdecomp}, $\text{ATT}_h = \theta_h + \varphi_0 - \sum_{j=1}^{h} \delta_j.$ By Lemma \ref{aeexpression}, we have $\varphi_0 = \sum_{j \in \mathcal{S}} A_j,$ where $A_j \in [\underline{A}_j, \overline{A}_j] \text{ for all } j \in \mathcal{S}.$ 

Let $r = s^{\star}$. By Assumption \ref{assump:rmmult}, $|\delta_j| \leq M|\delta_r|$ for all $j \in \{1,\ldots,h\}$. Hence, for each $j$, there is $M_{j}^{\star} \in [-M,M]$ such that $\delta_j = M_j^{\star} \delta_r$. (If $\delta_r = 0$, Assumption~\ref{assump:rmmult} implies $\delta_j = 0$ for all $j \in \{1, \ldots , h\}$, and the equality holds for any $M_{j}^{\star} \in [-M,M]$.) Using $\delta_r = \Delta_r - A_r$, we obtain $\delta_j = M_j^{\star}(\Delta_r - A_r).$ Because Assumption \ref{assump:rmmult} bounds each $|\delta_j|$ separately and imposes no cross-restriction, the true ratio vector $\boldsymbol{M}^{\star} \coloneqq (M_1^\star,\ldots,M_h^\star)'$ lies in $[-M,M]^{h}$, with its components unrestricted relative to one another. Define $\boldsymbol{m}\coloneqq(m_1,\ldots,m_h)' \in[-M,M]^h.$ 

\smallskip \noindent 
Substituting yields the following representation:
\begin{align*}
  f(\boldsymbol{A}, \boldsymbol{m}, r)
  &\coloneqq \theta_h + \sum_{j \in \mathcal{S}} A_j
    - \sum_{j=1}^h m_j(\Delta_r - A_r) = \theta_h + \sum_{j \in \mathcal{S} \setminus \{r\}} A_j
    + A_r - \Bigl(\textstyle\sum_{j=1}^h m_j\Bigr)(\Delta_r-A_r).
\end{align*}

\noindent Given $r \in \mathcal{S}$, to obtain the lower bound $\text{LB}_h(r)$ I solve:
\begin{align*}
  \min \quad
    & f(\boldsymbol{A}, \boldsymbol{m}, r) \\[1em]
  \text{subject to} \quad
    & \boldsymbol{A} \in \mathcal{A} \quad \text{ and } \quad  \boldsymbol{m} \in [-M, M]^h.
\end{align*}

\noindent Evaluating $f$ at the true DGP values---$\boldsymbol{A}\in\mathcal{A}$,
$\boldsymbol{M}^{\star}\in[-M,M]^h$, and $r=s^{\star}\in\mathcal{S}$---gives
$\text{ATT}_h=f(\boldsymbol{A},\boldsymbol{M}^{\star},s^{\star})\in f(\mathcal{F})$, where
$\mathcal{F}\coloneqq\mathcal{A}\times[-M,M]^h\times\mathcal{S}$. As $f$ is continuous
and $\mathcal{F}$ compact, its extrema are attained. Because $\mathcal F$ has a product structure,
$\min_{\mathcal{F}}f=\min_{r\in\mathcal{S}}\min_{(\boldsymbol{A},\boldsymbol{m})}f(\boldsymbol{A},\boldsymbol{m},r)$;
fix $r$ and compute the inner problem. \smallskip 

\noindent \textit{Minimization over $\boldsymbol{m}$.} The variables $\boldsymbol{m}$ enter $f$ only
through $-\bigl(\sum_j m_j \bigr)(\Delta_r-A_r)$, so minimizing $f$ maximizes
$\sum_{j=1}^h m_j (\Delta_r-A_r)$. By Lemma~\ref{separableresult} this separates
across $j$; if $\Delta_r-A_r\neq0$, each summand is maximized at
$m_j=\operatorname{sign}(\Delta_r-A_r)\,M$, while if $\Delta_r-A_r=0$ the summand
vanishes for every $m_j \in[-M,M]$. In either case, the maximized value of the sum is 
$hM|\Delta_r-A_r|$, so
\[
  \min_{\boldsymbol{m}\in[-M,M]^h}f(\boldsymbol{A},\boldsymbol{m},r)
  =\theta_h+\sum_{j\in\mathcal{S}}A_j-hM|\Delta_r-A_r|.
\]

\noindent \textit{Minimization over $\boldsymbol{A}$.} For $j\neq r$, $A_j$ enters linearly with
positive coefficient, so is minimized at $\underline{A}_j$ (Lemma~\ref{separableresult}).
The $j=r$ term is $g_r(a)\coloneqq a-hM|\Delta_r-a|$ for $a\in[\underline{A}_r,\overline{A}_r]$.
Since $a\mapsto|\Delta_r-a|$ is convex and $hM\ge0$, $-hM|\Delta_r-a|$ is concave, and
$a\mapsto a$ is affine and hence concave; therefore $g_r$ is concave on
$[\underline{A}_r,\overline{A}_r]$. For any $a\in(\underline{A}_r,\overline{A}_r)$, write
$a=\lambda\underline{A}_r+(1-\lambda)\overline{A}_r$ for some $\lambda\in(0,1)$. By concavity,
\[
  g_r(a)\ge\lambda\,g_r(\underline{A}_r)+(1-\lambda)\,g_r(\overline{A}_r)
  \ge\min\bigl\{g_r(\underline{A}_r),\,g_r(\overline{A}_r)\bigr\},
\]
so the minimum of $g_r$ over $[\underline{A}_r,\overline{A}_r]$ is attained at an endpoint.
Therefore
\[
  \text{LB}_h(r)\coloneqq\theta_h+\sum_{j\in\mathcal{S}\setminus\{r\}}\underline{A}_j
  +\min_{a\in\{\underline{A}_r,\overline{A}_r\}}\bigl\{a-hM|\Delta_r-a|\bigr\}.
\]
By the analogous argument, $h_r(a)\coloneqq a+hM|\Delta_r-a|$ is convex on
$[\underline{A}_r,\overline{A}_r]$ and hence attains its maximum at an endpoint, giving
\[
  \text{UB}_h(r)\coloneqq\theta_h+\sum_{j\in\mathcal{S}\setminus\{r\}}\overline{A}_j
  +\max_{a\in\{\underline{A}_r,\overline{A}_r\}}\bigl\{a+hM|\Delta_r-a|\bigr\}.
\]
Minimizing/maximizing over $r\in\mathcal{S}$ yields
$\text{ATT}_h\in\bigl[\min_{r\in\mathcal{S}}\text{LB}_h(r),\ \max_{r\in\mathcal{S}}\text{UB}_h(r)\bigr]\eqqcolon\mathcal{I}_{\text{ATT}_h}$. \smallskip

\noindent \textbf{Part 1(ii):} $\overline{\text{ATT}}_h \in \mathcal{I}_{\overline{\text{ATT}}_h}$.

\noindent By Corollary~\ref{corr:avgdecomp}, substituting the same expressions for
$\varphi_0$ and $\delta_j$,
\[
  f_{\text{avg}}(\boldsymbol{A},\boldsymbol{m},r)
  \coloneqq\tfrac1h\sum_{j=1}^h\theta_j+\sum_{j\in\mathcal{S}}A_j
  -\tfrac1h\sum_{j=1}^h(h-j+1)\,m_j (\Delta_r-A_r).
\]
The argument is identical to Part~1(i), except the weight on $m_j$ is
$(h-j+1)/h>0$. As all weights are strictly positive, each summand is again maximized at
$m_j=\operatorname{sign}(\Delta_r-A_r)\,M$ (and any $m_j\in[-M,M]$ is
optimal when $\Delta_r-A_r=0$); using $\sum_{j=1}^h(h-j+1)=h(h+1)/2$, the coefficient
$hM$ is replaced by $\tfrac{h+1}{2}M$. Hence, with
\[
  \text{LB}_{\text{avg}}(r)\coloneqq\tfrac1h\sum_{j=1}^h\theta_j
   +\sum_{j\in\mathcal{S}\setminus\{r\}}\underline{A}_j
   +\min_{a\in\{\underline{A}_r,\overline{A}_r\}}\Bigl\{a-\tfrac{h+1}{2}M|\Delta_r-a|\Bigr\},
\]
\[
  \text{UB}_{\text{avg}}(r)\coloneqq\tfrac1h\sum_{j=1}^h\theta_j
   +\sum_{j\in\mathcal{S}\setminus\{r\}}\overline{A}_j
   +\max_{a\in\{\underline{A}_r,\overline{A}_r\}}\Bigl\{a+\tfrac{h+1}{2}M|\Delta_r-a|\Bigr\},
\]
we obtain $\overline{\text{ATT}}_h\in\bigl[\min_{r \in \mathcal{S}}\text{LB}_{\text{avg}}(r),\ \max_{r \in \mathcal{S}}\text{UB}_{\text{avg}}(r)\bigr]\eqqcolon\mathcal{I}_{\overline{\text{ATT}}_h}$.

\noindent \textbf{Part 2:} $\mathcal{I}_{\text{ATT}_h}$ and $\mathcal{I}_{\overline{\text{ATT}}_h}$
are sharp 

\noindent I first establish sharpness for $\mathcal{I}_{\text{ATT}_h}$; the argument for
$\mathcal{I}_{\overline{\text{ATT}}_h}$ is identical, with $f$ replaced by
$f_{\text{avg}}$ and $hM$ replaced by $\tfrac{h+1}{2}M$, and is completed explicitly at
the end. Fix $\tau\in\mathcal{I}_{\text{ATT}_h}$.

\smallskip \noindent 
\textit{Every $\tau$ has a pre-image in $\mathcal{F}$.} For $r\in\mathcal{S}$,
$\mathcal{I}_r\coloneqq f\bigl(\mathcal{A}\times[-M,M]^h\times\{r\}\bigr)$ is the continuous
image of a connected set, hence an interval (Theorem 23.5 in \citealt{munkres2000}), with
endpoints attained by Weierstrass; thus $\mathcal{I}_r=[\text{LB}_h(r),\text{UB}_h(r)]$, and
$f(\mathcal{F})=\bigcup_{r\in\mathcal{S}}\mathcal{I}_r$. If $|\mathcal{S}|=1$ then the union is connected. If $|\mathcal{S}|>1$, take $r\neq s$; then
\begin{align*}
   \text{UB}_h(r)&-\text{LB}_h(s) = \\
   &\underbrace{\sum_{j\in\mathcal{S}\setminus\{r,s\}}(\overline{A}_j-\underline{A}_j)}_{\ge0}
  +\underbrace{\max_{a\in\{\underline{A}_r,\overline{A}_r\}}\{a+hM|\Delta_r-a|\}-\underline{A}_r}_{\ge\,hM|\Delta_r-\underline{A}_r|\ge0}
  +\underbrace{\overline{A}_s-\min_{a\in\{\underline{A}_s,\overline{A}_s\}}\{a-hM|\Delta_s-a|\}}_{\ge\,hM|\Delta_s-\overline{A}_s|\ge0}\ \ge 0,
\end{align*}
and symmetrically $\text{UB}_h(s)\ge\text{LB}_h(r)$, so $\mathcal{I}_r\cap\mathcal{I}_s\neq\emptyset$.
By the one-dimensional version of Helly's theorem, $\bigcap_{r\in\mathcal{S}}\mathcal{I}_r\neq\emptyset$, so by Theorem~23.3 in \cite{munkres2000} the union is connected: $\bigcup_{r}\mathcal{I}_r=[\min_r\text{LB}_h(r),\max_r\text{UB}_h(r)]=\mathcal{I}_{\text{ATT}_h}$. Hence $\tau$ has a pre-image $(\widetilde{\boldsymbol{A}},\widetilde{\boldsymbol{m}},\widetilde{r})\in\mathcal{F}$, so that
\[
  \theta_h+\widetilde{\varphi}_0-\sum_{j=1}^h\widetilde{\delta}_j=\tau,
  \quad
  \widetilde{\delta}_s=\Delta_s-\widetilde{A}_s\ (s\in\mathcal{S}),\quad
  \widetilde{\delta}_j=\widetilde{m}_j(\Delta_{\widetilde{r}}-\widetilde{A}_{\widetilde{r}})\ (j\in\{1,\ldots,h\}),
\]
and define the anticipation path
\[
  \widetilde{\varphi}_{-S}\coloneqq0,\qquad
  \widetilde{\varphi}_t\coloneqq\sum_{s=-(S-1)}^{t}\widetilde{A}_s,\quad t\in\{-(S-1),\ldots,0\},
  \qquad\text{so that}\qquad
  \widetilde{\varphi}_0=\sum_{s\in\mathcal{S}}\widetilde{A}_s.
\]

\noindent \textit{Construction of the DGP.} Throughout the construction, expectations without a subscript refer to the distribution corresponding to the observed data, whereas $\mathbb{E}_{\widetilde{P}}$ denotes expectation under the constructed DGP. Set $\widetilde{\delta}_j\coloneqq0$ for $j\in\{h+1,\ldots,T\}$. The estimands $\text{ATT}_h$ and $\overline{\text{ATT}}_h$ depend on the parallel trends violations only through $\delta_1,\ldots,\delta_h$ (Lemma~\ref{lemma:attKdecomp} and Corollary \ref{corr:avgdecomp}), so their values at $t>h$ do not affect the estimands; these periods are nonetheless constrained by Assumption~\ref{assump:rmmult}, which is satisfied by setting $\widetilde{\delta}_j=0$. This choice is available because, for $t>h$, the treated units' never-treated outcome $Y_t(0)$ is counterfactual and hence unrestricted by the observed data.

Let $c_t\coloneqq\mathbb{E}[Y_t-Y_{t-1}\mid X=0]$, and let $\widetilde{P}$ retain the observed joint distribution of $(Y_{-S},\ldots,Y_T,X)$. I specify the counterfactual conditional means as follows:
\[
  \mathbb{E}_{\widetilde{P}}[Y_t(0)\mid X=1]\coloneqq
  \begin{cases}
    \mathbb{E}[Y_{-S}\mid X=1], & t=-S,\\[3pt]
    \displaystyle\sum_{j=-(S-1)}^{t}\bigl(c_j+\widetilde{\delta}_j\bigr)+\mathbb{E}[Y_{-S}\mid X=1], & t\in\{-(S-1),\ldots,T\},
  \end{cases}
\]
so that $\mathbb{E}_{\widetilde{P}}[Y_t(0)-Y_{t-1}(0)\mid X=1]=c_t+\widetilde{\delta}_t$ for every $t\in\{-(S-1),\ldots,T\}$. Under the constructed DGP, I further impose $Y_t=Y_t(0)$ for comparison units and $Y_t=Y_t(1)$ for treated units. For treated units, let $Y_t(0)$ be $Y_t$ translated by a constant so that its conditional mean given $X=1$ equals the value displayed above, and assign the remaining potential outcomes arbitrarily. The resulting $\widetilde{P}$ reproduces the observed distribution, $\mathbb{E}_{\widetilde{P}}[Y_t\mid X=x]=\mathbb{E}[Y_t\mid X=x]$. \smallskip 

\noindent \textit{Verification.}

\smallskip 
\noindent (a) \textit{Assumptions.} At $t=-S$, $\varphi_{-S}(\widetilde{P})=\mathbb{E}[Y_{-S}\mid X=1]-\mathbb{E}_{\widetilde{P}}[Y_{-S}(0)\mid X=1]=0$ by construction, so Assumption~\ref{initialize} holds. For $-S<t\le0$, since $\Delta_j+c_j=\mathbb{E}[Y_j-Y_{j-1}\mid X=1]$ for $j\in\mathcal{S}$ and $\widetilde{\delta}_j=\Delta_j-\widetilde{A}_j$,
\begin{align*}
\varphi_t(\widetilde P)
&=\mathbb E[Y_t\mid X=1]-\mathbb E_{\widetilde P}[Y_t(0)\mid X=1] \\
&=\sum_{j=-(S-1)}^t(\Delta_j+c_j)-\sum_{j=-(S-1)}^t(c_j+\widetilde\delta_j)
 =\sum_{j=-(S-1)}^t(\Delta_j-\widetilde\delta_j) =\sum_{j=-(S-1)}^t\widetilde A_j=\widetilde\varphi_t.
\end{align*}
Hence the constructed anticipation increments equal $\widetilde{A}_s\in[\underline{A}_s,\overline{A}_s]$, so Assumption~\ref{abounds} holds, and $\varphi_0(\widetilde{P})=\widetilde{\varphi}_0$. For the parallel trends violations, comparison units realize $Y_t(0)=Y_t$ and $\widetilde{P}$ reproduces the observed distribution, so $\mathbb{E}_{\widetilde{P}}[Y_t(0)-Y_{t-1}(0)\mid X=0]=c_t$; therefore, for every $t\in\{-(S-1),\ldots,T\}$,
\[
  \delta_t(\widetilde{P})=\mathbb{E}_{\widetilde{P}}[Y_t(0)-Y_{t-1}(0)\mid X=1]-c_t
  =(c_t+\widetilde{\delta}_t)-c_t=\widetilde{\delta}_t.
\]
Finally, for $j\in\{1,\ldots,h\}$, $|\widetilde{\delta}_j|=|\widetilde{m}_j|\,|\widetilde{\delta}_{\widetilde{r}}|\le M|\widetilde{\delta}_{\widetilde{r}}|\le M|\widetilde{\delta}_{\widetilde{s}^{\star}}|$, where $\widetilde{s}^{\star}\in\arg\max_{s\in\mathcal{S}}|\widetilde{\delta}_s|$; and for $j\in\{h+1,\ldots,T\}$, $\widetilde{\delta}_j=0\le M|\widetilde{\delta}_{\widetilde{s}^{\star}}|$. Hence Assumption~\ref{assump:rmmult} holds for every $j\in\{1,\ldots,T\}$.

\smallskip 
\noindent(b) \textit{Observed data.} Since $\widetilde{P}$ preserves the observed joint distribution of $(Y_t,X)$, $\mathbb{E}_{\widetilde{P}}[Y_t\mid X=x]=\mathbb{E}[Y_t\mid X=x]$; hence $\Delta_s(\widetilde{P})=\Delta_s$ for all $s\in\mathcal{S}$, and $\theta_j(\widetilde{P})=\theta_j$ for every $j\in\{1,\ldots,h\}$ (in particular $\theta_h$).

\smallskip 
\noindent(c) $\text{ATT}_h(\widetilde{P})=\tau$. By Lemma~\ref{lemma:attKdecomp} applied to $\widetilde{P}$ and parts (a)--(b),
\[
  \text{ATT}_h(\widetilde{P})=\theta_h(\widetilde{P})+\varphi_0(\widetilde{P})-\sum_{j=1}^h\delta_j(\widetilde{P})
  =\theta_h+\widetilde{\varphi}_0-\sum_{j=1}^h\widetilde{\delta}_j
  =f(\widetilde{\boldsymbol{A}},\widetilde{\boldsymbol{m}},\widetilde{r})=\tau.
\]

\noindent Thus every $\tau\in\mathcal{I}_{\text{ATT}_h}$ is attained by an admissible DGP reproducing the observed reduced-form objects, so $\mathcal{I}_{\text{ATT}_h}$ lies in the identified set; with Part~1(i), $\mathcal{I}_{\text{ATT}_h}$ is the identified set for $\text{ATT}_h$.

\smallskip
\noindent\textit{Sharpness for $\overline{\text{ATT}}_h$.} Fix $\tau\in\mathcal{I}_{\overline{\text{ATT}}_h}$. By Part~1(ii) and the same connectedness and one-dimensional Helly argument applied to $f_{\text{avg}}$, $\tau$ has a pre-image $(\widetilde{\boldsymbol{A}},\widetilde{\boldsymbol{m}},\widetilde{r})\in\mathcal{F}$ with $f_{\text{avg}}(\widetilde{\boldsymbol{A}},\widetilde{\boldsymbol{m}},\widetilde{r})=\tau$. Construct $\widetilde{P}$ from this pre-image exactly as above. By (b) it reproduces $(\theta_1,\ldots,\theta_h)$, and by (a) it satisfies Assumptions~\ref{initialize}--\ref{assump:rmmult} with $\delta_j(\widetilde{P})=\widetilde{\delta}_j$ and $\varphi_0(\widetilde{P})=\widetilde{\varphi}_0$. Hence, by Corollary~\ref{corr:avgdecomp},
\[
  \overline{\text{ATT}}_h(\widetilde{P})=\tfrac1h\sum_{j=1}^h\theta_j+\widetilde{\varphi}_0
  -\tfrac1h\sum_{j=1}^h(h-j+1)\widetilde{\delta}_j
  =f_{\text{avg}}(\widetilde{\boldsymbol{A}},\widetilde{\boldsymbol{m}},\widetilde{r})=\tau.
\]
Therefore every element of $\mathcal{I}_{\overline{\text{ATT}}_h}$ is attained by an admissible DGP reproducing the observed reduced-form objects, and with Part~1(ii), $\mathcal{I}_{\overline{\text{ATT}}_h}$ is the identified set for $\overline{\text{ATT}}_h$.
\end{proofnolab} \smallskip 

\ProofOf{Theorem \ref{thm:attmultpretrendsa}}
\begin{proofnolab}
\noindent The argument is the case $h=1$ of the proof of Theorem~\ref{thm:multpost}. For a fixed horizon $h$, the bounds depend only on the post-treatment parallel trends violations through period $h$. Thus, when $h=1$, the relevant relative-magnitude restriction is precisely Assumption~\ref{assump:rmmultpretrends}. Lemma~\ref{lemma:attKdecomp} becomes $\text{ATT}_1=\theta_1+\varphi_0-\delta_1$, and
\[
\text{LB}(r)=\theta_1+\sum_{j\in\mathcal{S}\setminus\{r\}}\underline{A}_j+\min_{a\in\{\underline{A}_r,\overline{A}_r\}}\{a-M|\Delta_r-a|\},\quad
\text{UB}(r)=\theta_1+\sum_{j\in\mathcal{S}\setminus\{r\}}\overline{A}_j+\max_{a\in\{\underline{A}_r,\overline{A}_r\}}\{a+M|\Delta_r-a|\}.
\]
By the same outer-bound and sharpness argument, $\mathcal{I}^{A}_{\text{ATT}_1}=[\min_{r\in\mathcal{S}}\text{LB}(r),\,\max_{r\in\mathcal{S}}\text{UB}(r)]$ is the identified set for $\text{ATT}_1$.
\end{proofnolab} \smallskip

\ProofOf{Theorem \ref{thm:attmultpretrendsk}}
\begin{proofnolab}
\noindent To begin, I show that the $\text{ATT}_{1}$ lies within a set $\mathcal{I}^{k}_{\text{ATT}_{1}}$. Next, I show that this set is sharp, meaning that for every $\tau \in \mathcal{I}^{k}_{\text{ATT}_{1}}$, there exists a DGP consistent with the observed data and the maintained assumptions of the theorem under which $\text{ATT}_{1} = \tau$. \smallskip

\noindent \textbf{Part 1:} $\text{ATT}_{1} \in \mathcal{I}^{k}_{\text{ATT}_{1}}$. \smallskip    

\noindent Let $J(r)\coloneqq\{0,r,r-1\}$. Note that $J(0)=\{0,-1\}$. Define $\boldsymbol{k} \coloneqq (k^{\star}_{-S}, \ldots , k^{\star}_{0})^{\prime} \in \mathcal{K},$ where $\mathcal{K} \coloneqq  \prod_{j = -S}^{0} [\underline{k}_{j}, \overline{k}_{j}] \subseteq \mathbb{R}^{S+1}.$ By Lemma \ref{lemma:attbiasdecomp23}, we have: $\text{ATT}_{1} = \theta_1 + \varphi_0 - \delta_1.$ By Assumption \ref{assump:krelaxmultpre}, for each \(t\in\{-S,\ldots,0\}\),
$\varphi_t = k_t^\star\cdot \text{ATT}_1$ where $k_t^\star\in[\underline{k}_t,\overline{k}_t].$ Let $r = s^{\star}$. By Assumption \ref{assump:rmmultpretrends}, $|\delta_1| \le M|\delta_r|.$ Equivalently, there exists $M^{\star} \in [-M,M]$ such that $\delta_1 = M^{\star}\delta_r.$ (If $\delta_r = 0$, Assumption \ref{assump:rmmultpretrends} implies $\delta_1 = 0$, and the equality holds for any $M^{\star} \in [-M,M]$.) By Corollary \ref{prop:pretrenddecopmpsim}, $\delta_r = \Delta_r - (\varphi_r - \varphi_{r-1})$, so we obtain
\begin{align*}
\delta_1 
    &= M^{\star} (\Delta_r - (\varphi_r - \varphi_{r-1})) = M^{\star} \big(\Delta_r - (k^{\star}_{r} - k^{\star}_{r-1}) \cdot \text{ATT}_{1}\big).
\end{align*}
Therefore, we have the following representation:
\begin{align*}
    &\text{ATT}_{1} = \theta_1 + k_{0}^{\star} \cdot \text{ATT}_{1} - M^{\star} \bigl( \Delta_{r} - (k_{r}^{\star} - k_{r-1}^{\star}) \cdot \text{ATT}_{1} \bigr) \\
    \Longleftrightarrow \quad &\text{ATT}_{1} = \frac{\theta_1 - M^{\star} \Delta_{r}}{1 - k_{0}^{\star} - M^{\star}(k_{r}^{\star} - k_{r-1}^{\star})}.
\end{align*}
The second line follows by Assumption \ref{assump:nonzerodenomk}. For generic arguments $(k_j)_{j\in J(r)}\in\prod_{j\in J(r)}[\underline{k}_j,\overline{k}_j]$, $m\in[-M,M]$, and $r\in\mathcal{S}$, define
\[
f\bigl((k_j)_{j\in J(r)},m,r\bigr)\coloneqq\frac{\theta_1-m\Delta_r}{1-k_0-m(k_r-k_{r-1})}.
\]
Thus,
\[
\text{ATT}_{1}=f\bigl((k_j^{\star})_{j\in J(r)},M^{\star},r\bigr).
\]
Note that $f$ depends only on $((k_j)_{j\in J(r)},m,r)$; the remaining components of $\boldsymbol{k}$ do not enter the objective.
    For a given \(r\in\mathcal S\), define the relevant feasible set $\mathcal F_r \;\coloneqq\; \prod_{j\in J(r)}[\underline k_j,\overline k_j]\times[-M,M].$ Given $r \in \mathcal{S}$, to obtain the lower bound we solve the following optimization problem:
    \begin{align*}
\min\quad 
& f\big((k_j)_{j\in J(r)},m, r \big)
\\
\text{s.t.}\quad 
& \bigl( (k_j)_{j\in J(r)}, m \bigr) \in \mathcal{F}_{r}
\end{align*}

Analogously, we obtain the upper bound by maximizing the objective (or equivalently, minimizing the negation of the objective) subject to the same constraints. Since $f$ is continuous and $\mathcal{F}_{r}$ is compact, by the Weierstrass Extreme Value Theorem both a minimum and maximum are attained. I consider two cases: (1) $r \neq 0$, and (2) $r = 0$. I use the KKT conditions to assess interior candidates. \smallskip

\noindent \textbf{Case 1:} $\boldsymbol{r \neq 0}$. \smallskip

The Lagrangian is
\begin{align*}
\mathcal L_r
&=
f\big((k_j)_{j\in J(r)},\, m, r \big)
\;+\;
\sum_{j\in J(r)}
\Big[
\lambda_{j,-}(\underline k_j-k_j)
+\lambda_{j,+}(k_j-\overline k_j)
\Big]
\;+\;
\mu_-(-M-m)
+\mu_+(m-M),
\end{align*}
where \(\lambda_{j,\pm}\ge 0\) for all \(j\in J(r)\) and \(\mu_\pm\ge 0\).

For $r\neq 0$, the partial derivatives of $f$ are:
\begin{align*}
    \frac{\partial f}{\partial k_{0}}
    &= \frac{\theta_1 - m \Delta_{r}}
           {(1 - k_{0} - m(k_{r} - k_{r-1}))^{2}}, \\[0.5em]
    \frac{\partial f}{\partial k_{r}}
    &= \frac{m (\theta_1 - m \Delta_{r})}
           {(1 - k_{0} - m(k_{r} - k_{r-1}))^{2}}, \\[0.5em]
    \frac{\partial f}{\partial k_{r-1}}
    &= \frac{-m (\theta_1 - m \Delta_{r})}
           {(1 - k_{0} - m(k_{r} - k_{r-1}))^{2}}, \\[0.5em]
    \frac{\partial f}{\partial m}
    &= \frac{\theta_1(k_{r} - k_{r-1}) - \Delta_{r} (1-k_{0})}
           {(1 - k_{0} - m(k_{r} - k_{r-1}))^{2}}.
\end{align*}

I begin by assessing potential interior candidates. On the interior, all constraints are slack, and therefore by complementary slackness, all multipliers are zero. Hence, to have KKT stationarity, I need to set the above partial derivatives equal to zero. This is equivalent to setting the numerators of the above partial derivatives to zero. I consider several sub-cases below: \smallskip

\begin{enumerate}
    \item $\Delta_{r} \neq 0, \theta_1 \neq 0$. Here, we need: $\frac{\partial f}{\partial k_{0}} = \frac{\partial f}{\partial k_{r}} = \frac{\partial f}{\partial k_{r-1}} = 0 \; \Longleftrightarrow \; m = \frac{\theta_1}{\Delta_{r}}.$  Note that $\frac{\partial f}{\partial k_{r}} = \frac{\partial f}{\partial k_{r-1}} = 0$ can also occur if $m = 0$. However, if $\theta_1 \neq 0$, then we won't have $\frac{\partial f}{\partial k_{0}} = 0$, that condition would not be enough to yield stationarity in this case.   \smallskip

    In addition, we need: $\frac{\partial f}{\partial m} = 0 \; \Longleftrightarrow \; k_{r} - k_{r-1} = \frac{\Delta_{r} (1 - k_{0})}{\theta_1}.$ However, evaluating the denominator of $f$ when $m = \frac{\theta_1}{\Delta_{r}}$ and $k_{r} - k_{r-1} = \frac{\Delta_{r} (1 - k_{0})}{\theta_1}$:
    \begin{align*}
        1 - k_{0} - m(k_{r} - k_{r-1}) &= 1 - k_{0}- \frac{\theta_1}{\Delta_{r}} \cdot \frac{\Delta_{r} (1 - k_{0})}{\theta_1} = (1 - k_{0}) - (1 - k_{0}) = 0,
    \end{align*}
    which is ruled out by assumption. Therefore, $(m , (k_{r} - k_{r-1})) = \left( \frac{\theta_1}{\Delta_{r}} , \frac{\Delta_{r} (1 - k_{0})}{\theta_1} \right) $ is infeasible.

   \item $\Delta_{r} = 0, \theta_1 \neq 0$. In this case, if $\Delta_{r} = 0$, then the numerator of $\frac{\partial f}{\partial k_{0}}$ is $\theta_1 \neq 0$. Therefore, stationarity does not hold.

   \item $\Delta_{r} \neq 0, \theta_1 = 0$. Here, we need: $\frac{\partial f}{\partial k_{0}} = \frac{\partial f}{\partial k_{r}} = \frac{\partial f}{\partial k_{r-1}} = 0 \; \Longleftrightarrow \; m = 0.$ and:
    $\frac{\partial f}{\partial m} = 0 \; \Longleftrightarrow \; k_{0} = 1.$ However, evaluating the denominator of $f$ at $(k_{0}, m) = (1,0)$ yields:
    \begin{align*}
        1 - k_{0} - m(k_{r} - k_{r-1}) &= 1 - 1 - 0 = 0,
    \end{align*}
    which is ruled out by assumption. Therefore, $(k_{0}, m) = (1,0)$ is infeasible.

   \item $\Delta_{r} = 0, \theta_1 = 0$. In this case, regardless of the value of $(k_{0}, k_{r}, k_{r-1}, m)$, 
   $f(k_0,k_r,k_{r-1},m, r) = 0.$ Hence, the objective is flat in this case and all feasible points satisfy stationarity. WLOG, we can consider the corners.
\end{enumerate}
Here, we see that no interior points satisfy KKT stationarity within the conditions of this problem, except for the case where $\Delta_{r} = 0, \theta_1 = 0$, in which case all points satisfy stationarity. Next, I consider the edges and faces. 

\begin{enumerate}
    \item One free variable; others fixed at bounds
    \begin{itemize}
        \item $m$ is free: The numerator of $\frac{\partial f}{\partial m}$ is constant, as it only depends on the fixed $k_j$ values and the data. If this numerator is nonzero, then $f$ is monotone in $m$, which implies that $m = \pm M$ yields the optimum. If this numerator is zero, then $f$ is flat in $m$ so WLOG we can consider $m = \pm M$.

        \item $k_{0}$ is free: The numerator of $\frac{\partial f}{\partial k_{0}}$ is constant, as it only depends on the fixed $m$ value and the data. If this numerator is nonzero, then $f$ is monotone in $k_{0}$, which implies that $k_{0} \in \{\underline{k}_0, \overline{k}_0 \}$ yields the optimum. If this numerator is zero, then $f$ is flat in $k_{0}$ so WLOG we can consider $k_{0} \in \{\underline{k}_0, \overline{k}_0 \}$.

        \item $k_{r}$ is free: The numerator of $\frac{\partial f}{\partial k_{r}}$ is constant, as it only depends on the fixed $m$ value and the data. If this numerator is nonzero, then $f$ is monotone in $k_{r}$, which implies that $k_{r} \in \{\underline{k}_r, \overline{k}_r \}$ yields the optimum. If this numerator is zero, then $f$ is flat in $k_{r}$ so WLOG we can consider $k_{r} \in \{\underline{k}_{r}, \overline{k}_{r} \}$.

        \item $k_{r-1}$ is free: Analogous to previous case except with minus sign in partial derivative.
    \end{itemize}

    Therefore, when we consider edges defined by one free variable, $f$ is either monotone or flat in the free variable, so we can consider corners.

\item More than one free variable

Consider any face with more than one free variable. First suppose that $m$ is free.
For each fixed value of $m$, the sign of the derivative of $f$ with respect to each
free $k_j$ does not depend on $k_j$. Hence, the minimum over the free $k_j$
variables is attained at endpoints of those variables. Thus, the minimization over
the face can be written as a minimum over the finite collection of endpoint
configurations for the free $k_j$ variables:
\[
\min_{m\in[-M,M]}\;
\min_{\text{free } k_j \in [\underline{k}_j,\overline{k}_j]}
f(m,k_0,k_r,k_{r-1})
=
\min_{m\in[-M,M]}\;
\min_{\text{free } k_j \in \{\underline{k}_j,\overline{k}_j\}}
f(m,k_0,k_r,k_{r-1}).
\]
Equivalently, this is the minimum over all pairs consisting of $m$ and an endpoint configuration, so the order of the two minimizations can be interchanged:
\[
\min_{m\in[-M,M]}\;
\min_{\text{free } k_j \in \{\underline{k}_j,\overline{k}_j\}}
f(m,k_0,k_r,k_{r-1})
=
\min_{\text{free } k_j \in \{\underline{k}_j,\overline{k}_j\}}\;
\min_{m\in[-M,M]}
f(m,k_0,k_r,k_{r-1}).
\]
Now fix one such endpoint configuration for the free $k_j$ variables. Conditional
on this configuration, the sign of $\partial f/\partial m$ does not depend on
$m$. Hence $f$ is monotone or flat in $m$, so the minimum over $m\in[-M,M]$ is
attained at $m\in\{-M,M\}$.

If $m$ is fixed on the face, then the first step alone applies: conditional on
that fixed value of $m$, the minimum over the free $k_j$ variables is attained at
their endpoints. Therefore, in either case, the minimum on the face is attained at
a corner of the face. The argument for the maximum is identical, replacing minima
by maxima.
\end{enumerate}
Therefore, we have shown that when we consider edges and faces, the objective is either monotone or flat in the free variable(s), meaning we can consider the corners. \smallskip

Hence, for a given $r \in \mathcal{S} \setminus \{0\}$, we have the following lower bound:
$$\text{LB}(r) \coloneqq \min_{\substack{
k_j \in \{\underline{k}_j, \overline{k}_j \}, \; j \in J(r)\\
m\in\{-M,M\}}}
\frac{\theta_1 - m\,\Delta_r}{1 - k_0 - m\,(k_r - k_{r-1})}.
$$

The above argument is analogous for the maximum, just with the negated objective:

$$\text{UB}(r) \coloneqq \max_{\substack{
k_j \in \{\underline{k}_j, \overline{k}_j \}, \; j \in J(r)\\
m\in\{-M,M\}}}
\frac{\theta_1 - m\,\Delta_r}{1 - k_0 - m\,(k_r - k_{r-1})}.$$

\noindent 
Thus, for each $r\in\mathcal{S}\setminus\{0\}$, the minimum and maximum of $f$ over $\mathcal{F}_r$ are $\text{LB}(r)$ and $\text{UB}(r)$, respectively.

\noindent \textbf{Case 2:} $\boldsymbol{r = 0}$. When $r = 0$, the objective becomes the following:
$$f(k_0, k_{-1}, m, r) = \frac{\theta_1 - m\Delta_{0}}{1 - k_{0} - m(k_{0} - k_{-1})} = \frac{\theta_1 - m\Delta_{0}}{1 - k_{0}(1+m) + m k_{-1}}.$$

The partial derivatives are given by:
\begin{align*}
    \frac{\partial f}{\partial k_{0}} &= \frac{(1+m)(\theta_1 - m \Delta_{0})}{(1 - k_{0}(1+m) + m k_{-1})^{2}} \\[0.5em]
    \frac{\partial f}{\partial k_{-1}} &= \frac{- m(\theta_1 - m \Delta_{0})}{(1 - k_{0}(1+m) + m k_{-1})^{2}} \\[0.5em]
    \frac{\partial f}{\partial m} &= \frac{\theta_1 (k_{0} - k_{-1}) - \Delta_{0} (1- k_{0})}{(1 - k_{0}(1+m) + m k_{-1})^{2}}
\end{align*}

As before, I begin by assessing potential interior candidates. I consider several sub-cases below:
\begin{enumerate}
    \item $\Delta_{0} \neq 0, \theta_1 \neq 0$. Here, we need: $\frac{\partial f}{\partial k_{0}} = \frac{\partial f}{\partial k_{-1}} = 0 \; \Longleftrightarrow \; m = \frac{\theta_1}{\Delta_{0}}.$ In addition, we need: $\frac{\partial f}{\partial m} = 0 \; \Longleftrightarrow \; k_{0} - k_{-1} = \frac{\Delta_{0} (1- k_{0})}{\theta_1}.$ However, evaluating the denominator of $f$ under the above conditions:
    \begin{align*}
        1 - k_{0} - m(k_{0} - k_{-1}) = 1 - k_{0} - \frac{\theta_1}{\Delta_{0}} \cdot \frac{\Delta_{0} (1- k_{0})}{\theta_1} = 0,
    \end{align*}
   which is ruled out by assumption. Therefore, $(m , (k_{0} - k_{-1})) = \left( \frac{\theta_1}{\Delta_{0}} , \frac{\Delta_{0} (1 - k_{0})}{\theta_1} \right) $ is infeasible.

   \item $\Delta_{0} = 0, \theta_1 \neq 0$. If $\Delta_{0} = 0$ and $\theta_1 \neq 0$, then in order to get $\frac{\partial f}{\partial k_{0}} = 0,$ we need $m = -1$, and in order to get $\frac{\partial f}{\partial k_{-1}} = 0$, we need $m = 0$. Therefore, there is no choice of $m$ that simultaneously yields $\frac{\partial f}{\partial k_{0}} = \frac{\partial f}{\partial k_{-1}} = 0$, so stationarity does not hold.

   \item $\Delta_{0} \neq 0, \theta_1 = 0$. In this case, we can have $\frac{\partial f}{\partial k_{0}} = \frac{\partial f}{\partial k_{-1}} = 0$ if $m = 0$. In addition, to have $\frac{\partial f}{\partial m} = 0$, we need $k_{0} = 1$. However, evaluating the denominator of $f$ at $(k_{0}, m) = (1,0)$ yields:
   $$1 - k_{0} - m(k_{0} - k_{-1}) = 1- 1 - 0 = 0,$$
    which is ruled out by assumption. Therefore, $(k_{0}, m) = (1,0)$ is infeasible.

     \item $\Delta_{0} = 0, \theta_1 = 0$. In this case, regardless of the value of $(k_{0}, k_{-1}, m)$, 
   $f(k_0, k_{-1}, m, r) = 0.$ Hence, the objective is flat in this case and all feasible points satisfy stationarity. WLOG, we can consider the corners.
\end{enumerate}

Next, I consider edges. Fix any two of the three coordinates of $(k_{0}, k_{-1}, m)$ at one of their bounds, and let the 
remaining coordinate vary over its feasible interval. From the derivative expressions above, the numerator of each partial derivative does not depend on the corresponding variable:
\[
\frac{\partial f}{\partial k_{0}} 
  \propto (1+m)(\theta_1 - m\Delta_0),\qquad
\frac{\partial f}{\partial k_{-1}}
  \propto -m(\theta_1 - m\Delta_0),\qquad
\frac{\partial f}{\partial m}
  \propto \theta_1(k_{0} - k_{-1}) 
            - \Delta_0(1-k_{0}).
\]
The denominator in each case is a strictly positive square. Therefore, along any one–dimensional edge, the partial derivative with respect to the free variable either has a constant nonzero sign (so $f$ is strictly monotone in that variable) or is identically zero (so $f$ is flat in that variable). In either case, the minimum and maximum of $f$ along that edge are achieved at the endpoints of the interval for the free variable. \smallskip 

Consider a face where two coordinates vary freely. The same argument used for the $r\neq 0$ case applies: after one free coordinate is reduced to its endpoint values, the finite endpoint minimization can be interchanged with the minimization over the remaining free coordinate. Hence the extrema on the face are attained at corners. \smallskip

For $r=0$, the objective depends only on $(k_0,k_{-1},m)$, and the
argument above implies that both extrema occur at the corners of the
three–dimensional feasible box. Thus,
\[
\text{LB}(0)
\coloneqq 
\min_{\substack{
k_j \in \{\underline{k}_j, \overline{k}_j \}, \; j \in J(0)\\
m\in\{-M,M\}}}
\frac{\theta_1 - m\,\Delta_0}{1 - k_0 - m\,(k_0 - k_{-1})},
\]
and
\[
\text{UB}(0)
\coloneqq 
\max_{\substack{
k_j \in \{\underline{k}_j, \overline{k}_j \}, \; j \in J(0)\\
m\in\{-M,M\}}}
\frac{\theta_1 - m\,\Delta_0}{1 - k_0 - m\,(k_0 - k_{-1})}.
\]
Thus, the minimum and maximum of $f$ over $\mathcal{F}_0$ are $\text{LB}(0)$ and $\text{UB}(0)$, respectively.

Combining both cases, for each $r\in\mathcal{S}$, the minimum and maximum of $f$ over $\mathcal{F}_r$ are $\text{LB}(r)$ and $\text{UB}(r)$, respectively. Therefore,
\[
\text{ATT}_1
\in
\left[
\min_{r\in\mathcal{S}}\text{LB}(r),\,
\max_{r\in\mathcal{S}}\text{UB}(r)
\right]
\eqqcolon
\mathcal{I}_{\text{ATT}_1}^{k}.
\]

\noindent \textbf{Part 2:} $\mathcal{I}^{k}_{\text{ATT}_{1}}$ is sharp. \smallskip

Next, I show that the bounds in $\mathcal{I}_{\text{ATT}_{1}}^{k}$ are sharp. Let $\tau$ be an arbitrary element in $\mathcal{I}^{k}_{\text{ATT}_{1}}$. I first show that there exists $r \in \mathcal{S}$ and $((k_j)_{j\in J(r)},m)\in\mathcal F_r$ such that $f\bigl((k_j)_{j\in J(r)},m,r\bigr)=\tau$. For a given $r \in \mathcal{S}$, define:
\begin{align*}
    \mathcal{I}_{r} &\coloneqq \bigl\{ f((k_j)_{j \in J(r)}, m, r) : ((k_j)_{j \in J(r)}, m, r) \in \mathcal{F}_{r} \times \{r \}  \bigr\} = \bigl[  \text{LB}(r), \text{UB}(r) \bigr].
\end{align*}
The second equality follows by Theorem 23.5 in \cite{munkres2000}: the image of a connected space under a continuous map is connected. Since $\mathcal{F}_{r} \times \{r\}$ is connected and $f$ is continuous (given Assumption \ref{assump:nonzerodenomk}), it follows that $\mathcal{I}_{r}$ is connected. In $\mathbb{R}$, connected sets are intervals. By the Weierstrass Extreme Value Theorem, the minimum and maximum are attained, so the endpoints are attained---i.e. the interval is closed. Define the global feasible set $\mathcal F \coloneqq \bigcup_{r\in\mathcal S}\bigl(\mathcal F_r\times\{r\}\bigr)$. The image of the feasible set $\mathcal{F}$ under the objective function $f$ is given by:
\begin{align}
    f (\mathcal{F} ) &= f \left( \bigcup_{r \in \mathcal{S}} (\mathcal{F}_{r} \times \{ r \} )  \right) = \bigcup_{r \in \mathcal{S}} f (\mathcal{F}_{r} \times \{ r \} ) = \bigcup_{r \in \mathcal{S} } \bigl[  \text{LB}(r), \text{UB}(r) \bigr] =  \bigcup_{r \in \mathcal{S}} \mathcal{I}_{r}. \label{unionkconnected?}
\end{align}
The second equality follows from the fact that the image of the union is the union of the images. The third and fourth equalities follow from the derivation of $\mathcal{I}_{\text{ATT}_{1}}^{k}$. \smallskip

If the union in (\ref{unionkconnected?}) is connected, then it is equivalent to $\mathcal{I}_{\text{ATT}_{1}}^{k}$, which in turn implies that any $\tau \in \mathcal{I}_{\text{ATT}_{1}}^{k}$ has a pre-image $((k_j)_{j \in J(r)}, m, r)  \in \mathcal{F}$. Note that if $| \mathcal{S}| = 1$ (i.e. $\mathcal{S} = \{0\})$ then the union in (\ref{unionkconnected?}) being connected is trivial. Now consider $|\mathcal{S}| > 1$. Since we bound $m$ in $[-M,M]$ for $M \in \mathbb{R}_{\geq 0}$, we know that we will always have $0 \in [-M,M]$. Hence, for a feasible $((k_j)_{j \in J(r)})$, we always have the feasible tuple: $((k_j)_{j \in J(r)},0) \in \mathcal{F}_{r}.$ Evaluating $f$ at $m = 0$:
$$f((k_j)_{j \in J(r)},0, r) = \frac{\theta_1}{1 - k_{0}}. $$
Note that this expression only depends on $\theta_1$ and $k_{0}$, both of which do not vary with $r$. Since we assume that there are no feasible parameters that yield a zero denominator in the objective, note that $1 \notin [\underline{k}_0, \overline{k}_0]$. Therefore, this expression is well-defined for all $k_0 \in [\underline{k}_0, \overline{k}_0]$. Define: $\mathcal{C} \coloneqq \{  \frac{\theta_1}{1 - k_{0}} : k_{0} \in [\underline{k}_0, \overline{k}_0] \}.$ Take an arbitrary $c\in\mathcal{C}$ and an arbitrary $r\in\mathcal{S}$. By definition of $\mathcal{C}$, choose $\check{k}_0\in[\underline{k}_0,\overline{k}_0]$ such that $c=\frac{\theta_1}{1-\check{k}_0}.$ Next, define a vector $(\check{k}_j)_{j\in J(r)}$ by setting $\check{k}_0$ as above and choosing any $\check{k}_j\in[\underline{k}_j,\overline{k}_j]$ for all $j\in J(r)\setminus\{0\}.$ Then $((\check{k}_j)_{j\in J(r)},0,r)\in\mathcal{F}_r\times\{r\}$ and $f\bigl((\check{k}_j)_{j\in J(r)},0,r\bigr)=\frac{\theta_1}{1-\check{k}_0}=c,$ so $c\in\mathcal{I}_r$. \smallskip

\noindent Since $c$ was an arbitrary element in $\mathcal{C}$ and $r$ was arbitrary, we have shown that for all $r \in \mathcal{S} $, $\mathcal{C} \subseteq \mathcal{I}_{r}$. Hence, $\mathcal{C}$ is contained in each $\mathcal{I}_{r}$. Note that $\mathcal{C}$ is nonempty since $[\underline{k}_0, \overline{k}_0]$ is nonempty, and Assumption \ref{assump:nonzerodenomk} guarantees that $\frac{\theta_1}{1 - k_0}$ is well-defined. Therefore, we have: $\cap_{r \in \mathcal{S}} \mathcal{I}_{r} \neq \emptyset.$ By Theorem 23.3 in \cite{munkres2000}, the union in (\ref{unionkconnected?}) is connected:
$$\bigcup_{r \in \mathcal{S}} \mathcal{I}_{r} = \Bigl[  \min_{r \in \mathcal{S} } \text{LB}(r), \max_{r \in \mathcal{S}} \text{UB}(r) \Bigr] .$$

We have now established that $\mathcal{I}^{k}_{\text{ATT}_{1}}$ can be expressed as the image of the feasible set under $f$, which implies that every $\tau \in \mathcal{I}^{k}_{\text{ATT}_{1}}$  has a pre-image $((k_j)_{j \in J(r)}, m,r) \in \mathcal{F}$. Fix $\bigl((\widetilde{k}_j)_{j\in J(\widetilde{r})},\widetilde{m},\widetilde{r} \bigr)\in \mathcal{F} $
such that $f((\widetilde{k}_j)_{j\in J(\widetilde{r})},\widetilde{m},\widetilde{r})=\tau$.
Extend $(\widetilde{k}_j)_{j\in J(\widetilde{r})}$ to a full vector 
$\widetilde{\boldsymbol{k}}=(\widetilde{k}_{-S},\ldots,\widetilde{k}_0)\in\mathcal{K}$
by choosing arbitrary values $\widetilde{k}_s\in[\underline k_s,\overline k_s]$ for all $s\notin J(\widetilde{r})$.
\begin{equation}\label{taudefnkcase}
    f((\widetilde{k}_j)_{j\in J(\widetilde{r})},\widetilde{m},\widetilde{r} ) = \frac{\theta_1 - \widetilde{m} \Delta_{\tilde{r}}}{1 - \widetilde{k} _{0} - \widetilde{m} (\widetilde{k}_{\tilde{r}} - \widetilde{k} _{\tilde{r}-1})} = \tau
\end{equation}
where: $\widetilde{\varphi}_{s} \coloneqq \widetilde{k}_{s} \cdot \tau$ for all $s \in \{-S, \ldots , 0\}$, $\widetilde{\delta}_{s} \coloneqq \Delta_{s} - (\widetilde{k}_{s} - \widetilde{k}_{s-1} ) \cdot \tau$ for all $s \in \mathcal{S}$, and $\widetilde{\delta}_{1} \coloneqq \widetilde{m} (\Delta_{\tilde{r}} - (\widetilde{k}_{\tilde{r}} - \widetilde{k}_{\tilde{r}-1}) \cdot \tau)$ where $\tilde{r} \in \mathcal{S}$. \smallskip

\noindent \textit{Construction of the DGP.} Throughout the construction, expectations without a subscript refer to the distribution corresponding to the observed data, whereas $\mathbb{E}_{\widetilde{P}}$ denotes expectation under the constructed DGP. Let $c_t \coloneqq \mathbb{E}[Y_t - Y_{t-1} \mid X=0]$, and let $\widetilde{P}$ retain the observed joint distribution of $(Y_{-S},\ldots,Y_1,X)$. I specify the treated units' never-treated potential outcomes through the conditional means
\begin{equation}
\label{dgpkcase}
\resizebox{0.9\linewidth}{!}{$
\mathbb{E}_{\widetilde{P}}[Y_t(0)\mid X=1]
\coloneqq
\begin{cases}
\mathbb{E}[Y_{-S}\mid X=1] - \widetilde{k}_{-S}\, \tau,
  & t = -S, \\[6pt]
\displaystyle \sum_{j=-(S-1)}^{t} (c_j + \widetilde{\delta}_j)
  + \mathbb{E}[Y_{-S}\mid X=1]
  - \widetilde{k}_{-S}\, \tau,
  & t \in \{-(S-1),\ldots,0,1\},
\end{cases}
$}
\end{equation}
so that $\mathbb{E}_{\widetilde{P}}[Y_t(0)-Y_{t-1}(0)\mid X=1]=c_t+\widetilde{\delta}_t$ for every $t\in\{-(S-1),\ldots,1\}$. Under the constructed DGP, set $Y_t(0)=Y_t$ for comparison units and $Y_t(1)=Y_t$ for treated units. For treated units, define $Y_t(0)$ by translating $Y_t$ by a period-specific constant chosen so that its conditional mean given $X=1$ equals the value displayed above, and assign the remaining potential outcomes arbitrarily. By construction, the marginal distribution of $(Y_{-S},\ldots,Y_1,X)$ under $\widetilde P$ equals the observed distribution. In particular,
$\mathbb E_{\widetilde P}[Y_t\mid X=x] =\mathbb E[Y_t\mid X=x].$

 \smallskip
\noindent \textit{Verification.}

\smallskip
\noindent (a) \textit{Assumptions.} For comparison units, $Y_t(0)=Y_t$ for all $t\in\{-S,\ldots,1\}$. Since $\widetilde P$ reproduces the observed distribution, for every $t\in\{-(S-1),\ldots,1\}$, $\mathbb{E}_{\widetilde{P}}[Y_t(0)-Y_{t-1}(0)\mid X=0]=c_t$; hence
\[
\delta_t(\widetilde{P})=\mathbb{E}_{\widetilde{P}}[Y_t(0)-Y_{t-1}(0)\mid X=1]-c_t=(c_t+\widetilde{\delta}_t)-c_t=\widetilde{\delta}_t.
\]
For the anticipation effects, $\varphi_t(\widetilde{P})=\mathbb{E}[Y_t\mid X=1]-\mathbb{E}_{\widetilde{P}}[Y_t(0)\mid X=1]$. At $t=-S$,
\[
\varphi_{-S}(\widetilde{P})=\mathbb{E}[Y_{-S}\mid X=1]-\bigl(\mathbb{E}[Y_{-S}\mid X=1]-\widetilde{k}_{-S}\tau\bigr)=\widetilde{k}_{-S}\tau=\widetilde{\varphi}_{-S}.
\]
For $-S<t\le0$, since $\Delta_j+c_j=\mathbb{E}[Y_j-Y_{j-1}\mid X=1]$ for $j\in\mathcal{S}$ and $\widetilde{\delta}_j=\Delta_j-(\widetilde{k}_j-\widetilde{k}_{j-1})\tau$,
\begin{align*}
\varphi_t(\widetilde{P})
&=\mathbb{E}[Y_t\mid X=1]-\mathbb{E}_{\widetilde{P}}[Y_t(0)\mid X=1]\\
&=\sum_{j=-(S-1)}^{t}(\Delta_j+c_j)-\sum_{j=-(S-1)}^{t}(c_j+\widetilde{\delta}_j)+\widetilde{k}_{-S}\tau
=\sum_{j=-(S-1)}^{t}(\widetilde{k}_j-\widetilde{k}_{j-1})\tau+\widetilde{k}_{-S}\tau\\
&=(\widetilde{k}_t-\widetilde{k}_{-S})\tau+\widetilde{k}_{-S}\tau=\widetilde{k}_t\tau=\widetilde{\varphi}_t.
\end{align*}
Thus $\varphi_s(\widetilde{P})=\widetilde{k}_s\tau=\widetilde{\varphi}_s$ for all $s\in\{-S,\ldots,0\}$; Assumption~\ref{assump:krelaxmultpre} is verified at the end, once $\text{ATT}_1(\widetilde{P})=\tau$ has been established. Since $\widetilde{r}\in\mathcal{S}$ and $\delta_{\widetilde{r}}(\widetilde{P})=\widetilde{\delta}_{\widetilde{r}}$,
\[
|\delta_1(\widetilde{P})|=|\widetilde{\delta}_1|=|\widetilde{m}|\,|\widetilde{\delta}_{\widetilde{r}}|\le M|\widetilde{\delta}_{\widetilde{r}}|\le M|\widetilde{\delta}_{\widetilde{s}^{\star}}|,
\]
where $\widetilde{s}^{\star}\in\arg\max_{s\in\mathcal{S}}|\widetilde{\delta}_s|$; hence Assumption~\ref{assump:rmmultpretrends} holds.

\smallskip
\noindent (b) \textit{Observed data.} Since $\widetilde{P}$ preserves the observed joint distribution of $(Y_t,X)$, $\mathbb{E}_{\widetilde{P}}[Y_t\mid X=x]=\mathbb{E}[Y_t\mid X=x]$; hence $\Delta_s(\widetilde{P})=\Delta_s$ for all $s\in\mathcal{S}$ and $\theta_1(\widetilde{P})=\theta_1$.

\smallskip
\noindent (c) $\text{ATT}_1(\widetilde{P})=\tau$. By Lemma~\ref{lemma:attbiasdecomp23} applied to $\widetilde{P}$ and parts (a)--(b),
\[
\text{ATT}_1(\widetilde{P})=\theta_1(\widetilde{P})+\varphi_0(\widetilde{P})-\delta_1(\widetilde{P})=\theta_1+\widetilde{\varphi}_0-\widetilde{\delta}_1.
\]
By \eqref{taudefnkcase} and Assumption~\ref{assump:nonzerodenomk} (which makes the denominator nonzero),
\[
\tau=\frac{\theta_1-\widetilde{m}\Delta_{\widetilde{r}}}{1-\widetilde{k}_0-\widetilde{m}(\widetilde{k}_{\widetilde{r}}-\widetilde{k}_{\widetilde{r}-1})}
\iff
\tau=\theta_1+\widetilde{k}_0\tau-\widetilde{m}\bigl(\Delta_{\widetilde{r}}-(\widetilde{k}_{\widetilde{r}}-\widetilde{k}_{\widetilde{r}-1})\tau\bigr)=\theta_1+\widetilde{\varphi}_0-\widetilde{\delta}_1,
\]
using $\widetilde{\varphi}_0=\widetilde{k}_0\tau$ and the definition of $\widetilde{\delta}_1$. Hence $\text{ATT}_1(\widetilde{P})=\tau$.

\smallskip
\noindent Finally, with $\text{ATT}_1(\widetilde{P})=\tau$ established, $\varphi_s(\widetilde{P})=\widetilde{k}_s\tau=\widetilde{k}_s\cdot\text{ATT}_1(\widetilde{P})$ for $s\in\{-S,\ldots,0\}$; since $\widetilde{k}_s\in[\underline{k}_s,\overline{k}_s]$, Assumption~\ref{assump:krelaxmultpre} holds under $\widetilde{P}$ with $k_s^{\star}=\widetilde{k}_s$.

\smallskip
Therefore, the constructed DGP $\widetilde{P}$ yields the observed data, satisfies the maintained assumptions, and gives $\text{ATT}_1(\widetilde{P})=\tau$; so $\tau$ is in the identified set for $\text{ATT}_1$. Since $\tau$ was an arbitrary element of $\mathcal{I}^{k}_{\text{ATT}_1}$, $\mathcal{I}^{k}_{\text{ATT}_1}$ is contained in the identified set for $\text{ATT}_1$. \smallskip

By Parts 1 and 2, $\mathcal{I}_{\text{ATT}_{1}}^{k}$ is the identified set for the $\text{ATT}_{1}$.
\end{proofnolab} \smallskip

\ProofOf{Corollary \ref{corr:avgdecomp}}
\begin{proofnolab}
\noindent Applying Lemma~\ref{lemma:attKdecomp} to each $j\in\{1,\ldots,h\}$ and averaging:
\begin{align*}
  \overline{\text{ATT}}_h
  &= \frac{1}{h}\sum_{j=1}^h \text{ATT}_j
   = \frac{1}{h}\sum_{j=1}^h\Bigl(\theta_j + \varphi_0 - \sum_{i=1}^j\delta_i\Bigr) = \frac{1}{h}\sum_{j=1}^h\theta_j
     + \varphi_0
     - \frac{1}{h}\sum_{j=1}^h\sum_{i=1}^j\delta_i.
\end{align*}
Exchanging the order of summation:
\[
  \sum_{j=1}^h\sum_{i=1}^j\delta_i
  = \sum_{i=1}^h\sum_{j=i}^h\delta_i
  = \sum_{i=1}^h(h-i+1)\,\delta_i.
\]
Substituting yields the result.
\end{proofnolab} \smallskip 

\ProofOf{Corollary \ref{corr:stagcohort}}
\begin{proofnolab}
\noindent Fix $g\in\mathcal G$ and restrict attention to the subpopulation with $G \in \{g, \infty\}$. Define $X^{(g)} \coloneqq \mathbbm{1}[G=g]$ on this subpopulation and re-index calendar time by $r=t-g$. The event time support is $\{-S_g, \ldots , t_{\text{max}}-g\}$, treatment begins at $r=1$, and units with $X^{(g)}=0$ follow the never-treated path. After the substitutions
\[
(X, Y_r (1), Y_r (0), \varphi_r, \delta_r, \Delta_r, \theta_h, S, \mathcal{S}, T)
\rightarrow
(X^{(g)}, Y_{g+r}(g), Y_{g+r}(\infty), \varphi_{g,r}, \delta_{g,r}, \Delta_{g,r}, \theta_{g,h}, S_g, \mathcal{S}_g, t_{\text{max}}-g)
\]
with the bounds $[\underline{A}_r, \overline{A}_r]$ unchanged, Assumptions \ref{assump:staginit}, \ref{assump:stagbounds}, and \ref{assump:stagrm} are the cohort-$g$ analogs of Assumptions \ref{initialize}, \ref{abounds}, and \ref{assump:rmmult}. Since $h\leq t_{\text{max}}-g$, $\text{ATT}(g,h)$ and $\theta_{g,h}$ are well-defined. Therefore Theorem \ref{thm:multpost} yields
\[
\mathcal I_{g,h}
=
\Bigl[
\min_{r\in\mathcal S_g} l_{g,h}(r),
\;
\max_{r\in\mathcal S_g} u_{g,h}(r)
\Bigr],
\]
with $l_{g,h}(r)$ and $u_{g,h}(r)$ as defined in the Corollary statement. Sharpness follows from
the sharpness conclusion of Theorem~\ref{thm:multpost} under the same
re-indexing.
\end{proofnolab} \smallskip 

\ProofOf{Theorem \ref{thm:stagagg}}
\begin{proofnolab}
\noindent The proof proceeds in two steps. Since $\mathcal{G}_h \coloneqq \{g \in \mathcal{G} : g+h \leq t_{\text{max}} \}$, $\text{ATT}(g,h)$ is well-defined for every $g \in \mathcal{G}_h$.

\smallskip
\noindent\textbf{Step 1: The joint identified set equals the Cartesian product.}

\smallskip
\noindent Define the joint parameter vector $\boldsymbol{\tau} \coloneqq (\tau_g)_{g\in\mathcal{G}_h} \in\mathbb R^{|\mathcal{G}_h|}.$ Let $\mathcal{D}$ denote the collection of DGPs that (i) generate the observed data and (ii) satisfy Assumptions~\ref{assump:staginit}--\ref{assump:stagrm}. Define the joint identified set
$\mathcal I_{\text{joint}} \coloneqq \Bigl\{ \boldsymbol{\tau}\in\mathbb R^{|\mathcal{G}_h|}
: \exists\,D\in\mathcal D \text{ such that } \text{ATT}(g,h)=\tau_g \; \forall g\in\mathcal{G}_h
\Bigr\}.$
\noindent I claim that $\mathcal I_{\text{joint}} = \prod_{g\in\mathcal{G}_h} \mathcal I_{g,h}.$ \smallskip 

\noindent\textbf{($\subseteq$).}
Let $\boldsymbol{\tau}\in\mathcal I_{\text{joint}}$. By definition, there exists a DGP consistent with the observed data and Assumptions~\ref{assump:staginit}--\ref{assump:stagrm} such that
$\text{ATT}(g,h)=\tau_g$ for all $g\in\mathcal{G}_h$. Fix any $g \in \mathcal{G}_h$. Restricting the DGP to the subpopulation with $G \in \{g,\infty\}$ yields a DGP consistent with the cohort $g$ observed data, satisfying the maintained assumptions for cohort $g$, and $\text{ATT}(g,h) = \tau_g$. Therefore, by Corollary~\ref{corr:stagcohort},
$\tau_g\in\mathcal I_{g,h}.$ Since $g \in \mathcal{G}_h$ was arbitrary, $\boldsymbol{\tau} \in \prod_{g\in\mathcal{G}_h} \mathcal I_{g,h}.$ Since $\boldsymbol{\tau}$ was an arbitrary element of $\mathcal{I}_{\text{joint}}$, it follows that $\mathcal I_{\text{joint}} \subseteq \prod_{g\in\mathcal{G}_h} \mathcal I_{g,h}.$

\smallskip
\noindent\textbf{($\supseteq$).}
Let $\boldsymbol{\tau} \in \prod_{g\in\mathcal{G}_h} \mathcal I_{g,h},$ so that $\tau_g\in[l_{g,h},u_{g,h}]$ for every $g \in \mathcal{G}_{h}$. By the sharpness part of Corollary \ref{corr:stagcohort}, there exists an admissible DGP on the subpopulation with $G \in \{g, \infty\}$ that reproduces the observed data for that subpopulation, satisfies the maintained assumptions for cohort $g$, and yields $\text{ATT}(g,h)=\tau_g$.

\smallskip
\noindent These constructions are cohort-specific. Although the bounds $[\underline{A}_r, \overline{A}_r]$ and $M$ are common across cohorts, they apply separately to each cohort and impose no restrictions linking $(\varphi_{g,r}, \delta_{g,r})$ across cohorts. Each cohort-level construction modifies only the counterfactual outcomes for units in cohort $g$, while preserving the observed distribution, including that of the never-treated group. Hence, the cohort-specific constructions can be imposed simultaneously. Therefore, there exists a joint DGP that is consistent with the observed data, satisfies Assumptions~\ref{assump:staginit}--\ref{assump:stagrm}, and satisfies $\text{ATT}(g,h)=\tau_g$ for all $g\in\mathcal{G}_h.$ 

\smallskip
\noindent Therefore $\boldsymbol{\tau} \in
\mathcal I_{\text{joint}}.$ Since $\boldsymbol{\tau}$ was an arbitrary element of $\prod_{g\in\mathcal{G}_h} \mathcal I_{g,h}$, it follows that $\mathcal I_{\text{joint}} \supseteq \prod_{g\in\mathcal{G}_h} \mathcal I_{g,h}.$

\smallskip
\noindent\textbf{Step 2: The identified set for $\text{ATT}_h^w$ is the weighted Minkowski sum.}

\smallskip
\noindent From Step 1, $\mathcal I_{\text{joint}} = \prod_{g\in\mathcal{G}_h} \mathcal I_{g,h} =  \prod_{g\in\mathcal{G}_h} [l_{g,h},u_{g,h}].$ Consider the image of the joint identified set under the linear map $L:\boldsymbol\tau
\mapsto \sum_{g\in\mathcal{G}_h} w_g\tau_g.$ I first show that $\mathcal I_{\text{ATT}_h^w} = L(\mathcal I_{\text{joint}}) = L\!\left( \prod_{g\in\mathcal{G}_h} [l_{g,h},u_{g,h}] \right).$

\smallskip \noindent
For the inclusion $L\!\left( \prod_{g\in\mathcal{G}_h} [l_{g,h},u_{g,h}] \right) \subseteq \mathcal I_{\text{ATT}_h^w},$ let $\tau^w \in L\!\left( \prod_{g\in\mathcal{G}_h} [l_{g,h},u_{g,h}] \right).$
Then there exists $\boldsymbol{\tau} \in \prod_{g\in\mathcal{G}_h} [l_{g,h},u_{g,h}]$ such that
$L(\boldsymbol{\tau}) =\tau^w. $ By Step 1, $\boldsymbol{\tau}$ is realized by some DGP in $\mathcal D$. Under that DGP, $\text{ATT}_h^w = L(\boldsymbol{\tau}) = \tau^w,$ so $\tau^w \in \mathcal I_{\text{ATT}_h^w}.$ \smallskip

\noindent For the reverse inclusion $\mathcal I_{\text{ATT}_h^w} \subseteq L\!\left( \prod_{g\in\mathcal{G}_h} [l_{g,h},u_{g,h}] \right),$ let $\tau^w \in \mathcal I_{\text{ATT}_h^w}.$
Then, by definition of $\mathcal I_{\text{ATT}_h^w}$, there exists a DGP in $\mathcal D$ such that
$\text{ATT}_h^w = \tau^w.$ Let
$\boldsymbol{\tau} = (\text{ATT}(g,h))_{g\in\mathcal{G}_h}$ denote the vector of cohort-specific effects under this same DGP. Hence, $\boldsymbol{\tau} \in \mathcal{I}_{\text{joint}}$, so by Step 1, $\boldsymbol{\tau} \in \prod_{g\in\mathcal{G}_h} [l_{g,h},u_{g,h}]$. Moreover, by the definitions of $L$ and $\text{ATT}_{h}^{w}$, 
$L(\boldsymbol{\tau}) = \tau^w.$ Hence
$\tau^w \in L\!\left( \prod_{g\in\mathcal{G}_h} [l_{g,h},u_{g,h}] \right).$ 
Hence
\begin{align}
\mathcal I_{\text{ATT}_h^w} = L\!\left( \prod_{g\in\mathcal{G}_h} [l_{g,h},u_{g,h}] \right) = \left\{\sum_{g\in\mathcal{G}_h} w_g\tau_g : \tau_g\in[l_{g,h},u_{g,h}]\ \forall g\in\mathcal{G}_h \right\}.
\label{eqn:stagimage_agg}
\end{align}
The set on the right side of~\eqref{eqn:stagimage_agg} is the weighted Minkowski sum 
$\sum_{g\in\mathcal{G}_h} w_g\,\mathcal I_{g,h}.$ I now show that, given $w_g \geq 0$, this set is the interval $[\sum_{g \in \mathcal{G}_h} w_g l_{g,h}, \sum_{g \in \mathcal{G}_h} w_g u_{g,h}].$ I first establish its lower and upper endpoints and then show that every value between them is attained.

\smallskip
\noindent\textit{Lower bound.}
Setting $\tau_g=l_{g,h}$ for all $g$ gives
$\sum_{g \in \mathcal{G}_h} w_g\tau_g = \sum_{g \in \mathcal{G}_h} w_g l_{g,h},$
so $\sum_{g \in \mathcal{G}_h} w_g l_{g,h}\in\mathcal I_{\text{ATT}_h^w}$.
Moreover, for any admissible $\boldsymbol{\tau}$, $\tau_g\geq l_{g,h}$ for all $g \in \mathcal{G}_h$. Since $w_g\geq0$ for all $g \in \mathcal{G}_h$,
$\sum_{g \in \mathcal{G}_h} w_g\tau_g \geq \sum_{g \in \mathcal{G}_h} w_g l_{g,h}.$ Thus $\sum_{g \in \mathcal{G}_h} w_g l_{g,h}$ is the minimum of
$\mathcal I_{\text{ATT}_h^w}$.

\smallskip
\noindent\textit{Upper bound.}
Setting $\tau_g=u_{g,h}$ for all $g$ gives
$\sum_{g \in \mathcal{G}_h} w_g\tau_g = \sum_{g \in \mathcal{G}_h} w_g u_{g,h},$
so $\sum_{g \in \mathcal{G}_h} w_g u_{g,h}\in\mathcal I_{\text{ATT}_h^w}$.  Moreover, for any admissible $\boldsymbol{\tau}$, $\tau_g\leq u_{g,h}$ for all $g \in \mathcal{G}_h$. Since $w_g\geq0$ for all $g \in \mathcal{G}_h$, $\sum_{g \in \mathcal{G}_h} w_g\tau_g \leq\sum_{g \in \mathcal{G}_h} w_g u_{g,h}.$ Thus $\sum_{g \in \mathcal{G}_h} w_g u_{g,h}$ is the maximum of
$\mathcal I_{\text{ATT}_h^w}$.

\smallskip
\noindent\textit{All intermediate values.}
Let
$\tau^w \in \left[ \sum_{g \in \mathcal{G}_h} w_g l_{g,h}, \; \sum_{g \in \mathcal{G}_h} w_g u_{g,h}
\right].$ Then there exists $\lambda\in[0,1]$ such that $\tau^w =(1-\lambda)\sum_{g \in \mathcal{G}_h} w_g l_{g,h} + \lambda\sum_{g \in \mathcal{G}_h} w_g u_{g,h}.$ Define, for each $g\in\mathcal{G}_h$, $\tau_g = (1-\lambda)l_{g,h} +\lambda u_{g,h}$, and let $\boldsymbol{\tau} \coloneqq (\tau_g)_{g \in \mathcal{G}_h}$.
Then $\tau_g\in[l_{g,h},u_{g,h}]$ for all $g \in \mathcal{G}_h$, so $\boldsymbol{\tau} \in \prod_{g \in \mathcal{G}_h} [l_{g,h},u_{g,h}]. $ Furthermore,
\begin{align*}
\sum_{g \in \mathcal{G}_h} w_g\tau_g = \sum_{g \in \mathcal{G}_h} w_g\bigl[(1-\lambda)l_{g,h}+\lambda u_{g,h}\bigr] =
(1-\lambda)\sum_{g \in \mathcal{G}_h} w_g l_{g,h}
+
\lambda\sum_{g \in \mathcal{G}_h} w_g u_{g,h} = \tau^w.
\end{align*}
Since $\boldsymbol{\tau} \in \prod_{g\in\mathcal G_h} [l_{g,h},u_{g,h}]$ and $L(\boldsymbol{\tau})=\tau^w,$ we have $\tau^w \in L\!\left( \prod_{g\in\mathcal G_h} [l_{g,h},u_{g,h}] \right).$ By the equality established above,
$L\!\left( \prod_{g\in\mathcal G_h} [l_{g,h},u_{g,h}] \right) = \mathcal I_{\text{ATT}_h^w}.$
Therefore $\tau^w\in\mathcal I_{\text{ATT}_h^w}.$

\smallskip
\noindent Hence $\mathcal I_{\text{ATT}_h^w}
=
\sum_{g \in \mathcal{G}_h} w_g\,\mathcal I_{g,h}
=
\left[
\sum_{g \in \mathcal{G}_h} w_g l_{g,h},
\;
\sum_{g \in \mathcal{G}_h} w_g u_{g,h}
\right].$
\end{proofnolab} \smallskip

\ProofOf{Proposition \ref{prop:aboundsbenchmark}}
\begin{proofnolab}
\noindent For any $t \in \{-(S-1),\ldots,0\}$, Corollary \ref{prop:pretrenddecopmpsim} gives the decomposition $\Delta_t=\delta_t+(\varphi_t-\varphi_{t-1}).$ First consider $t \in \{-(S-1),\ldots,t_{\text{ann}}\}$. By Assumption \ref{initialize}, $\varphi_{-S}=0$, and by the maintained no anticipation restriction prior to period $t_{\text{ann}}+1$, $\varphi_t=0$ for all $t\in\{-S,\ldots,t_{\text{ann}}\}$. Hence, $\varphi_t-\varphi_{t-1}=0$ for all $t\in\{-(S-1),\ldots,t_{\text{ann}}\}.$ Thus Assumption \ref{abounds} holds in the pre-announcement window with $[\underline A_t,\overline A_t]=[0,0]$.

Now consider $t\in\{t_{\text{ann}}+1,\ldots,0\}$. Rearranging the pre-trend decomposition yields
$\varphi_t-\varphi_{t-1}=\Delta_t-\delta_t$. By assumption, $|\delta_t|\leq \Delta^{\text{ann}}$, so $\delta_t\in[-\Delta^{\text{ann}},\Delta^{\text{ann}}]$. Therefore,
$\Delta_t-\delta_t \in [\Delta_t-\Delta^{\text{ann}},\Delta_t+\Delta^{\text{ann}}].$
Equivalently, $\varphi_t-\varphi_{t-1} \in [\Delta_t-\Delta^{\text{ann}},\Delta_t+\Delta^{\text{ann}}]$ for all $t\in\{t_{\text{ann}}+1,\ldots,0\}.$

Combining the pre-announcement and post-announcement cases gives \eqref{eq:cal2bounds}, so Assumption \ref{abounds} holds with the stated bounds. \end{proofnolab}

\subsection*{Appendix \ref{appendix:discussion23}}

\ProofOf{Lemma \ref{lem:feasiblesets23}} 

\begin{proofnolab}
\noindent Fix $a\in[\underline{A}_{0},\overline{A}_{0}]$. Write $\text{LB}(a)=\theta_1+a-M|\Delta_0-a|$ and  $\text{UB}(a)=\theta_1+a+M|\Delta_0-a|.$ 
If $M=0$, then $\text{LB}(a)=\text{UB}(a)=\theta_1+a$, so $\mathcal{I}^{A}_{\text{ATT}_1}=
\bigl[\theta_1+\underline{A}_0,\ \theta_1+\overline{A}_0\bigr],$
which corresponds to (iii). Going forward, assume $M>0$.

\smallskip\noindent
\textbf{Part 1: Which endpoint minimizes $\text{LB}(\cdot)$.} For $a\le \Delta_0$,
\[
\text{LB}(a)=\theta_1+a-M(\Delta_0-a)=\theta_1+(1+M)a-M\Delta_0,
\]
which is increasing in $a$ since $M>0$. For $a\ge \Delta_0$,
\[
\text{LB}(a)=\theta_1+a-M(a-\Delta_0)=\theta_1+(1-M)a+M\Delta_0,
\]
which is increasing in $a$ if $M<1$, constant if $M=1$, and decreasing in $a$ if $M>1$. Therefore:

\begin{itemize}[leftmargin=*]
\item If $\overline{A}_0\le \Delta_0$, then both endpoints satisfy $a\le\Delta_0$, so $\text{LB}$ is increasing over $[\underline{A}_0,\overline{A}_0]$ and the minimum is attained at $\underline{A}_0$.
\item If $\underline{A}_0\ge \Delta_0$, then both endpoints satisfy $a\ge\Delta_0$, so the minimum is attained at $\underline{A}_0$ when $M\le 1$ and at $\overline{A}_0$ when $M>1$ (with a tie when $M=1$).
\item If $\underline{A}_0\le \Delta_0\le \overline{A}_0$, then $\text{LB}(\underline{A}_0)=\theta_1+\underline{A}_0(1+M)-M\Delta_0$ and $\text{LB}(\overline{A}_0)=\theta_1+\overline{A}_0(1-M)+M\Delta_0.$ Thus $\text{LB}(\underline{A}_0)\le \text{LB}(\overline{A}_0)$ iff $\Delta_0\ge c_L$, where $c_L \coloneqq \frac{\underline{A}_0(1+M)-\overline{A}_0(1-M)}{2M}.$
\end{itemize}

\noindent
\textbf{Part 2: Which endpoint maximizes $\text{UB}(\cdot)$.} For $a\le \Delta_0$,
\[
\text{UB}(a)=\theta_1+a+M(\Delta_0-a)=\theta_1+(1-M)a+M\Delta_0,
\]
which is increasing in $a$ if $M<1$, constant if $M=1$, and decreasing if $M>1$. For $a\ge \Delta_0$,
\[
\text{UB}(a)=\theta_1+a+M(a-\Delta_0)=\theta_1+(1+M)a-M\Delta_0,
\]
which is increasing in $a$. Therefore:

\begin{itemize}[leftmargin=*]
\item If $\underline{A}_0\ge \Delta_0$, then both endpoints satisfy $a\ge\Delta_0$ and $\text{UB}$ is increasing, so the maximum is attained at $\overline{A}_0$.
\item If $\overline{A}_0\le \Delta_0$, then both endpoints satisfy $a\le\Delta_0$, so the maximum is attained at $\overline{A}_0$ when $M\le 1$ and at $\underline{A}_0$ when $M>1$ (with a tie when $M=1$).
\item If $\underline{A}_0\le \Delta_0\le \overline{A}_0$, then
$\text{UB}(\underline{A}_0)=\theta_1+\underline{A}_0(1-M)+M\Delta_0$ and $\text{UB}(\overline{A}_0)=\theta_1+\overline{A}_0(1+M)-M\Delta_0.$ Thus $\text{UB}(\overline{A}_0)\ge \text{UB}(\underline{A}_0)$ iff $\Delta_0\le c_U$, where $c_U \coloneqq \frac{\overline{A}_0(1+M)-\underline{A}_0(1-M)}{2M}.$
\end{itemize} 

\noindent
\textbf{Part 3: Conditions for $\boldsymbol{\underline{A}_{0} \leq \Delta_0 \leq \overline{A}_{0}}$.}
Suppose $\underline{A}_0\le \Delta_0\le \overline{A}_0$. Parts 1 and 2 imply the following endpoint selection rules in this region:
\begin{align}
\arg\min_{a\in\{\underline{A}_0,\overline{A}_0\}}\text{LB}(a)
&=
\begin{cases}
\underline{A}_0, & \text{if } \Delta_0\ge c_L,\\
\overline{A}_0, & \text{if } \Delta_0< c_L,
\end{cases}
\label{eq:LBselect}\\[4pt]
\arg\max_{a\in\{\underline{A}_0,\overline{A}_0\}}\text{UB}(a)
&=
\begin{cases}
\overline{A}_0, & \text{if } \Delta_0\le c_U,\\
\underline{A}_0, & \text{if } \Delta_0> c_U,
\end{cases}
\label{eq:UBselect}
\end{align}
(with ties at $\Delta_0=c_L$ or $\Delta_0=c_U$). The location of $c_L$ and $c_U$ relative to $[\underline{A}_0,\overline{A}_0]$ depends on $M$. Note that:
\[
c_L-\underline{A}_0=\frac{(\underline{A}_0-\overline{A}_0)(1-M)}{2M},
\qquad
c_U-\overline{A}_0=\frac{(\overline{A}_0-\underline{A}_0)(1-M)}{2M}.
\]
Hence, if $M<1$, then $c_L\le \underline{A}_0$ and $c_U\ge \overline{A}_0$, so every $\Delta_0\in[\underline{A}_0,\overline{A}_0]$ satisfies $\Delta_0\ge c_L$ and $\Delta_0\le c_U$. By \eqref{eq:LBselect}--\eqref{eq:UBselect}, the minimizing endpoint for $\text{LB}$ is $\underline{A}_0$ and the maximizing endpoint for $\text{UB}$ is $\overline{A}_0$, so the cross-endpoint configuration $[\text{LB}(\underline{A}_0),\,\text{UB}(\overline{A}_0)]$
obtains. Notice that the conditions for same endpoint bounds would be contradicted here. 

If $M>1$ and $\underline{A}_0 < \overline{A}_0$, then $c_L>\underline{A}_0$ and $c_U<\overline{A}_0$. Since $c_U-c_L=(\overline{A}_0-\underline{A}_0)/M\ge 0$, values of $\Delta_0$ satisfying $\underline{A}_{0} \leq \Delta_0 \leq \overline{A}_{0}$ are divided into the following cases: 
\[
\underline{A}_0 \le \Delta_0 < c_L,\qquad
c_L \le \Delta_0 \le c_U,\qquad
c_U < \Delta_0 \le \overline{A}_0.
\]
Applying \eqref{eq:LBselect}--\eqref{eq:UBselect} in each subregion yields:
\begin{itemize}[leftmargin=*]
\item If $\underline{A}_0 \le \Delta_0 < c_L$, then $\text{LB}$ is minimized at $\overline{A}_0$ and $\text{UB}$ is maximized at $\overline{A}_0$, so $[\text{LB}(\overline{A}_0),\,\text{UB}(\overline{A}_0)]$ obtains.
\item If $c_L \le \Delta_0 \le c_U$, then $\text{LB}$ is minimized at $\underline{A}_0$ and $\text{UB}$ is maximized at $\overline{A}_0$, so $[\text{LB}(\underline{A}_0),\,\text{UB}(\overline{A}_0)]$ obtains.
\item If $c_U < \Delta_0 \le \overline{A}_0$, then $\text{LB}$ is minimized at $\underline{A}_0$ and $\text{UB}$ is maximized at $\underline{A}_0$, so $[\text{LB}(\underline{A}_0),\,\text{UB}(\underline{A}_0)]$ obtains.
\end{itemize}
Finally, when $M=1$ we have $c_L=\underline{A}_0$ and $c_U=\overline{A}_0$, so the cross-endpoint configuration obtains.  Hence, when $\underline{A}_{0} \leq \Delta_0 \leq \overline{A}_{0}$, the same-endpoint cases are only feasible when $M>1$.

\smallskip\noindent
\textbf{Part 4: Combine parts 1-3.}
The identified set is obtained by pairing the minimizing endpoint for $\text{LB}$ with the maximizing endpoint for $\text{UB}$. This yields the three feasible configurations in (i)--(iii). \smallskip 

\noindent Finally, the remaining configuration
$[\text{LB}(\overline{A}_0),\,\text{UB}(\underline{A}_0)]$ is infeasible. If
$\Delta_0\leq \underline{A}_0$, then the maximum of $\text{UB}$ is attained at
$\overline{A}_0$, so $\underline{A}_0$ cannot attain the upper endpoint. If
$\overline{A}_0\leq \Delta_0$, then the minimum of $\text{LB}$ is attained at
$\underline{A}_0$, so $\overline{A}_0$ cannot attain the lower endpoint. It remains
to consider $\underline{A}_0\leq \Delta_0\leq \overline{A}_0$. In this case, the
reversed configuration would require $\Delta_0<c_L$ and $\Delta_0>c_U$
simultaneously, which is impossible because
$c_U-c_L=\frac{\overline{A}_0-\underline{A}_0}{M}\geq 0.$
Hence (iv) is infeasible when $\underline{A}_0<\overline{A}_0$.
\end{proofnolab} \smallskip

\ProofOf{Proposition \ref{corr:Awidth}} 

\begin{proofnolab}
\noindent Assume $M>0$. When only parallel trends is relaxed (i.e.\ $\underline{A}_0=\overline{A}_0=0$), the identified set has width $W_{PT}=2M|\Delta_0|.$ By Lemma \ref{lem:feasiblesets23}, under joint deviations the identified set either (a) has same-endpoint form
$[\text{LB}(a),\text{UB}(a)]$ for some $a\in\{\underline{A}_0,\overline{A}_0\}$, or (b) has cross-endpoint form
$[\text{LB}(\underline{A}_0),\text{UB}(\overline{A}_0)]$.

\smallskip\noindent
\textbf{(i) Same-endpoint bounds.}
If both bounds are attained at the same endpoint $a\in\{\underline{A}_0,\overline{A}_0\}$, then
$W_{PTAE}=\text{UB}(a)-\text{LB}(a)=2M|\Delta_0-a|.$ Since $M>0$, $W_{PTAE}<W_{PT}$ is equivalent to $|\Delta_0-a|<|\Delta_0|$. Squaring both sides yields
$(\Delta_0-a)^2<\Delta_0^2
\quad\Longleftrightarrow\quad
a(a-2\Delta_0)<0,$
which is equivalent to $a\in(0,2\Delta_0)$ when $\Delta_0>0$ and to $a\in(2\Delta_0,0)$ when $\Delta_0<0$ (and is never satisfied when $\Delta_0=0$).

\smallskip\noindent
\textbf{(ii) Cross-endpoint bounds.}
If the identified set is $[\text{LB}(\underline{A}_0),\text{UB}(\overline{A}_0)]$, then
$W_{PTAE}
=
\text{UB}(\overline{A}_0)-\text{LB}(\underline{A}_0)
=
(\overline{A}_0-\underline{A}_0)
+M\bigl(|\Delta_0-\overline{A}_0|+|\Delta_0-\underline{A}_0|\bigr).$
We compare this expression to $W_{PT}=2M|\Delta_0|$ on each feasible sub-case in Lemma \ref{lem:feasiblesets23}.
\begin{itemize}
    \item[] \textbf{Case (ii.a): $\overline{A}_0\le \Delta_0$ and $M<1$.}
Then $|\Delta_0-\overline{A}_0|=\Delta_0-\overline{A}_0$ and $|\Delta_0-\underline{A}_0|=\Delta_0-\underline{A}_0$, so
$W_{PTAE}=(\overline{A}_0-\underline{A}_0)+M\bigl(2\Delta_0-(\overline{A}_0+\underline{A}_0)\bigr).$
Thus $W_{PTAE}<W_{PT}$ is equivalent to
$\overline{A}_0(1-M)-\underline{A}_0(1+M)+2M(\Delta_0-|\Delta_0|)<0,$ i.e.
$\Delta_0-|\Delta_0|
<
\frac{\underline{A}_0(1+M)-\overline{A}_0(1-M)}{2M}.$

\item[] \textbf{Case (ii.b): $\underline{A}_0\ge \Delta_0$ and $M<1$.}
Then $|\Delta_0-\overline{A}_0|=\overline{A}_0-\Delta_0$ and $|\Delta_0-\underline{A}_0|=\underline{A}_0-\Delta_0$, so
$W_{PTAE}=(\overline{A}_0-\underline{A}_0)+M\bigl((\overline{A}_0+\underline{A}_0)-2\Delta_0\bigr).$ Thus $W_{PTAE}<W_{PT}$ is equivalent to
$\overline{A}_0(1+M)-\underline{A}_0(1-M)-2M(\Delta_0+|\Delta_0|)<0,$ i.e.
$\Delta_0+|\Delta_0|
>
\frac{\overline{A}_0(1+M)-\underline{A}_0(1-M)}{2M}.$

\item[] \textbf{Case (ii.c): $\underline{A}_0\le \Delta_0\le \overline{A}_0$ and
$\frac{\underline{A}_0(1+M)-\overline{A}_0(1-M)}{2M}\le\Delta_0\le\frac{\overline{A}_0(1+M)-\underline{A}_0(1-M)}{2M}$.}
Then $|\Delta_0-\overline{A}_0|=\overline{A}_0-\Delta_0$ and $|\Delta_0-\underline{A}_0|=\Delta_0-\underline{A}_0$, hence
$W_{PTAE}=(\overline{A}_0-\underline{A}_0)+M(\overline{A}_0-\underline{A}_0)=(\overline{A}_0-\underline{A}_0)(1+M).$ Therefore $W_{PTAE}<W_{PT}$ is equivalent to
$(\overline{A}_0-\underline{A}_0)(1+M)<2M|\Delta_0|
\quad\Longleftrightarrow\quad
|\Delta_0|>\frac{(\overline{A}_0-\underline{A}_0)(1+M)}{2M}.$
\end{itemize}

\noindent This establishes the stated conditions.
\end{proofnolab} \smallskip

\ProofOf{Corollary \ref{corr:0inAset23}} 

\begin{proofnolab}
\noindent Assume $0\in[\underline{A}_0,\overline{A}_0]$ and $M>0$. Recall that in the $\mathcal{S}=\{0\}$ case,
$$\mathcal{I}^{A}_{\text{ATT}_1}
=
\Bigl[
\min\{\text{LB}(\underline{A}_0),\text{LB}(\overline{A}_0)\},\ 
\max\{\text{UB}(\underline{A}_0),\text{UB}(\overline{A}_0)\}
\Bigr],$$
where $\text{LB}(a)=\theta_1+a-M|\Delta_0-a|$ and $\text{UB}(a)=\theta_1+a+M|\Delta_0-a|$. Since $0\in[\underline{A}_0,\overline{A}_0]$, we have
$
\min\{\text{LB}(\underline{A}_0),\text{LB}(\overline{A}_0)\}
\leq \text{LB}(0)$ and $\max\{\text{UB}(\underline{A}_0),\text{UB}(\overline{A}_0)\}
\geq \text{UB}(0).
$ Hence,
$[\text{LB}(0), \text{UB}(0)] \subseteq \mathcal{I}_{\text{ATT}_{1}}^{A}.$ It follows that:
$$2M|\Delta_0| = \text{UB}(0) - \text{LB}(0)  \leq \max\{\text{UB}(\underline{A}_0),\text{UB}(\overline{A}_0)\} - \min\{\text{LB}(\underline{A}_0),\text{LB}(\overline{A}_0)\},$$
that is,
$2M |\Delta_0| \leq W_{PTAE}.$ Thus $W_{PTAE}<W_{PT}$ is impossible.
\end{proofnolab}

\subsection*{Appendix \ref{appendix:assumpnonzero}}

\ProofOf{Lemma \ref{lemma:nonzerokmultpretrends}}
\begin{proofnolab}
\noindent Suppose $K \geq 0$ and $M \geq 0$, and $k^{\star}_{s} \in [-K,K]$ for all $s \in \{-S, \ldots , 0\}$. To begin, I prove sufficiency:
$K < \frac{1}{1+2M} \; \Longrightarrow \; 1 - k_0 - m(k_r - k_{r-1}) \neq 0$ for all $((k_j)_{j \in J(r)}, m) \in \prod_{j \in J(r)}[-K,K] \times [-M,M]$. 

Suppose $K < \frac{1}{1+2M}$. Consider a feasible $((k_j)_{j \in J(r)}, m) \in \prod_{j \in J(r)}[-K,K] \times [-M,M]$. Then, we have $|k_0| \leq K$, $|m| \leq M$. Since $k_r, k_{r-1} \in [-K,K]$, we have:
$(k_r - k_{r-1}) \in [-2K, 2K] \quad \Longrightarrow \quad |k_r - k_{r-1}| \leq 2K.$
Then, we have:
\begin{align*}
    | k_0 + m(k_r - k_{r-1} ) | &\leq |k_0 | + |m (k_r - k_{r-1}) | \\
    &= |k_0| + |m| \cdot |k_r - k_{r-1} | \\
    &\leq K + M \cdot 2K \\
    &= K(1+2M) \\
    &< 1.
\end{align*}
The first line follows from the triangle inequality. The third and fifth lines follow by assumption. Therefore, $k_0 + m(k_r - k_{r-1}) \in (-1,1)$, which precludes $k_0 + m(k_r - k_{r-1}) = 1$. This in turn implies that
$1 - k_0 - m(k_r - k_{r-1}) \neq 0$
for all $((k_j)_{j \in J(r)}, m) \in \prod_{j \in J(r)}[-K,K] \times [-M,M]$. \smallskip

\noindent Next, I prove necessity:
$1 - k_0 - m(k_r - k_{r-1}) \neq 0 \; \Longrightarrow \; K < \frac{1}{1+2M} $
for all $((k_j)_{j \in J(r)}, m) \in \prod_{j \in J(r)}[-K,K] \times [-M,M]$.

I prove this direction via contrapositive. Suppose $K \geq 1/(1+2M)$. Consider the feasible tuple:
$$((k_j)_{j \in J(r)}, m) = \begin{cases} 
      (1/(1+2M), 1/(1+2M), -1/(1+2M), M) & \text{ if } r \neq 0 \\
     (1/(1+2M), -1/(1+2M), M) & \text{ if } r = 0
   \end{cases}$$

Then, we have:
\begin{align*}
    1 - k_0 - m(k_r - k_{r-1}) &= 1 - \frac{1}{1+2M} - M \left(\frac{1}{1+2M} + \frac{1}{1+2M} \right) \\
    &= 1 - \frac{1}{1+2M} - \frac{2M}{1+2M} \\
    &= \frac{1+2M}{1+2M} - \frac{1+2M}{1+2M} \\
    &= 0.
\end{align*}
Hence, if $K \geq 1/(1+2M)$, then a feasible tuple yields a zero denominator. Therefore, if the nonzero denominator condition holds, then it must be the case that $K < 1/(1+2M)$.
    
\end{proofnolab} \smallskip

\ProofOf{Footnote \ref{fn:kexample}}
\begin{proofnolab}
    Suppose $K \geq 0$ and $M \geq 0$, and $k^{\star}_{s} \in [0,K]$ for all $s \in \{-S, \ldots , 0 \}$. To begin, I prove sufficiency:
$K < \frac{1}{1+M} \; \Longrightarrow \; 1 - k_0 - m(k_r - k_{r-1}) \neq 0$
for all $((k_j)_{j \in J(r)}, m) \in \prod_{j \in J(r)}[0,K] \times [-M,M]$.

Suppose $K < 1/(1+M)$. Consider a feasible tuple $((k_j)_{j \in J(r)}, m) \in \prod_{j \in J(r)}[0,K] \times [-M,M]$. Since $k_r, k_{r-1} \in [0,K]$,
$k_r -k_{r-1} \in [-K,K] \; \Longrightarrow \quad |k_r - k_{r-1}| \leq K.$

Therefore,
\begin{align*}
    |k_0 + m(k_r - k_{r-1}) | &\leq |k_0| + |m (k_r - k_{r-1})| \\
    &= |k_0| + |m| \cdot |k_r - k_{r-1} | \\
    &\leq K + M \cdot K \\
    &= K(1+M) \\
    &< 1.
\end{align*}
The first line follows by the triangle inequality. The third and fifth lines follow by assumption. 

Therefore, $k_0 + m(k_r - k_{r-1}) \in (-1,1)$, which precludes $k_0 + m(k_r - k_{r-1}) = 1$. This in turn implies that
$1 - k_0 - m(k_r - k_{r-1}) \neq 0$
for all $((k_j)_{j \in J(r)}, m) \in \prod_{j \in J(r)}[0,K] \times [-M,M]$. \smallskip

\noindent Next, I prove necessity:
$1 - k_0 - m(k_r - k_{r-1}) \neq 0 \; \Longrightarrow \; K < \frac{1}{1+M}$
for all $((k_j)_{j \in J(r)}, m) \in \prod_{j \in J(r)}[0,K] \times [-M,M]$.

I prove this direction via contrapositive. Suppose $K \geq 1/(1+M)$. Consider the feasible tuple:
$$((k_j)_{j \in J(r)}, m) = \begin{cases} 
      (1/(1+M), 1/(1+M),0,M) & \text{ if } r \neq 0 \\
      (1/(1+M),0,M) & \text{ if } r = 0 
   \end{cases}$$

Then, we have:
\begin{align*}
    1 - k_0 - m(k_r - k_{r-1}) &= 1 - \frac{1}{1+M} - M\left( \frac{1}{1+M} \right) = \frac{1+M}{1+M} - \frac{1}{1+M} - \frac{M}{1+M} = 0.
\end{align*}
Hence, if $K \geq 1/(1+M)$, then a feasible tuple yields a zero denominator. Therefore, if the nonzero denominator condition holds, then it must be the case that $K < 1/(1+M)$. 

\end{proofnolab}

\subsection*{Appendix \ref{appendix:inference}}

\ProofOf{(\ref{eqn:simullowercredbands})}
\begin{proofnolab}
\noindent Consider a finite grid of anticipation sensitivity values $\mathcal{H} = \{ \boldsymbol{a}_j : j \in \mathcal{J} \}$ where $\mathcal{J}$ is a finite index set. For each $j \in \mathcal{J}$, let $M_j$ be a random variable that denotes the value of the breakdown frontier evaluated at $\boldsymbol{a}_j$. Let $D$ denote the random variable distributed according to the posterior $\Pi(\cdot \mid Z)$ defined as follows:
    $D \coloneqq \max_{j \in \mathcal{J}} \max \left\{ \frac{\hat{m}_j - M_j}{s_j}, 0   \right\},$
    and define $c^{D}_{1 - \alpha}$ as the $(1-\alpha)$-quantile of $D$:
    $c^{D}_{1-\alpha} \coloneqq  \inf \bigl\{ c \in \mathbb{R} : \Pi(D \leq c \mid Z) \geq 1-\alpha  \bigr\}.$
    The simultaneous lower credible band is defined as follows:
    $\hat{L}_{j} \equiv \hat{L}(\boldsymbol{a}_j) = \hat{m}_j - c^{D}_{1-\alpha} \cdot s_j, \quad j \in \mathcal{J}.$
    Since $D \geq 0$ almost surely, then $c^{D}_{1 - \alpha} \geq 0$. Define the event:
    $A \coloneqq \bigl\{ M_j \geq \hat{L}_j \text{ for all } j \in \mathcal{J}  \bigr\}.$
    Assuming $s_j > 0$ for each $j \in \mathcal{J}$:
    \begin{align*}
        &M_j \geq \hat{L}_j \Longleftrightarrow \quad M_j \geq \hat{m}_j - c^{D}_{1-\alpha} \cdot s_j \Longleftrightarrow \quad \hat{m}_j - M_j \leq c^{D}_{1-\alpha} \cdot s_j \Longleftrightarrow \quad \frac{\hat{m}_j - M_j}{s_j} \leq c^{D}_{1-\alpha}.
    \end{align*}
    Hence,
    $A \Longleftrightarrow \left\{ \frac{\hat{m}_j - M_j}{s_j} \leq c^{D}_{1-\alpha} \text{ for all } j \in \mathcal{J}  \right\} \Longleftrightarrow \left\{ \max_{j \in \mathcal{J}} \frac{\hat{m}_j - M_j}{s_j} \leq c^{D}_{1-\alpha}  \right\}.$
    Since $c^{D}_{1 - \alpha} \geq 0$, 
    $\left\{ \max_{j \in \mathcal{J}} \frac{\hat{m}_j - M_j}{s_j} \leq c^{D}_{1-\alpha}  \right\} \Longleftrightarrow \left\{ \max_{j \in \mathcal{J}} \max \left\{ \frac{\hat{m}_j - M_j}{s_j}, 0 \right\} \leq c^{D}_{1-\alpha}  \right\} \Longleftrightarrow \{ D \leq c^{D}_{1-\alpha} \}.$
    Therefore, we have:
    $\Pi(A \mid Z) = \Pi(D \leq c^{D}_{1-\alpha} \mid Z) \geq 1-\alpha,$
    where the last inequality follows by the definition of a quantile. If the posterior CDF of $D \mid Z$ is continuous at $c^{D}_{1-\alpha}$, then
    $\Pi(A \mid Z) = \Pi(D \leq c^{D}_{1-\alpha} \mid Z) = 1-\alpha.$
\end{proofnolab} \smallskip

\ProofOf{(\ref{eqn:credsets})}

\begin{proofnolab}
\noindent Define the event:
     $A \coloneqq \{ \mathcal{I}^{\boldsymbol{a}}_{\text{ATT}_{1}}(\boldsymbol{a}, M; P) \subseteq \hat{I}_{1-\alpha}(\boldsymbol{a}, M)  \}.$
     Let $C$ and $c^{C}_{1-\alpha}$ be defined as in Appendix \ref{subsec:credints}. Define: $\hat{I}(c) \coloneqq [\hat{L} - c, \hat{U} + c]$. Hence, $\hat{I}_{1-\alpha}(\boldsymbol{a}, M)  \equiv \hat{I}(c^{C}_{1-\alpha})$. 
      For any $c \geq 0$, $\mathcal{I}^{\boldsymbol{a}}_{\text{ATT}_{1}}(\boldsymbol{a}, M; P) \subseteq \hat{I}(c)$ is equivalent to $C \leq c$.  Therefore:
     $A \Longleftrightarrow \{ C \leq c^{C}_{1-\alpha} \}.$
     Hence,
     $\Pi(A \mid Z) = \Pi(C \leq c^{C}_{1-\alpha} \mid Z).$
     By the definition of $c^{C}_{1-\alpha}$ as the $(1-\alpha)$-quantile of $C \mid Z$, we have:
     $\Pi(C \leq c^{C}_{1-\alpha} \mid Z) \geq 1-\alpha.$
     If the posterior CDF of $C \mid Z$ is continuous at $c^{C}_{1-\alpha}$, then 
     $\Pi(C \leq c^{C}_{1-\alpha} \mid Z) = 1-\alpha.$ \end{proofnolab}

\subsection*{Appendix \ref{appendix:auxresults}}

\ProofOf{Lemma \ref{aeexpression}}

\begin{proofnolab}
     \noindent Suppose Assumptions \ref{initialize} and \ref{abounds} hold.  We have:
\begin{equation}\label{eqn:AsDefn}
    A_s = \varphi_s - \varphi_{s-1}.
\end{equation}

\noindent Re-arranging, 
 $\varphi_s = A_s + \varphi_{s-1}.$ Therefore, we can recursively express the anticipation effect at period $s$ as:
$\varphi_s 
    = A_s + A_{s-1} + \ldots + A_{-(S-1)} 
    + \varphi_{-S} = \sum_{j = -(S-1)}^{s} A_j 
    + \varphi_{-S}.$
By Assumption \ref{initialize}, this becomes:
$\varphi_{s} = \sum_{j = -(S-1)}^{s} A_j,$ where by Assumption \ref{abounds}, each $A_j$ term is contained in $[\underline{A}_{j}, \overline{A}_{j}]$ for $j \in \{ -(S-1), \ldots , 0 \}$.
\end{proofnolab} \smallskip

\ProofOf{Lemma \ref{separableresult}}

\begin{proofnolab}
\noindent For each $j \in \{1, \ldots , J\}$, let $x_{j}^{\text{min}} \in \arg\min_{x_j \in \mathcal{X}_j} f_j (x_j)$. This exists since we assume that the minimum is attained. Define the feasible vector:
    $\boldsymbol{x}^{\text{min}} \coloneqq (x_{1}^{\text{min}}, \ldots , x_{J}^{\text{min}})^{\prime} \in \mathcal{X}_{1} \times \ldots \times \mathcal{X}_{J} = \mathcal{X}.$
    Fix any $\boldsymbol{x} = (x_1, \ldots , x_J)^{\prime} \in \mathcal{X}$. For each component $j$,
    $f_j (x_j) \geq \min_{x_j \in \mathcal{X}_j} f_j (x_j) = f_j (x_{j}^{\text{min}}).$
    Summing over $j = 1, \ldots , J$ yields:
    $\sum_{j=1}^{J} f_j (x_j) \geq \sum_{j=1}^{J} f_j (x_{j}^{\text{min}}),$
    where the above expression holds for all $\boldsymbol{x} \in \mathcal{X}$. Hence, the RHS of the above inequality is a lower bound on the set:
    $\biggl\{ \sum_{j=1}^{J} f_j (x_j) : \boldsymbol{x} \in \mathcal{X} \biggr\}.$
    Therefore,
    \begin{equation}\label{eqn:separablefirstdirection}
        \min_{\boldsymbol{x} \in \mathcal{X}} \sum_{j=1}^{J} f_j (x_j) \geq  \sum_{j=1}^{J} f_j (x_{j}^{\text{min}}).
    \end{equation}
    However, also note that since $\boldsymbol{x}^{\text{min}} \in \mathcal{X}$, it is one of the feasible points in the minimization. Therefore, the minimum over $\mathcal{X}$ cannot exceed the objective evaluated at that point:
    \begin{equation}\label{eqn:separableseconddirection}
        \min_{\boldsymbol{x} \in \mathcal{X}} \sum_{j=1}^{J} f_j (x_j) \leq  \sum_{j=1}^{J} f_j (x_{j}^{\text{min}}).
    \end{equation}
    Combining (\ref{eqn:separablefirstdirection}) and (\ref{eqn:separableseconddirection}), we get:
    $\min_{\boldsymbol{x} \in \mathcal{X}} \sum_{j=1}^{J} f_j (x_j) =  \sum_{j=1}^{J} f_j (x_{j}^{\text{min}}) = \sum_{j=1}^{J} \min_{x_j \in \mathcal{X}_j} f_j (x_j),$
    as desired.  
    
    The maximization statement follows analogously by choosing $x_{j}^{\text{max}} \in \arg\max_{x_j \in \mathcal{X}_j} f_j (x_j)$.  \end{proofnolab} \smallskip

\singlespacing
\bibliographystyle{econometrica}
\bibliography{references}

\makeatletter\@input{mainaux.tex}\makeatother